\documentclass[a4paper,11pt]{article}
\pdfoutput=1

\usepackage{comment}
\usepackage{epsfig,amssymb,euscript,xspace,xcolor}
\usepackage{amsfonts,amssymb,amsmath,mathtools,empheq,amsthm,graphicx,paralist}
\usepackage{mathrsfs,float}  
\usepackage{pgf,tikz,pgfplots}
\usetikzlibrary{arrows}
\usepackage{tikz,caption,subcaption,marvosym} 
\usetikzlibrary{decorations.markings,arrows,snakes}
\usepackage[belowskip=-15pt,aboveskip=0pt]{caption}
\usepackage{enumitem} 
\usepackage{tcolorbox}
\usepackage{xcolor}
\usepackage{color}
\usepackage{framed}

\usepackage[utf8]{inputenc}

\definecolor{lightblue}{rgb}{.1,.4,.5}
\definecolor{brown1}{rgb}{.64,.43,.138}

\usepackage{jheppub}

\preprint{MPP-2025-154}

\date{\today}
\title{Mastering Cosmological Amplitudes Using Generalized Ramanujan's Theorem}

\begin{document}
\vspace*{0.1ex}
\centerline{\it ~} 
\author{Prashanth Raman,}
\author{Qinglin Yang}
%\author[a,e,f]{,\,Prashanth Raman}
%\author[a,d]{,\,Chi Zhang}
\affiliation[]{Max-Planck-Institut f\"{u}r Physik, %Werner-Heisenberg-Institut, 
Boltzmannstr. 8,
%}
%\centerline{\it~ 
85748 Garching, Germany.}

\emailAdd{praman@mpp.mpg.de}
\emailAdd{qlyang@mpp.mpg.de}
%\emailAdd{prashanthr@imsc.res.in}
%\emailAdd{zhangchi@itp.ac.cn}
\date{\today}
\abstract{We present a systematic method for computing cosmological amplitudes, including {\it in-in} correlators and wavefunction coefficients, in FRW spacetime. Specializing to cases with conformally-coupled external scalars and massive scalar exchanges, we introduce a decomposition into massive family trees, which capture the nested time structure common to these observables. We then evaluate these building blocks using the {\it Method of Brackets} (MoB), a multivariate extension of {\it Ramanujan’s master theorem} that operates directly on the integrand, translating integrals into discrete summations via a compact set of algebraic rules. This yields infinite series representations valid across the full space of external momenta and internal energies. We also develop Feynman-like diagrammatic rules that map interaction graphs to summand structures, enabling efficient and scalable computation. The resulting expressions make time evolution manifest, smoothly interpolate to the conformal limit, and are well suited for both numerical evaluation and analytic analysis of massive field effects in cosmology.}

\maketitle

\newpage

\section{Introduction}
Modern cosmology rests on the $\Lambda$CDM framework, which provides an empirically successful description of the universe’s evolution from the hot Big Bang to its present accelerated expansion \cite{Frieman:2008sn, 10.1093/ptep/ptac097}. Observations of the cosmic microwave background (CMB), large-scale structure, and the uniform expansion of the universe reveal that it is remarkably homogeneous and isotropic on large scales. These symmetries are naturally encoded in the Friedmann–Robertson–Walker (FRW) class of spacetimes, which are the most general cosmological solutions to Einstein’s equations consistent with spatial homogeneity and isotropy \cite{Planck:2018}. As such, FRW geometries form the foundational backdrop for virtually all cosmological modeling, from the earliest quantum fluctuations to the late-time cosmic acceleration~\cite{PeeblesRatra:2003}.

Despite the success of the standard model of cosmology, several foundational questions remain unresolved—among them the horizon, flatness, and monopole problems, as well as the origin of primordial density fluctuations. These issues are elegantly addressed by the paradigm of cosmic inflation, a period of accelerated expansion in the early universe~\cite{Guth:1980zm,  Linde:1981mu}. Inflation stretches quantum fluctuations to cosmic scales, imprinting a nearly scale-invariant spectrum that seeds large-scale structure \cite{Achucarro:2022qrl}. This process transforms the early universe into a natural high-energy laboratory for quantum fields evolving on a dynamical spacetime background. Since the inflationary epoch lies far beyond direct experimental access, its imprints—encoded in primordial correlation functions—provide a rare observational window into ultra-high-energy physics~\cite{Maldacena:2002vr, Baumann:2009ds, Chen:2010}.

In recent years, cosmological collider physics has emerged as a powerful framework for probing high-energy physics through the imprints of massive fields during the inflationary epoch~\cite{Chen:2012ge,Noumi:2012vr,Arkani-Hamed:2015bza,Chen:2016hrz,Chen:2016nrs,Chen:2016uwp,Lee:2016vti,An:2017hlx}. Central to this program is the study of cosmological correlation functions—commonly referred to as {\it cosmological correlators}—and the associated {\it wavefunction coefficients} of the universe, which encode rich information about primordial fluctuations and their interactions. These observables not only access energy scales far beyond those probed by colliders, but also provide an arena for studying quantum field theory in curved spacetime backgrounds \cite{Anninos:2014lwa,Stefanyszyn:2023qov,Penedones:2023uqc,Loparco:2023rug,Loparco:2023akg,Marolf:2010nz,Cespedes:2023aal, Melville:2024ove}.

While cosmological correlators encode this rich physics, the computation of cosmological observables remains a formidable challenge, even at tree level. Unlike flat-space scattering amplitudes—which are constrained by energy conservation, Lorentz invariance, and analyticity—cosmological correlators must be computed within the Schwinger–Keldysh or {\it in-in} formalism \cite{Schwinger, Keldysh:1964ud, Weinberg_2005, Calzetta_Hu_2008, Chen:2017ryl}. This requires performing nested-time integrals over the entire bulk evolution of the universe. The lack of time translation symmetry in cosmological spacetimes eliminates many simplifications familiar from flat space: correlators are sensitive to both growing and decaying modes and must be evaluated with full knowledge of the time-dependent background. Moreover, even at tree level, cosmological correlators exhibit nontrivial analytic structures—such as branch cuts—due to spontaneous particle production \cite{Myhrvold:1983hx, Mishima:1986in}. These features, which only arise at loop level in flat-space amplitudes, reflect the intrinsically non-equilibrium nature of cosmology. The situation becomes even more intricate when massive particles are exchanged: the resulting integrals involve Hankel functions of non-integer order and complicated nested-time orderings, making closed-form expressions difficult to obtain. At loop level, the situation becomes even more complex: the curvature of the background alters virtual propagation, infrared divergences can accumulate over long inflationary durations, the lack of global energy conservation obstructs traditional momentum-space cutting techniques, and secular growth can render some diagrams ill-defined unless properly resummed \cite{Cespedes:2023aal}. Various strategies have been developed to manage these difficulties. At tree level, approaches including bootstrapping correlators via boundary kinematic equations \cite{ Arkani-Hamed:2018kmz, Baumann:2019oyu, Pimentel:2022fsc, Jazayeri:2022kjy,Hogervorst:2021uvp,Baumann:2022jpr}, Mellin space reformulations \cite{Sleight:2019hfp, Sleight:2019mgd, Sleight:2020obc, Sleight:2021plv} , numerical evaluation methods \cite{Werth:2023pfl,Werth:2024aui, Pinol:2023oux},  dispersive integral techniques \cite{Liu:2024xyi} and partial Mellin-Barnes representations \cite{Xianyu:2023ytd}, off-shell cutting rules founded on unitarity and analyticity \cite{Wang:2022eop,DuasoPueyo:2023kyh,Goodhew:2020hob,Jazayeri:2021fvk,Melville:2021lst,Goodhew:2021oqg,Baumann:2021fxj,Meltzer:2021zin,DiPietro:2021sjt,Tong:2021wai} and spectral representations \cite{Xianyu:2022jwk,Werth:2024mjg} provide powerful computational frameworks. Dressing rules that relate loop level cosmological integrands to their flat-space counterparts have recently been proposed, suggesting the possibility of systematically mapping complex loop integrals into more tractable forms \cite{Chowdhury:2023arc,Chowdhury:2025ohm}. 

Simultaneously, the application of techniques from modern scattering amplitude theory has led to a deeper structural understanding of {\it cosmological amplitudes}. It is now appreciated that flat-space amplitudes appear as residues on the total energy pole of wavefunction coefficients, while factorization on partial energy poles reveals lower-point amplitudes and wavefunction coefficients \cite{Raju:2012zr}. These analytic structures—accessible through analytic continuation in external energies—suggest that many of the organizing principles of flat-space amplitudes can be imported into cosmology. Furthermore, the intersection of cosmological correlators and wavefunctions coefficients with scattering amplitude theory has led to novel geometric insights. {\it Positive Geometry} frameworks have been extended to cosmology, leading to the development of {\it cosmological polytopes} and the {\it Cosmohedron} for the wavefunction and the {\it Correlatron} for {\it in-in} correlators~\cite{Arkani-Hamed:2017fdk,Benincasa:2019vqr,Arkani-Hamed:2018bjr,Benincasa:2020aoj,Benincasa:2024leu,Arkani-Hamed:2024jbp,Figueiredo:2025daa}. These structures geometrize the analytic properties of cosmological amplitudes and offer new perspectives on their singularities, factorization, and recursion. In parallel, differential equation techniques have revealed that cosmological amplitudes often satisfy rich algebraic structures, and led to the {\it kinematic flow} picture with time emerging as an auxiliary parameter \cite{Arkani-Hamed:2023kig,Arkani-Hamed:2023bsv, Baumann:2024mvm,Baumann:2025qjx,Gasparotto:2024bku,Capuano:2025ehm,Glew:2025ypb}. In \cite{Chowdhury:2023arc} it was shown that even at two loops, in-in correlators in conformally-coupled $\phi^4$ theory retain the same transcendental structure as flat-space amplitudes, hinting at deeper simplicity in the loop-level cosmological regime and in de Sitter space, these integrals with $n$ interaction vertices reduce to polylogarithmic functions of weight $n$, making them amenable to tools from the theory of Feynman integrals, such as the symbol \cite{Hillman:2019wgh}. These approaches also complement recent advances in understanding cosmological amplitudes as special classes of Euler–Mellin integrals \cite{Matsubara-Heo:2023ylc}, connecting them to rich mathematical structures such as GKZ systems and twisted cohomology \cite{berkesch2014euler,De:2023xue,De:2024zic,Fevola:2024nzj,Chen:2024glu,Grimm:2025zhv}.
Together, these developments hint at a more unified, geometric understanding of perturbative cosmology, akin to the modern amplitudes program in flat space.

Motivated by these developments, in this work we revisit the computation of cosmological amplitudes with generic polynomial interactions of conformal scalars as external states and generic massive exchanges in FRW power-law backgrounds. The computation of cosmological amplitudes using the {\it in-in } formalism involves sums and differences of basic time integrals. Taking inspiration from the {\it family tree decomposition} proposed in \cite{Fan:2024iek} and the Euler-Mellin representation \cite{He:2024olr} for the conformal scalars, we propose an alternate basis for cosmological amplitudes with massive exchanges, which we call {\it massive family trees}. Despite being an over complete basis, each massive family tree admits a Mellin representation after trading the time integrals for some auxillary variables \cite{He:2024olr} with integrand being determined directly from the underlying graph, analogously to the Feynman parametrization of flat space Feynman integrals which involve the Symanzik graph polynomials \cite{ItzyksonZuber,Smirnov2004, Weinzierl2022}. 

We then employ a technique for evaluating Mellin transforms known as the Method of Brackets~\cite{Gonzalez:2008xm,Gonzalez:2010nm}. This empirical algorithm, originally developed for evaluating Mellin integrals, is well suited for cosmological applications due to its ability to directly act on Euler–Mellin-type integrals at the integrand level. It enables efficient series expansions for analysis across all physically relevant kinematic regions. Despite the nested-time structure intrinsic to cosmological amplitudes not being manifest in the Euler-Mellin representation, when the Method of Brackets is applied, this structure becomes manifest at level of the resulting series solutions, allowing for systematic generalizations to higher-order interactions and more complex massive exchanges. Using this approach, we obtain explicit series solutions for tree-level cosmological integrals that are also useful for numerical evaluations. Our approach offers a unified and practical method for computing the core ingredients of cosmological observables in the presence of massive fields.

\paragraph{Main results} The main results of the paper are summarized below:
\begin{itemize}
\item We introduce an alternative basis to decompose any tree-level {\it in-in} correlator or wavefunction coefficient, which we call {\it massive family trees} and these are listed in \eqref{eq:twositechain}-\eqref{mft4}. 
\item We give explicit series solutions for these basis elements at any multiplicity, generic polynomial interactions and on FRW backgrounds in the physical region $\{w_1>w_i,k_{i,j}\}$. We do this by giving Feynman rules for writing down the summand of the series solutions directly from the graph. The relevant equations are \eqref{eq:resultmass}-\eqref{eq:T111} and in \eqref{gr1}-\eqref{gr4}. These rules simplify when the tree graph is a $n$-site chain. The interested reader may see  for the expressions \eqref{gnsr}.

\item We provide several examples, to demonstrate how to get the full in-in correlators and wavefunction coefficients from family trees. For one massive exchange these results are in \eqref{sec:two site massive} and for two and three massive exchanges the results are given in Appendix \ref{app:results}.

\end{itemize}

\paragraph{Outline}The rest of our paper is organized as the follows. In the rest of this section, we present basic definitions and notations/conventions we use throughout this work and very briefly review the rules for evaluation of tree level correlators and wavefunction coefficients using the {\it in-in} formalism. We then review the Ramanujan's master theorem and method of brackets(MoB) in section \ref{sec:RMT}, with several inviting examples  that also appear in cosmological calculations. In section \ref{sec:massless correlators}, we will demonstrate that by using MoB for conformally-coupled scalars, the series solutions we obtain, agree with results previously obtained in the literature. In section, \ref{sec:massive correlators}, we will turn to generic massive exchanges, we propose massive family trees as a basis for cosmological amplitudes and argue why they admit a Euler-Mellin representation.  We then use the MoB to get series solutions for \emph{all} these integrals and finally we explicitly discuss our solution for the 2-site massive chain. We end with some open questions and future directions in section \ref{sec:summary}. The appendices \ref{app:Hardy}, \ref{sec:basic massive solution derivation} and \ref{app:results} contain more mathematical details about Ramanujan's master theorem and MoB provides a more solid, record explicit derivations of our main results and also more explicit MoB series solutions, for the building blocks of cosmological amplitudes  with up to 3-massive exchanges respectively.

\subsection{Review and Notations}
Before ending this section, we review some basic facts about cosmological perturbation theory for self interacting scalars, that we will need. For more details, readers can refer to \cite{Weinberg_2005, Chen:2017ryl}. We will begin with the following action with a massless scaler field $\varphi$ and a massive scalar field $\Phi$
\begin{equation}\label{eq:action}
    S[\Phi,\varphi]=-\int{\rm d}^dx \, {\rm d}\tau\sqrt{-g}\left(\frac12(\partial_\mu\Phi)^2+\frac12(\hat{m}^2{+}\xi R) \Phi^2+\frac12(\partial_\mu\varphi)^2+\frac12\xi R \varphi^2{+}\mathcal{L}_{int}[\Phi,\varphi]\right),
\end{equation}
with a FRW power-law metric background
\begin{equation}
{\rm d}s^2=a(\tau)^2(-{\rm d}\tau^2+{\rm d}x^2)=\left(\frac{\tau}{\tau_0}\right)^{2\rho}(-{\rm d}\tau^2+{\rm d}x^2).
\end{equation}
with $\tau_0$ being the renormalized time scale, $R$ being the Ricci tensor, Hubble parameter $H$ normalized to be $1$, and general polynomial interaction between two types of scalar fields
\begin{equation}    \mathcal{L}_{int}=\sum_{n=3}^{\infty}\mathcal{L}_n,\ \ \mathcal{L}_n=\sum_{k=0}^n\frac{\tilde{\lambda}_{n-j,j}}{j!(n{-}j)!}\Phi^j\varphi^{n{-}j}
\end{equation}
Choices of power $\rho$ in the metric account for different cosmological backgrounds, {\it e.g.} $\rho=-1$ for de Sitter spacetime, $\rho={-}1{+}\epsilon$ for inflation, and  $\rho=0$ for Minkowski spacetime, {\it etc.}. We will focus on the special case where $\xi$ is chosen as $\frac{d{-}1}{4d}$, referred to as conformal coupling. After a Weyl transformation $\varphi\to (a(\tau))^{\frac{1{-}d}{2}}\varphi$ and $\Phi\to (a(\tau))^{\frac{1{-}d}{2}}\Phi$, the action is conformally equivalent to that of a scalar theory in Minkowski spacetime with time dependent masses and couplings as
\begin{equation}\label{eq:actionp2}
    S[\varphi,\Phi]=-\int {\rm d}^dx \, {\rm d}\tau \left(\frac12 (\partial \varphi)^2+\frac12(\partial \Phi)^2+\frac12\mu(\tau)\Phi^2+\mathcal{L}_{int}[\varphi,\Phi,\tau]\right),
\end{equation}
where the mass of field $\Phi$ reads
\begin{equation}
    \mu(\tau)=a(\tau)^2\hat{m}^2
\end{equation} 
and the interaction couplings $\lambda_{n-j,j}(\tau)=\tilde\lambda_{n-j,j} (-\tau)^{q_n{-}1}$, with twist exponents  $q_n=(d{+}1{-}\frac{n(d{-}1)}{2})\rho{+}1$. Since massive states do not survive to the end of the inflation, we only focus on the processes with only conformally-coupled external states $\varphi$, with generic massive exchanges $\Phi$.

In cosmology, two kinds of {\it cosmological amplitudes}  are mainly considered, namely {\it wavefunction coefficients} $\psi_n({\bf k}_1,\cdots,{\bf k}_n) $ from the definition 
\begin{align}
    \log{\Psi[\hat\varphi] } = {-}i \sum_{n\ge 2}\tfrac{1}{n!} \int \prod\limits_{i=1}^n \left[\frac{d^d{\bf k}_i \, \hat{\varphi}({\bf k}_i)}{(2\pi)^d}\right] \psi_n({\bf k}_1,\cdots,{\bf k}_n) (2 \pi)^d \delta^{(d)}\left(\sum_i{\bf k}_i\right )
\end{align}
with the wavefunction of the universe as
\begin{equation}
    \Psi[\hat\varphi]=\int_{\varphi(\tau=-\infty)=0}^{\varphi(\tau=0)=\hat\varphi}\mathcal{D}\,\varphi \mathcal{D}\Phi\ e^{\text{i} \,  S[\varphi,\Phi]}, 
\end{equation}
and {\it correlators} $\langle \varphi_1\cdots\varphi_N\rangle$, formally defined as
\begin{equation}\langle\varphi(x_1)\cdots\varphi(x_n)\rangle=\frac{\int\mathcal{D}\varphi\prod_{i=1}^n\varphi(x_i)|\Psi[\varphi]|^2}{\int\mathcal{D}\varphi|\Psi[\varphi]|^2}.
\end{equation}
 Computation of these cosmological amplitudes is perturbatively done by using a Schwinger-Keldysh / {\it in-in} formalism \cite{ Schwinger, Keldysh:1964ud, Weinberg_2005, Calzetta_Hu_2008, Chen:2017ryl}. Analogously to flat space amplitudes, these also involve summing over different kinds of ``Feynman diagrams"  albeit with different rules due to the time-dependent background.
 
We will restrict ourselves to the cases where the Feynman diagram, denoted by $\mathcal{E}$, is a tree. For cases with loop topologies, extra integration for the loop momenta also need to be performed, and these also involve many subtleties, such as dealing with UV/IR divergences, finding symmetry-preserving regularization schemes and renomalization on a time dependent background which is beyond the scope of this paper.  

 We now directly provide Feynman rules for both the wavefunction coefficients and the {\it in-in} correlators. For wavefunction coefficients, a tree graph $\mathcal{E}$ is obtained by evaluating 
\begin{equation}    \psi_n(\mathbf{k}_1,\cdots,\mathbf{k}_n;\mathcal{E})=(-\text{i})^n\int_{-\infty}^0\prod_{v=1}^n[{\rm d}\tau_v\ \lambda_v(\tau_v) B(w_v,\tau_v)]\prod_{e\in\mathcal{E}}G_e(k_e,\tau_{v_e},\tau_{v_{e^\prime}}).
\end{equation}
where $\mathcal{E}$ is the topology of the diagram, $w_v=|\mathbf{k}_v|$ and $k_e=|\sum_e\mathbf{k}|$ are the energies flowing through a node or an edge. The {\it bulk-to-boundary} propagators are conformally-coupled states 
\begin{equation}\label{eq:btobound1}       B(w_1,\tau_1)=\text{e}^{\text{i}w_1\tau_1}
\end{equation}
and the {\it bulk-to-bulk} propagators for massive states is given by
\begin{equation}\label{eq:btobulk0}
\begin{aligned}   &G_e(k_e,\tau_{v_{e}},\tau_{v_{e^\prime}})=\frac{\pi}4\text{e}^{-\pi\nu}(\tau_1\tau_2)^{1/2}\left(\text{H}_{-\text{i}\nu}^{(2)}(-k\tau_1)\text{H}_{\text{i}\nu}^{(1)}(-k\tau_2)\theta(\tau_2{-}\tau_1)\right.\\
    &\phantom{aaaaaaaaaaaaaaaaaa}\left.{+}\text{H}_{-\text{i}\nu}^{(2)}(-k\tau_2)\text{H}_{\text{i}\nu}^{(1)}(-k\tau_1)\theta(\tau_1{-}\tau_2)+\text{i}\text{H}_{-\text{i}\nu}^{(2)}(-k\tau_2)\text{H}_{-\text{i}\nu}^{(2)}(-k\tau_1)\right)
\end{aligned}
\end{equation}
where $\nu=\sqrt{\hat{m}^2-\frac14}$, $\hat{m}^2=m^2{-}2$ related to the actual mass $m$ of the scalar $\Phi$, and $\text{H}^{(j)}_{\text{i}\nu}(x)$ with $j=1,2$ are first and second type Hankel functions respectively. They are solutions to the Klein-Gordon equation for a massive scalar field on FRW background. In particular, for the case when $\hat{m}{=}0$, {\it i.e.} $m^2{=}2$ or $\nu\to\frac{\text{i}}2$, which is called the conformally-coupled scalar, the mode functions degenerate to exponential functions. Another case of interest is the massless case which corresponds to $m^2{=}0$ or $\nu\to\frac{3~\text{i}}2$ for which the mode functions also degenerate into  exponential functions with  a linear prefactor. However, the results for massless scalars can be translated to those for conformal scalars by the use of  weight shifting operators \cite{Baumann:2019oyu}. Therefore in this work, we will frequently refer to conformally-coupled scalars as ``massless" as well for simplicity.

In this work we will restrict ourselves to conformally-coupled scalars as external states as alluded to earlier. We also recall that $\lambda_{v}(\tau_v)=\tilde{\lambda}_{n-j,j}(-\tau_v)^{q_v{-}1}$, and $q_v$ stands for the twist on node $v$. We can set $\tilde{\lambda}_{n{-}j,j}:=1$ since they only contribute as overall factors. For our analysis, we will assume $q_v$ are general parameters.

For the cosmological correlators, each tree-level graph $\mathcal{E}$ is computed by evaluating 
\begin{align}\label{eq:correlators}    \mathcal{T}_n(\mathbf{k}_1,\cdots,\mathbf{k}_n;\mathcal{E})=({-}\text{i})^n\sum_{a_v=\pm} \int_{-\infty}^0\prod_{v=1}^n[{\rm d}\tau_v\,  a_v\ \lambda_v(\tau_v) D_{a_v}(w_v,\tau_v)]\prod_{e\in\mathcal{E}}D^{\nu_e}_{a_{v_e}a_{v_{e^\prime}}}(k_e,\tau_{v_e},\tau_{v_{e^\prime}}),
\end{align}
using the {\it in-in} formalism. The bulk-to-boundary propagators  and
bulk-to-bulk propagators read \cite{Chen:2017ryl} 
\begin{equation}\label{eq:btobound2}
    D_\pm(w,\tau):=\text{e}^{\pm \text{i}w\tau}
\end{equation}
and
\begin{align}
    &D^{\nu}_{-+}(k,\tau_1,\tau_2):=\frac\pi{4}e^{-\pi \nu}(\tau_1\tau_2)^{1/2}\text{H}_{-\text{i}\nu}^{(2)}(-k\tau_2)\text{H}_{\text{i}\nu}^{(1)}(-k\tau_1),\ D_{+-}^{\nu}(k,\tau_1,\tau_2)=(D^{\nu}_{-+}(k,\tau_1,\tau_2))^\star\label{eq:btobulk1}\\
    &D^{\nu}_{\pm\pm}(k,\tau_1,\tau_2):=D^{\nu}_{\pm\mp}(k,\tau_1,\tau_2)\theta(\tau_2{-}\tau_1)+D^{\nu}_{\mp\pm}(k,\tau_1,\tau_2)\theta(\tau_1{-}\tau_2)\label{eq:btobulk2}
\end{align}

Note that the cosmological amplitudes i.e., both the wavefunction coefficients and the {\it in-in} correlators involve sums and differences of integrals with fixed time ordering (given by a product of theta functions) and consisting of products of exponentials and Hankel functions. We will call these building blocks {\it cosmological integrals}.
These are in general A-hypergeometric functions \cite{berkesch2014euler} which are solutions to GKZ differential equations \cite{gelfand_hypergeometric_1989}.
The main goal of this work is to show that these integrals can be evaluated  efficiently using the method of brackets (MoB) \cite{Gonzalez:2008xm, Gonzalez:2010nm} which we will review in the next section.

\section{Ramanujan's master theorem and Method of Brackets}
\label{sec:RMT}
The Method of brackets (MoB) is a method to evaluate multivariate Mellin integrals \cite{Gonzalez:2008xm,Gonzalez:2010nm}. The method is an extension of Ramanujan's Master theorem (RMT) to several variables and is also related to the negative dimensional integration method \cite{NDIM1,NDIM2,NDIM3}. The method has already found several applications in physics, such as for the evaluation of Feynman integrals \cite{Gonzalez:2011nq, Gonzalez:2015msa, Ananthanarayan:2021not, Ghosh:2022net}. We shall first describe the RMT and then its generalization the MoB.
\subsection{Ramanujan's master theorem}
The Ramanujan's master theorem gives us a way to evaluate the Mellin transform of functions by Taylor expanding the integrand around the origin. We know that almost always such a procedure leads to divergent results and does not directly work without some sort of analytic continuation. The RMT gives us the result by an analytic continuation of the Taylor coefficients of the expansion of the function around the origin $x=0$.
\begin{tcolorbox}
\noindent {\bf Ramanujan's master theorem:} Consider, a single variable function which has a Taylor expansion around $x=0$ given by $f(x)=\sum\limits_{n=0}^\infty \frac{\left(-x\right)^n}{n!} a(n)$, with $a(0)\neq 0$. Then under some growth conditions on $a(n)$ we have
\begin{eqnarray}
 \mathcal{M} \left\{ f \right\} \left(s\right)&=&\int_{0}^{\infty}dx~ x^{s-1}~ f(x), \nonumber \\
 &=&~a(-s)~\Gamma\left(s\right) , \quad {\rm for} \,\, Re(s) >0
\end{eqnarray}
where $a(-s)$ is interpreted as a {\it natural} analytic continuation of the sequence $\{a(n)\}_{n \in \mathbb{N}}$ to complex values $s \in \mathbb{C}$.
\end{tcolorbox}
\noindent Conditions on the growth of the Taylor coefficients are needed to ensure uniqueness of  the analytic continuation, and this assumption is required for the theorem to hold. As an example,if $f(x)= e^{-x}$ then we have $a(n) = 1 , \forall n ~\in \mathbb{N}$ which can be analytically continued trivially to $a(s)=1, \forall s ~\in \mathbb{C}$  and we indeed have $\mathcal{M} \left\{ e^{-x} \right\} \left(s\right)= \Gamma(s)$.

But, we could also have chosen $a_1(s)= \cos{ (2 \pi s)}, \forall s ~\in \mathbb{C}$ to be the analytic continuation of $a(n)=1, \forall n ~\in \mathbb{N}$ and this would have given us the incorrect result for  $\mathcal{M} \left\{ e^{-x} \right\} \left(s\right)$. It is well a known result, called {\it Carlson's theorem} that when one imposes certain growth conditions on $a(n)$ then uniqueness of analytic continuation can be ensured,. For more details see Appendix.\ref{app:Hardy} and also \cite{ZB}.
 
Let us now outline the procedure for using the RMT in a convenient form below \cite{Gonzalez:2008xm, Gonzalez:2010nm}. The basic idea involves assigning a object called a {\it bracket} $\langle a \rangle$ to any parameter $a$, which is inspired by the RMT. It is a symbol associated with a divergent integral \[\langle a\rangle =\int_{0}^{\infty}dx ~ x^{a-1}.\] The formal rules for operating with these brackets is described below.
Let us, consider
\begin{equation}
I=\int_{0}^{\infty} x^{s-1} f(x) ,
\end{equation}
with $f(x)= \sum_{n=0}^{\infty} \frac{(-1)^n}{n!} a_n x^{r\, n}$, $r\in\mathbb{R}$, then 
\begin{itemize}
    \item {\bf Rule 1:} We shall formally assign a {\it Bracket series} to the integral $I$ as follows 
   \begin{equation}\label{rmt2p}
    \int_{0}^{\infty} \sum_{n=0}^{\infty} a(n) x^{r \, n+s-1} = \sum_{n=0}^{\infty} a(n) \langle r\, n+s \rangle.
 \end{equation}
 \item {\bf Rule 2:} A bracket series is assigned a value 

\begin{equation} \label{rmt2}
\sum_{n} \phi_n a(n) \langle r \, n+s \rangle = \frac{1}{|r|} a(-n^{*}) \Gamma(-n^{*}).
\end{equation}
where, $\phi_n= \frac{(-1)^n}{n!}$ is the {\it indicator} of the index $n$ and $n^{*}$ is the solution of the linear equation in the bracket $r \, n +s=0$ which is called the {\it bracket equation}. 
\end{itemize}  

\noindent The generalization to several variables is called the method of brackets (MoB). In what follows we shall denote by $\phi_{m_1,\cdots,m_n}= \phi_{m_1}\cdots \phi_{m_n}$ the multi-index indicator. We abbreviate this to $\phi_{\{m\}}$ for brevity and $\{m\}$ is understood to be the collection of all the indices that appear in the respective summations. 
\subsection{Rules for using the MoB} 
We now mention the simple rules for using MoB below. Let us consider the Mellin integral.
\begin{equation}
I=\int_{\mathbb{R}_{+}^n} \prod_{i=1}^n  d x_i \, x_i^{\alpha_i-1} \,f(x_1,\cdots,x_n)
\end{equation}
where $f(x_1,\cdots,x_n)=\prod_{k=1}^tf_k(x_1,\cdots,x_n)$ is a product or composition of multivariate functions $f_k$.
\begin{itemize}
\item {\bf Rule 1:} As we did in the single variable case, we expand the integrand around the origin. This would need multiple intermediate steps to expand each $f_k(x_1,\cdots,x_n)$ around origin as
\begin{equation}
f_k(x_1,\cdots,x_n)=\sum_{m_{j,k}=0}^\infty \,\, \phi_{\{m\}}\,\, a_k(\{m\}) \,\,\prod_{i=1}^n x_i^{\beta_{i,k} m_{i,k}}
\end{equation}
and the expansion finally formally looks like 
\begin{equation}
f(x_1,\cdots,x_n)=\sum_{m_{j,k}=0}^\infty \,\, \phi_{\{m\}}\,\, \prod_{k=1}^ta_k(\{m\}) \,\,\prod_{i=1}^n x_i^{\beta_{i,k} m_{i,k}},
\end{equation}
with $\beta_{ij}\in\mathbb{R}$, $\phi_\{m\}$ is the multi-index indicator, and $a\{m\}$ the coefficients that depend on all indices $m_{j,k}$.  
\item {\bf Rule 2:} As an intermediate step, if we encounter a term that needs to be multinomially expanded then it is convenient to use the following rule
\begin{equation}\label{eq:bracketrule2}
(a_1+\cdots+a_k)^{-\alpha}=\sum_{n_1=0}^{\infty} \sum_{n_2=0}^{\infty}\cdots\sum_{n_k=0}^{\infty}\phi_{n_1,\dots,n_k} \,a_1^{n_1}\cdots a_k^{n_k} \frac{\langle\alpha+n_1+\cdots n_k \rangle}{\Gamma(\alpha)}
\end{equation}
Here the bracket $\langle\alpha+n_1+\cdots n_k \rangle$ has the meaning as the single variable case, and $a_i$ can be furthermore functions of integration variables $x_i$.

\item {\bf Rule 3:} Let us suppose, after fully expanding the integrand around origin and gathering the power of all $x_i$, we have $r$ indices $m_i$ with $1\leq i\leq r$ to sum over, and $x_i$ has a power of $\beta_{i1} m_1+\cdots+\beta_{ir} m_r$. Furthermore, suppose during that expansion we introduce $s-n$ brackets from \eqref{eq:bracketrule2} as $\prod_{j=n+1}^s \langle \beta_{j1} m_1+\cdots+\beta_{jr} m_r +\alpha_j\rangle$.  We then replace the integration of $x_i$ by assigning formally a {\it bracket series} as
    \begin{align}\label{braseries}
    &\int_{\mathbb{R}_{+}^n} \prod_{i=1}^n  d x_i \sum_{m_1,\dots,m_r=0}^\infty \phi_{\{m\}}\,\, a(\{m\})  \,\,\prod_{i=1}^n x_i^{\beta_{i1} m_1+\cdots+\beta_{ir} m_r+\alpha_i-1}   \prod_{j=n+1}^s \langle \beta_{j1} m_1{+}\cdots{+}\beta_{jr} m_r {+}\alpha_j\rangle \nonumber \\
    &=
\sum_{m_1,\dots,m_r=0}^\infty \phi_{\{m\}}\,\, a(\{m\}) \, \prod_{j=1}^s\langle \beta_{j1}m_1+\cdots \beta_{jr}m_r+\alpha_j\rangle 
      \end{align}
We define $\mathbf{B}=\{\beta\}_{i,j}$, $\vec{m}=\{m_1,\cdots,m_r\}^T$ and $\vec{\alpha}=\{\alpha_1,\cdots,\alpha_r\}^T$ and we call the linear system $\mathbf{B}.\vec{m}=-\vec{\alpha}$ to be the {\it bracket equations}.
\item {\bf Rule 4:} In general, we have $s$ linear equations in $r$ variables. We define the {\it rank } = $r-s$, as the difference between the number of sums and the number of brackets. 
\noindent \subitem {\bf  rank $=0$ \,:} \,  In this case, the bracket equations have a unique solution which we call $\{m^*\}$, and we assign the following value to the bracket series in eq (2.3)
\begin{equation} \label{rmt3}
I=\frac{1}{|det \, \mathbf{B}|}a(\{-m^*\}) \Gamma(-m_1^*) \cdots \Gamma(-m_r^*)
\end{equation} 
\noindent \subitem {\bf  rank $>0$ \,:} In this case, the bracket equations do not have a unique solution. We can solve them by choosing a subset of free indices, which in this work is always denoted as $\sigma \subset \{1,\cdots,r\}$ with $|\sigma|=r{-}s$, and $\bar{\sigma}=\{1,\dots,r\}~ \backslash ~\sigma$ is the complement of $\sigma$ containing all solved indices. We denote the solutions of the bracket equations with the choice of free indices $\sigma$ by $m_i^\star$ for $i\in{\bar{\sigma}}$. There are several  namely $\left( {r \atop s} \right)$ choices for set $\sigma$. We have to consider all of them. For each of these choices, we get a $r-s$ fold series, called the {\it basis series} defined as follows:
\begin{equation}\label{rmt4}
I_{\sigma}= \frac{1}{|det ~\mathbf{B}_{\sigma}|}\sum_{i \in \sigma} \phi_{\{m_{\sigma}\}}\, a(\{m_{\bar{\sigma}}^*\}) \prod_{i \in \bar{\sigma}} \Gamma(-m_{i}^*)
\end{equation}
where, and $\mathbf{B}_{\sigma}$ is the submatrix of $\mathbf{B}$ with the columns labeled by $\sigma$ removed. \\

The basis series \eqref{rmt4} for the cosmological integrals we shall consider generally look like 
\begin{equation}
    I_{\sigma}=\sum_{m_\sigma}\phi_{\{m\}}\, b(\{m\}) \prod_{i\in\sigma} A_{i,\sigma}^{m_i}
\end{equation}
where $A_{i,\sigma}$ are dimensionless ratios of variables that $I_\sigma$ depends on, and $b(\{m\})$ are coefficient of the power series. 

\item In practice there are several regions corresponding to the different ranges the parameters $\{A_{i,\sigma}\}$ can take.

The final rule is that all the basis series that converge in a given region  have to be added up to get a series solution valid for that range of parameters. The series which diverge are discarded.

For almost all the cases we shall encounter in this work, the region where a basis series $I_{\sigma}$ converges corresponds directly to  $\{|A_{i,\sigma}|<1 \}$ \footnote{However, this is not necessarily true in general as if the coefficients $b(\{m\})$ grow rapidly then the series could be divergent see \eqref{ser31} and \eqref{ser32} in the next subsection for an example of this.}. This allows us to reorganize our computation more efficiently. Since we can apriori figure out which solutions to the bracket equations would contribute to a given region by directly examining the exponents of the $\prod_{i\in\sigma} A_{i,\sigma}^{m_i}$ before we solve the bracket equations for all allowed cases. This proves to be very useful for us indeed as we shall see later.

\end{itemize}
\subsection{Simple applications for MoB}
In this subsection, we present some simple examples of MoB, whose results will also be important for the following discussion of cosmological amplitudes.
\paragraph{Massless contact functions}
The simplest example, which we already looked at earlier corresponds to the Gamma function, is the contact functions for conformally-coupled cosmological amplitudes.  
\begin{equation}\label{eq:contactmassless}
    (-\text{i})\int_0^{\infty}{\rm d}\tau\ {\rm e}^{-\text{i}\omega\tau}\tau^{q{-}1}=\frac{-\text{i}}{(\text{i}\omega)^q}\Gamma(q)
\end{equation}
The integral can be trivially performed and converges to the right hand side when $\Re(q)>0$ and $\Im(\omega)<0$. Here, we instead adopt MoB to derive its result as our first exercise. Following the rules described in the previous section, we expand the integrand at $\tau=0$ as
\begin{equation}
    (-\text{i})\int_0^{\infty}{\rm d}\tau\sum_n\phi_n(\text{i}\omega)^n\tau^{n{+}q{-}1}
\end{equation}
This gives us following \eqref{braseries} the bracket series
\begin{equation}
    (-\text{i})\int_0^{\infty}{\rm d}\tau\ {\rm e}^{-\text{i}\omega\tau}\tau^{q{-}1}= \sum_{n=0}^{\infty}\phi_n(\text{i}\omega)^n\langle n{+}q\rangle
\end{equation}
with a single bracket equation $n+q=0$ and plugging in the solution of the equation into \eqref{rmt3} gives us $\frac{-\text{i}}{(\text{i}\omega)^q}\Gamma(q)$.

\paragraph{Massive contact functions}
In our later discussions about cosmological amplitudes, we will encounter the following integral
\begin{equation}
    \int_0^{\infty}{\rm d}\tau\ {\rm e}^{-\text{i}w \tau}\tau^{q-1}J_\nu(k\tau)
\end{equation}
which is closely related to the contact functions for massive scalar propagators, where $J_{v}(x)$ is the Bessel function of the first kind. Following the procedure of MoB, we expand both the exponential  and Bessel functions at $\tau=0$ 
\begin{align}
&{\rm e}^{-\text{i}w \tau} \tau^{q-1} =\sum_{n_1}\phi_{n_1} (\text{i}w)^{n_1} \tau^{n+q-1}\nonumber\\
&J_\nu(k \tau)=\sum_{n_2}\phi_{n_2}\frac1{\Gamma(n_2{+}\nu{+}1)}\left(\frac 
{k \tau}{2}\right)^{2n_2{+}\nu} 
\end{align}
and we obtain the following by taking a product of the expansions and assigning a bracket series following \eqref{braseries}
\begin{equation}    \int_0^{\infty}{\rm d}\tau\sum_{n_1,n_2}\phi_{n_1,n_2}\frac{(\text{i}w)^{n_1}(k/2)^{2n_2{+\nu}}}{\Gamma(n_2{+}\nu{+}1)}\tau^{ 2n_2{+}n_1{+}\nu{+}q{-}1}  = \sum_{n_1,n_2}\phi_{n_1,n_2}\frac{(\text{i}w)^{n_1}(k/2)^{2n_2{+\nu}}}{\Gamma(n_2{+}\nu{+}1)}\langle 2n_2{+}n_1{+}\nu{+}q\rangle  \, ,
\end{equation}
Now we have two choices for solving the bracket equation $2n_2{+}n_1{+}\nu{+}q=0$ either $n^*_1=-2 n_2{-}\nu{-}q$ or $n^*_2=\frac{-n_1{-}\nu{-}q}{2}$ which keeps $n_2$ or $n_1$ free respectively. Using \eqref{rmt4} gives us the two basis series with 1-fold sums as
\begin{align}\label{eq:massivecontact1}
    \sum_{n_2}\phi_{n_2}\Gamma(-n^*_1)\frac{(\text{i}w)^{n^*_1}(k/2)^{2n_2{+\nu}}}{\Gamma(n_2{+}\nu{+}1)}&=\frac{2^{{-}\nu}k^\nu}{(\text{i}w)^{q{+}\nu}}\sum_{n}\frac{2^{-2n}\ \Gamma(2n{+}q{+}\nu)}{n!\Gamma(n{+}\nu{+}1)}\left(\frac{k}{w}\right)^{2n}
\end{align}
with $n_2$ free, which converges for $|\frac kw|<1$, and
\begin{align}
    &{\frac12}\sum_{n_1}\phi_{n_1}\Gamma(-n^*_2)\frac{(\text{i}w)^{n_1}(k/2)^{2n^*_2{+\nu}}}{\Gamma(n_2^*{+}\nu{+}1)}=\frac12\left(\frac{2}{k}\right)^q\sum_{n}\frac{(-1)^n}{n!}\frac{\Gamma(\frac12(n{+}q{+}\nu))}{\Gamma(\frac12(2{-}n{-}q{+}\nu))}\left(\frac{\text{i}w}{k}\right)^n
\end{align}
with $n_1$ free, and this series expression works for the region $|k/w|>1$.
The sums in the above expressions can  be performed to rewrite the result in terms of Gauss hypergeometric functions:\\
\begin{equation}
\begin{cases}
 g_1(q,\nu,k,w)= \frac{2^{{-}\nu}k^\nu}{(\text{i}w)^{q{+}\nu}}\frac{\Gamma(q{+}\nu)}{\nu!} \ _2F_1\left(\frac12(q{+}\nu),\frac12(q{+}\nu{+}1),1{+}\nu,\frac{k^2}{w^2}\right) & ~\rm{for}~ |\frac{k}{w}|<1 \nonumber  \\
  g_2(q,\nu,k,w)= h(q,1,\nu,k,w)- \text{i}w \,h(q+1,3,\nu,k,w) & ~\rm{for}~ |\frac{w}{k}|>1 \nonumber
\end{cases}
\end{equation}
with 
\[h(q,j,\nu,k,w)=\frac12\left(\frac{2}{k}\right)^q\frac{\Gamma(\frac{q{+}\nu}2)}{\Gamma(\frac{2{-}q{+}\nu}2)}\ _2F_1\left(\frac{q{-}\nu}2,\frac{q{+}\nu}2,\frac j2,\frac{w^2}{k^2}\right).\] 

In the context of cosmology, $w=\sum_i|\mathbf{k}_i|$ of all momentum $\mathbf{k}_i$ flowing through bulk-to-boundary propagators, while $k=|\sum_i\mathbf{k}_i|$ is the energy on the corresponding edge. Therefore, we always have $w>k$ in the physical region since these have to satisfy the triangle inequality, and \eqref{eq:massivecontact1} will be important for us in Sec.\ref{sec:two site massive}. For convenience, it is worth mentioning that by the two identities for hypergeometric functions
\begin{align}
    &\text{(Pfaff transformation)}\ \ \phantom{aaaa}_2F_1(a,b,c,z)=(1{-}z)^{-a}\ _2F_1\left(a,c{-}b,c,\frac{z}{z{-}1}\right)\nonumber\\
    &\text{(Quadratic transformation)}\ \ _2F_1(a,b,2b,z)=(1{-}z)^{-\frac{a}2}\ _2F_1\left(\frac{a}2,b{-}\frac{a}2,b{+}\frac12,\frac{z^2}{4z{-}4}\right)\nonumber
\end{align}
it can be shown that $g_1$ function has two equivalent expressions as
\begin{equation}\label{eq:masscontact01}
    g_1(q,\nu,k,w)=\frac{\Gamma(q{+}\nu)}{(\text{i}(w{+}k))^{q{+}\nu}\nu!}\left(\frac{k}2\right)^\nu\ _2F_1\left(q{+}\nu,\frac12{+}\nu,1{+}2\nu,\frac{2k}{k{+}w}\right)
\end{equation}
and
\begin{equation}\label{eq:masscontact02}
    g_1(q,\nu,k,w)=\frac{\Gamma(q{+}\nu)}{(\text{i}(w{-}k))^{q{+}\nu}\nu!}\left(\frac{k}2\right)^\nu\ _2F_1\left(q{+}\nu,\frac12{+}\nu,1{+}2\nu,\frac{2k}{k{-}w}\right)
\end{equation}

We end the discussion of this example by giving the explicit result for the simplest massive contact function
\begin{equation}\label{eq:massivecontact2}
\begin{aligned}
  \mathcal{F}^{(j)}(p,w,k,\nu)&:=(-\text{i})\int_{-\infty}^0{\rm d}\tau\ {\rm e}^{\text{i}w \tau}(-\tau)^{p-1}\text{H}^{(j)}_{(-1)^{j-1}~\text{i}\nu}(-k\tau)\\
  &=(-\text{i})\times \frac{(-1)^j}{\sinh(\pi \nu)}\left(g_1(p,(-1)^j~\text{i} \nu,k,w)-{\rm e}^{\pi \nu}\,g_1(p,(-1)^{j-1}~\text{i} \nu,k,w)\right)  
\end{aligned}
\end{equation}    
where \[\text{H}_\nu^{(j)}(x)=(-1)^{j}\frac{J_{-\nu}(x){-}{\rm e}^{\scriptsize{(-1)^j}~\text{i}\pi \nu}J_\nu(x)}{\sin(\pi \nu)}\] for $j=1,2$ are the Hankel functions of first and second kind respectively.

\paragraph{Euler-Mellin representation for Hankel functions}
We consider the following definite integral
\begin{equation}\label{eq:0}
    \int_0^{\infty}{\rm d}s\ {\rm e}^{-2\text{i}xs}(s(1+s))^{\mu-\frac12},
\end{equation}
which is convergent in region $|\Re(\mu)|<\frac12$. Following the basic rules \eqref{rmt2p} and \eqref{eq:bracketrule2} of MoB, we have
\begin{align}
    &\int_0^{\infty}{\rm d}s\ {\rm e}^{-2\text{i}xs}(s(1+s))^{\mu-\frac12}\nonumber\\&=\int_0^{\infty}{\rm d}s\left(\sum_{n_1}\phi_{n_1}\frac{(2\text{i}xs)^{n_1}}{n_1!}\right)\  s^{\mu-\frac12}\ \left(\sum_{n_2,n_3}\phi_{n_2,n_3}\frac{1^{n_2}s^{n_3}}{\Gamma({-}\mu{+}\frac12)}\right)\langle {-}\mu{+}\frac12{+}n_2{+}n_3\rangle\nonumber\\    &=\sum_{n_1,n_2,n_3}\frac{(2\text{i}x)^{n_1}}{\Gamma({-}\mu{+}\frac12)}\langle {-}\mu{+}\frac12{+}n_2{+}n_3\rangle\langle \mu{+}\frac12{+}n_1{+}n_3\rangle
\end{align}
In the first line, we expand the exponential function directly by its series expression, and expand factor $(1{+}s)^{\mu{-}\frac12}$ by bracket relation \eqref{eq:bracketrule2}, which introduces three indices $n_1,n_2,n_3$ and one bracket equation for $n_2$ and $n_3$. In the second line, we collect power of $s$ as $s^{\mu{-}\frac12{+}n_1{+}n_3}$ and turn it also into a bracket equation by bracket rule \eqref{rmt2p}.

Secondly, we solve the bracket equations. We have two equations depending on three indices, thus any two indices can be expressed by linear function of the third index. Correspondingly, we have three summand for this integral, whose bracket solutions are
\begin{align}    
     &\text{Series 1}: (n_1,n_3) ,\ \ n_1^*=n_2{-}2\mu,\ n_3^*={-}\frac12{-}n_2{+}\mu  \\
      &\text{Series 2}: (n_2,n_3) ,\ \ n_2^*=n_1{+}2\mu,\ n^*_3={-}\frac12{-}n_1{-}\mu \\
      &\text{Series 3}: (n_1,n_2) ,\ \ n^*_1={-}\frac12{-}n_3{-}\mu,\ n^*_2={-}\frac12{-}n_3{+}\mu 
\end{align}
and corresponding summand are 
\begin{align}   
     &\text{Series 1}: \ \ \sum_{n_3}\phi_{n_2}\frac{\Gamma(-n_1^*)\Gamma(-n_3^*)(2\text{i}x)^{n_1^*}}{\Gamma(-\mu{+}\frac12)}=\sum_n\phi_n\frac{(2\text{i}x)^{n{-}2\mu}\ \Gamma(\frac12{+}n{-}\mu)\Gamma({-}n{+}2\mu)}{\Gamma(-\mu{+}\frac12)}  \\
      &\text{Series 2}: \ \ \sum_{n_1}\phi_{n_1}\frac{\Gamma(-n_2^*)\Gamma(-n_3^*)(2\text{i}x)^{n_1}}{\Gamma(-\mu{+}\frac12)}=\sum_n\phi_n \frac{(2\text{i}x)^{n}\Gamma(\frac12{+}n{+}\mu)\Gamma({-}n{-}2\mu)}{\Gamma(-\mu{+}\frac12)}\\
      &\text{Series 3}: \ \ \sum_{n_3}\phi_{n_3}\frac{\Gamma(-n_1^*)\Gamma(-n_2^*)(2\text{i}x)^{n_1^*}}{\Gamma(-\mu{+}\frac12)}=\sum_n\phi_n\frac{(2\text{i}x)^{-\frac12{-}\mu{-}n}\ \Gamma(\frac12{+}n{-}\mu)\Gamma(\frac12{+}n{+}\mu)}{\Gamma(-\mu{+}\frac12)}  \label{ser31}
\end{align}
All the 3 basis series are one-fold infinite sums. As power series, the Series 1 and 2 have the same argument $x$. Therefore, we add up these two series as result in $|x|<1$ as
\begin{align}\label{eq:1p}
    &\sum_n\phi_n\frac{(2\text{i}x)^{n{-}2\mu}\ \Gamma(\frac12{+}n{-}\mu)\Gamma({-}n{+}2\mu)}{\Gamma(-\mu{+}\frac12)} + \sum_n\phi_n \frac{(2\text{i}x)^{n}\Gamma(\frac12{+}n{+}\mu)\Gamma({-}n{-}2\mu)}{\Gamma(-\mu{+}\frac12)}\nonumber\\
    &=\frac{{\rm e}^{-\text{i}\pi \mu}{\rm e}^{ix}}{(8x)^\mu}\Gamma(1{-}\mu)\Gamma(2\mu)J_{-\mu}(x){-}\frac{{\rm e}^{ix}}{(8x)^\mu}\Gamma(1{-}\mu)\Gamma(2\mu)J_{\mu}(x)\ \ \ (\text{if}\ |\Re(\mu)|<\frac12)\nonumber\\
    &=\left\{\begin{aligned}
        &-\frac {\text{i}}2\frac{\sqrt{\pi}\ \Gamma(\mu{+}\frac12)}{(2x)^\mu}{\rm e}^{-\text{i}\pi \mu{+}\text{i} x}\text{H}_{\mu}^{(2)}(x)\ \ \  (\text{if}\ x>0)\\
        &\frac{\text{i}}2\frac{\sqrt{\pi}\ \Gamma(\mu{+}\frac12)}{({-}2x)^\mu}{\rm e}^{\text{i}\pi\mu{+}\text{i} x}\text{H}_{\mu}^{(1)}(-x)\ \ \  (\text{if}\ x<0)
    \end{aligned}\right.
\end{align}
In other words, we have Euler-Mellin representation for Hankel functions in $x>0$ and $|\Re(\mu)|<\frac12$ region as
    \begin{equation}\label{eq:1}
    {\text H}^{(2)}_{\mu}(x)=\frac{{\text{i}}}{\sqrt{\pi}}\frac{2^{\mu{+}1}{\rm e}^{{\text{i}}\mu\pi}}{\Gamma[\mu{+}1/2]} x^\mu {\rm e}^{-\text{i} x}\int^\infty_0{\rm d}s\ {\rm e}^{-2\text{i}xs}(s(1+s))^{\mu{-}\frac12}
\end{equation}
and similarly for $\text{H}_\mu^{(1)}(x)$ by its complex conjugate. Furthermore we should mention that, originally sum of our MoB Series 1 and 2 only works for $|x|<1$. However, after expressing the series result by Hankel functions in \eqref{eq:1p}, we in fact perform the correct analytic continuation for the series to $|x|>1$ following the definition of Hankel functions, and \eqref{eq:1} works for the whole $x>0$ (or $x<0$ from \eqref{eq:1p}) region. Recall that from the propagators \eqref{eq:btobulk1}, index of Hankel functions considered in this work reads
\begin{equation}
    \mu=-\text{i}\sqrt{\hat{m}^2{-}\frac14}
\end{equation}
and the condition $|\Re(\mu)|<\frac12$ is guaranteed by considering the physical condition $\hat{m}>0$. This will be our beginning point for massive cosmological integrals.

On the other hand, Series $3$ has argument $\frac1x$, therefore is a series expansion for the integral near $|x|>1$. This series, however, does not yield a finite result if we directly sum it, since the summand grows as $n!$ for large $n$, and in fact offers an asymptotic expansion at $x\to\infty$ for the original integral \footnote{Following the usual rules of MoB we would just discard this series. It is rather curious that in this case even though the series is divergent it is still asymptotic to the original integral and if this phenomenon is more generically true then it could be very useful of method of regions type analysis \cite{SMIRNOV_1995,Beneke:2023wmt}.}. To see this one can directly evaluate the series at any numerical point $|x|>1$, which agrees very well with the numerical value of the integral if we include the first few terms, but rather quickly diverges as we include more and more terms. We can also resum the series by using  Borel resummation  method for the divergent series \cite{citeulike:13807058}. We first divide the summand by $n!$ to get the Borel transform of the series
\begin{equation} \label{ser32}
    \sum_n\frac{(-1)^n}{(n!)^2}\frac{(2\text{i}x)^{-\frac12{-}\mu{-}n}\ \Gamma(\frac12{+}n{-}\mu)\Gamma(\frac12{+}n{+}\mu)}{\Gamma(-\mu{+}\frac12)}=(2\text{i}x)^{{-}\frac12{-}\mu}\Gamma(\frac12{+}\mu)\  _2F_1(\frac12{-}\mu,\frac12{+}\mu,1,\frac{\text{i}}{2x}) 
\end{equation}
and then the final result is the Laplace transform of the Borel transformed series given by
\begin{align}    (2\text{i}x)^{{-}\frac12{-}\mu}&\Gamma\left(\frac12{+}\mu\right)\int_0^{\infty}{\rm d}t\ {\rm e}^{{-}t}\ _2F_1(\frac12{-}\mu,\frac12{+}\mu,1,\frac{it}{2x})\nonumber\\
&=\left\{\begin{aligned}
        &{-}\frac {\text{i}}2\frac{\sqrt{\pi}\ \Gamma(\mu{+}\frac12)}{(2x)^\mu}{\rm e}^{-\text{i}\pi \mu{+}\text{i} x}\text{H}_{\mu}^{(2)}(x)\ \ \  (\text{if}\ x>0)\\
        &\frac{\text{i}}2\frac{\sqrt{\pi}\ \Gamma(\mu{+}\frac12)}{({-}2x)^\mu}{\rm e}^{\text{i}\pi\mu{+}\text{i} x}\text{H}_{\mu}^{(1)}(-x)\ \ \  (\text{if}\ x<0)
    \end{aligned}\right.
\end{align}
which meets \eqref{eq:1}.

\section{Basic time integrals from massless cosmological amplitudes}
\label{sec:massless correlators}
After a basic introduction of the algorithm, we will now turn to practical calculation of cosmological amplitudes. As a warm-up, in this section we will firstly focus on the wavefunction coefficients/correlators, for conformally-coupled scalars in power law FRW-metric background \cite{Arkani-Hamed:2017fdk,Arkani-Hamed:2018bjr,Hillman:2019wgh,Arkani-Hamed:2023kig,Arkani-Hamed:2023bsv,De:2023xue,Benincasa:2024leu}, whose action is conformally equivalent to a massless scalar theory living in Minkowski spacetime namely
\begin{equation}\label{eq:actionp}
    S[\varphi]=-\int {\rm d}^dx \, {\rm d}\tau \left(\frac12 (\partial \varphi)^2+\sum_{k>2}\frac{\lambda_k(\tau)}{k!}\varphi^k\right),
\end{equation}
which also corresponds to a massless exchange in the model \eqref{eq:action} i.e., $\hat{m}=0$ or equivalently  $\nu\to \frac{\text{i}}2$. The Hankel functions degenerate to exponential functions, when we take $\nu\to \frac{\text{i}}2$. This in turn, leads to a simplification of the propagators given in \eqref{eq:btobulk0}, \eqref{eq:btobulk1} and \eqref{eq:btobulk2}. Thus, we get the following {\it bulk-bulk} propagators 
\begin{equation}\label{eq:waveprop}    G(k,\tau_1,\tau_2)=\frac1{2k}\left[\text{e}^{-\text{i}k(\tau_1{-}\tau_2)}\theta(\tau_1{-}\tau_2)+\text{e}^{\text{i}k(\tau_1{-}\tau_2)}\theta(\tau_2{-}\tau_1)-\text{e}^{ik(\tau_1{+}\tau_2)}\right]
\end{equation}
and
\begin{equation}\label{eq:correprop}
\begin{aligned}
    & D_{\pm \pm}(k;\tau_1,\tau_2)=\frac{1}{2k}\left[ e^{\mp \text{i} \, k (\tau_1-\tau_2)}\theta(\tau_1-\tau_2)+e^{\pm \text{i} \, k (\tau_1-\tau_2)} \theta(\tau_2-\tau_1) \right],\\
    & D_{\pm \mp}(k;\tau_1,\tau_2)=\frac{1}{2k} e^{\pm \text{i} \, k (\tau_1-\tau_2)},
\end{aligned}
\end{equation}
for the wavefunction and the {\it in-in} correlator respectively.
{\it Bulk-to-boundary} propagators between a bulk point at $\tau$ and the boundary $\tau=0$ carrying energy  $w_v$, on the other hand, are still \eqref{eq:btobound1} or \eqref{eq:btobound2}. Finally, $\lambda_v(\tau_v)$ is coupling constant at node $v$ in the diagram, which yields an extra factor $\lambda_v(\tau_v)\propto (-\tau)^{q_v{-}1}$ in the integrand. The cosmological amplitudes are then translated to nested-time integrals with exponential functions at each node. Without loss of generality, it is very natural to see that these nested-time integrals can be decomposed into basic building blocks \cite{Fan:2024iek,He:2024olr}, which are called {\it basic time integrals}, and defined as the follows:
\begin{equation}\label{eq:Tdef}	
P[\mathcal{N}(1\cdots n)]=(-\text{i})^n\int_{-\infty}^{0}\prod_{\ell=1}^n\left[\mathrm{d\tau}_{\ell}(-\tau_{\ell})^{q_\ell-1}e^{ \text{i}\omega_\ell \tau_\ell}\right]\prod_{(j,k)\in\mathcal{N}}\theta_{j,k}.
\end{equation}
where $\omega_\ell$ are linear combinations of energies flowing through the node\footnote{Generally speaking, $\omega_i$ in \eqref{eq:Tdef} will be a linear combination of $w_i+\sum_j\mathcal{N}_{i,j}k_{i,j}$, where $\mathcal{N}_{i,j}=1$ if we have $\theta(\tau_i-\tau_j)$, $\mathcal{N}_{i,j}=-1$ if we have $\theta(\tau_j-\tau_i)$, and $\mathcal{N}_{i,j}=0$ otherwise.}, and $\mathcal{N}$ is the time ordered structure of the directed diagram. For simplicity, we denote $\theta_{i,j}:=\theta(\tau_i{-}\tau_j)$ from here on. These integrals have been extensively studied recently in many different contexts \cite{Xianyu:2023ytd,Arkani-Hamed:2023bsv,Arkani-Hamed:2023kig,Fan:2024iek,He:2024olr,Baumann:2024mvm,Baumann:2025qjx}, both for tree level and loop level. Especially near the inflationary limit $q_i\to\epsilon$. These integrals become multi-polylogarithmic functions after a series expansion in $\epsilon$, enjoying rich symbology structures \cite{Hillman:2019wgh,Arkani-Hamed:2023kig,Arkani-Hamed:2023bsv} as well as connections to  positive geometries \cite{Arkani-Hamed:2017fdk,Arkani-Hamed:2024jbp,Figueiredo:2025daa}. In the following discussion, we will keep the twists $q_i$ as general parameters, for which all these integrals are finite generalized hypergeometric functions. We will also denote basic time integrals as directed graphs for simplicity in this section. For instance
\begin{center}
    \raisebox{-1em}{\begin{tikzpicture}
				\coordinate (X1) at (0,0);
				\coordinate (X2) at (1,0);
				\coordinate (X3) at (2,0);;
				\node[below] at (X1) {\small{$1$}};
				\node[below] at (X2) {\small{$2$}};
				\node[below] at (X3) {\small{$3$}};
				\draw[line width=1.5pt,->] (X1)--(0.6,0);
				\draw[line width=1.5pt] (0.5,0)--(X2);
				\draw[line width=1.5pt,->] (X2)--(1.6,0);
				\draw[line width=1.5pt] (1.5,0)--(X3);
				\path[fill=black] (X1) circle[radius=0.1];
				\path[fill=black] (X2) circle[radius=0.1];
				\path[fill=black] (X3) circle[radius=0.1];
		\end{tikzpicture}}\quad ,
\end{center}
stands for the  basic time integral
\begin{equation}
    (-\text{i})^{3}\int_{-\infty}^0\prod_{\ell=1}^3\left[\mathrm{d\tau}_{\ell}(-\tau_{\ell})^{q_\ell-1}e^{ \text{i}\omega_\ell \tau_\ell}\right]\theta_{3,2}\theta_{2,1} \quad ,
\end{equation} 
with $q_i$ and $\omega_i$ all general  parameters. 

\subsection{A review of recursive structures for massless family tree integrals}
The main aim of this section is to compute general basic time integrals \eqref{eq:Tdef} from MoB. The method is, however, not directly applicable if we begin with their nested-time integral definition \eqref{eq:Tdef}, since it is not a Mellin integral. In \cite{He:2024olr}, differential equations for the basic time integrals were considered in general, and an {\it Euler-Mellin representation} 
%\cite{Matsubara-Heo:2023ylc,berkesch2014euler,Aomoto:2011ggg,yoshida2013hypergeometric} 
was derived for basic time integrals, in terms of a graph polynomial. In this subsection, we will firstly review these results from \cite{Fan:2024iek,He:2024olr}, which will be our starting point for applying MoB in the next subsection.

Due to the identity relation $\theta_{j,i}{+}\theta_{i,j}{=}1$ for Heaviside theta functions, basic time integrals enjoy multiple linear relations. For example, the following relation
\begin{equation}\label{eq:exampletwoadj}
\begin{aligned}
   \raisebox{-11pt}{\begin{tikzpicture}[scale=0.75]
		\coordinate (X1) at (0,0);
		\coordinate (X2) at (1.5,0);
        \coordinate (X3) at (3,0);
		\node[below] at (X1) {\small{$2$}};
		\node[below] at (X2) {\small{$1$}};
        \node[below] at (X3) {\small{$3$}};
		\draw[line width=1.5pt,->] (X1)--(0.85,0);
        \draw[line width=1.5pt] (0.7,0)--(X2);
        \draw[line width=1.5pt,->] (X3)--(2.2,0);
        \draw[line width=1.5pt] (2.5,0)--(X2);
		\path[fill=black] (X1) circle[radius=0.11];
		\path[fill=black] (X2) circle[radius=0.11];
        \path[fill=black] (X3) circle[radius=0.11];
\end{tikzpicture}}=\raisebox{-11pt}{\begin{tikzpicture}[scale=0.75]
		\coordinate (X1) at (0,0);
		\coordinate (X2) at (1.5,0);
        \coordinate (X3) at (3,0);
		\node[below] at (X1) {\small{$1$}};
		\node[below] at (X2) {\small{$3$}};
        \node[below] at (X3) {\small{$2$}};
		\draw[line width=1.5pt,->] (X2)--(0.7,0);
        \draw[line width=1.5pt] (0.75,0)--(X1);
        \draw[line width=1.5pt,->] (X3)--(2.2,0);
        \draw[line width=1.5pt] (2.5,0)--(X2);
		\path[fill=black] (X1) circle[radius=0.11];
		\path[fill=black] (X2) circle[radius=0.11];
        \path[fill=black] (X3) circle[radius=0.11];
\end{tikzpicture}}+\raisebox{-11pt}{\begin{tikzpicture}[scale=0.75]
		\coordinate (X1) at (0,0);
		\coordinate (X2) at (1.5,0);
        \coordinate (X3) at (3,0);
		\node[below] at (X1) {\small{$1$}};
		\node[below] at (X2) {\small{$2$}};
        \node[below] at (X3) {\small{$3$}};
		\draw[line width=1.5pt,->] (X2)--(0.7,0);
        \draw[line width=1.5pt] (0.75,0)--(X1);
        \draw[line width=1.5pt,->] (X3)--(2.2,0);
        \draw[line width=1.5pt] (2.5,0)--(X2);
		\path[fill=black] (X1) circle[radius=0.11];
		\path[fill=black] (X2) circle[radius=0.11];
        \path[fill=black] (X3) circle[radius=0.11];
\end{tikzpicture}}
\end{aligned}.
\end{equation}
is a natural consequence of the relation $\theta_{1,2}\theta_{1,3}=\theta_{1,2}\theta_{2,3}+\theta_{1,3}\theta_{3,2}$. By repeatedly using the identity $\theta_{i,j}=1{-}\theta_{i,j}$, one can systematically reverse the time orderings on certain edges in a time integral. This process allows the original diagram to be decomposed into a sum of basis diagrams, each of which consists of (a product of) partially ordered diagrams—meaning that each node has at most one incoming directed edge. These basis integrals, though not linearly independent, are referred to as {\it family tree} integrals in the context of \cite{Xianyu:2023ytd}. The right-hand side of Eq.\eqref{eq:exampletwoadj} shows two examples of such family trees, while the left-hand side does not represent a family tree. By definition, a family tree must include a node with only outgoing edges, referred to as the ancestor node—denoted as node 1 throughout this work. If two nodes $i$ and $j$ are connected by a directed path with $i$ pointing to $j$,we refer to $j$ as a descendant of $i$.

Furthermore, for simplicity in discussing the massive case in the next section, and without loss of generality, we restrict our attention to family tree integrals in which the propagator for node~1—the \textit{ancestor}—contains only \( \theta_{1,2} \); that is, the ancestor has only \textit{one} outgoing edge. 

%More generally, any family tree integral can always be decomposed into a basis of family trees satisfying this condition. This can be achieved by starting with an arbitrary family tree integral, selecting a path from the ancestor node to any of its leaf nodes, and applying the theta function-identity recursively to reverse all arrows along that path.

Let us, consider the example in \eqref{eq:fig1}. On the left hand side we have a family tree with node $4$ as ancestor and more than one $\theta_{4,i}$ in its integrand. Nodes $1,2,3,5$ are descendants of node $4$, while $1,3$ are descendants of node $2$.  By applying identity $1{-}\theta_{i,j}=\theta_{j,i}$ we finally arrive at the right hand side, with ancestor node $1$ and only $\theta_{1,2}$ in its nested-time structure, accompanied with sum of simpler lower-site family trees.  

\begin{equation}\label{eq:fig1}
    \raisebox{-25pt}{\begin{tikzpicture}[scale=0.35]
				\coordinate (X1) at (3,4);
				\coordinate (X2) at (0,0);
				\coordinate (X3) at (4.5,2);
				\coordinate (X4) at (3,0);
				\coordinate (X5) at (6,0);
				\node[above] at (X1) {\small{$4$}};
				\node[below] at (X2) {\small{$5$}};
				\node[above right] at (X3) {\small{$2$}};
				\node[below] at (X4) {\small{$3$}};
				\node[below] at (X5) {\small{$1$}};
				\draw[line width=1.5pt,->] (X1)--(1.2,1.6);
				\draw[line width=1.5pt] (1.5,2)--(X2);
				\draw[line width=1.5pt,->] (X1)--(4.05,2.6);
				\draw[line width=1.5pt] (3.75,3)--(X3);
				\draw[line width=1.5pt,->] (X3)--(3.6,0.8);
				\draw[line width=1.5pt] (3.75,1)--(X4);
				\draw[line width=1.5pt,->] (X3)--(5.4,0.8);
				\draw[line width=1.5pt] (5.25,1)--(X5);
				\path[fill=black] (X1) circle[radius=0.2];
				\path[fill=black] (X2) circle[radius=0.2];
				\path[fill=black] (X3) circle[radius=0.2];
				\path[fill=black] (X4) circle[radius=0.2];
				\path[fill=black] (X5) circle[radius=0.2];
		\end{tikzpicture}}=\raisebox{-25pt}{\begin{tikzpicture}[scale=0.35]
				\coordinate (X1) at (3,4);
				\coordinate (X2) at (0,0);
				\coordinate (X3) at (4.5,2);
				\coordinate (X4) at (3,0);
				\coordinate (X5) at (6,0);
				\node[above] at (X1) {\small{$4$}};
				\node[below] at (X2) {\small{$5$}};
				\node[above right] at (X3) {\small{$2$}};
				\node[below] at (X4) {\small{$3$}};
				\node[below] at (X5) {\small{$1$}};
				\draw[line width=1.5pt,->] (X1)--(1.2,1.6);
				\draw[line width=1.5pt] (1.5,2)--(X2);
				\draw[line width=1.5pt,->] (X3)--(3.75,3);
				\draw[line width=1.5pt] (4.05,2.6)--(X1);
				\draw[line width=1.5pt,->] (X3)--(3.6,0.8);
				\draw[line width=1.5pt] (3.75,1)--(X4);
				\draw[line width=1.5pt,->] (X5)--(5.25,1);
				\draw[line width=1.5pt] (5.4,0.8)--(X3);
				\path[fill=black] (X1) circle[radius=0.2];
				\path[fill=black] (X2) circle[radius=0.2];
				\path[fill=black] (X3) circle[radius=0.2];
				\path[fill=black] (X4) circle[radius=0.2];
				\path[fill=black] (X5) circle[radius=0.2];
		\end{tikzpicture}}+ \text{sum of (products of lower-site family trees)}
\end{equation}

After expanding all basic time integrals on (products of) family tree integrals, in \cite{He:2024olr}, it was shown that for a family tree integral, its solution from the differential equation method can be expressed by an individual Euler-Mellin integral. The basic rules to write down the integrand is the following: 
\begin{enumerate}
    \item For any family tree integral, we have an overall constant factor $(-\text{i})^{n+\tilde{q}_1}\Gamma(\tilde{q}_1)$, where $\tilde{q}_j$ is a sum of all the twists for node $j$ and all its descendants in the tree.
    \item For every node $j$ in the diagram except the ancestor node $1$, assign a fold of integration:
    \begin{equation}
        \int_{0}^{1}\frac{{\rm d}\alpha_{j}}{\alpha_{j}}\alpha_{j}^{\tilde{q}_j}
    \end{equation}
    
    \item Define the graph polynomial $G(\omega,\alpha)$ as
    \begin{equation}\label{eq:Pdef}
G(\omega,\alpha)=\sum_{j=1}^{n}\omega_{j} \, \bar{\alpha}_j
    \end{equation}
    where $\bar{\alpha}_j$ is defined by product of $\alpha_j$ and all $\alpha_k$ that $j$ is descendant of $k$. For instance, for the family-tree in the left hand side of \eqref{eq:threesitechain}, $\bar{\alpha}_1=\alpha_1$, $\bar{\alpha}_2=\alpha_1\alpha_2$, $\bar{\alpha}_3=\alpha_1\alpha_2\alpha_3$. 
    
    Finally, we set $\alpha_1:=1$ as there is no integration over it.
    \item The {\it Euler-Mellin representation} for family tree is 
    \begin{equation}\label{eq:Mellinmassless}
        \begin{aligned}
            P[\mathcal{N}(1\cdots n)]=(-\text{i})^{n+\tilde{q}_{1}}\Gamma(\tilde{q}_{1})\int_{0}^{1}\prod_{j=2}^{n}\frac{{\rm d}\alpha_{j}}{\alpha_{j}}\alpha_{j}^{\tilde{q}_j}G(\omega,\alpha)^{-\tilde{q}_{1}} 
        \end{aligned}
        \end{equation}
        where $\mathcal{N}$ is the nested-time order of the family tree.
\end{enumerate}
We explicitly write down the Euler-Mellin representation for two family tree integrals as
\begin{equation}\label{eq:threesitechain}
\begin{aligned}
   \raisebox{-11pt}{\begin{tikzpicture}[scale=0.75]
		\coordinate (X1) at (0,0);
		\coordinate (X2) at (1.5,0);
        \coordinate (X3) at (3,0);
		\node[below] at (X1) {\small{$1$}};
		\node[below] at (X2) {\small{$2$}};
        \node[below] at (X3) {\small{$3$}};
		\draw[line width=1.5pt,->] (X1)--(0.9,0);
        \draw[line width=1.5pt] (0.75,0)--(X2);
        \draw[line width=1.5pt,->] (X2)--(2.4,0);
        \draw[line width=1.5pt] (2,0)--(X3);
		\path[fill=black] (X1) circle[radius=0.11];
		\path[fill=black] (X2) circle[radius=0.11];
        \path[fill=black] (X3) circle[radius=0.11];
\end{tikzpicture}}=(-\text{i})^{3+\tilde{q}_{1}}\!\!\!\int_{0}^{1}{\rm d} \alpha_{3} \, \alpha_{3}^{q_3-1}\!\!\!\int_{0}^{1}{\rm d}  \alpha_{2} \, \alpha_{2}^{\tilde{q}_{2}-1}\!\left[\omega_{1}\!+\!(\omega_{2}\!+\!\omega_{3}\alpha_{3})\alpha_{2}\right]^{-\tilde{q}_{1}}\Gamma(\tilde{q}_{1}),
\end{aligned}
\end{equation}
\begin{equation}
    \raisebox{-25pt}{\begin{tikzpicture}[scale=0.6]
		\coordinate (X4) at (0,0);
		\coordinate (X1) at (0,1.5);
        \coordinate (X2) at (1.3,-0.75);
        \coordinate (X3) at (-1.3,-0.75);
		\node[above] at (X1) {\small{$1$}};
		\node[below] at (X2) {\small{$3$}};
        \node[below] at (X3) {\small{$4$}};
        \node[below] at (X4) {\small{$2$}};
		\draw[line width=1.5pt,black,->] (X1)--(0,0.5);
        \draw[line width=1.5pt,black] (0,1)--(X4);
        \draw[line width=1.5pt,->] (X4)--(0.87,-0.5);
        \draw[line width=1.5pt] (0.65,-0.375)--(X2);
        \draw[line width=1.5pt,->] (X4)--(-0.87,-0.5);
        \draw[line width=1.5pt] (-0.65,-0.375)--(X3);
		\path[fill=black] (X1) circle[radius=0.11];
		\path[fill=black] (X2) circle[radius=0.11];
        \path[fill=black] (X3) circle[radius=0.11];
        \path[fill=black] (X4) circle[radius=0.11];
\end{tikzpicture}}=(-\text{i})^{4{+}\tilde{q}_1}\int_0^1\prod_{i=3}^4{\rm d}\alpha_i \alpha_i^{q_i{-}1}\int_0^1{\rm d}\alpha_2\alpha_2^{\tilde{q}_2{-}1}[\omega_1{+\alpha_2(\omega_2{+}\alpha_3\omega_3{+}\alpha_4\omega_4)}]^{-\tilde{q}_1}\Gamma(\tilde{q}_1).
\end{equation}
For more details about these representation readers can refer to \cite{He:2024olr}.

\subsection{Series representation for family trees from MoB}
With the Euler-Mellin integral, as our starting point, we can now apply MoB after a simple change of variables $\alpha_i\rightarrow \frac{s_i}{1+s_i}$, to family tree integrals in the massless model.

\paragraph{Series representation for family chain integrals}
As the first and a simple warm-up example, we look into the $n$-site directed chain graphs as the following
\begin{equation}\label{eq:masleschain}
P[(1\cdots N)]=\raisebox{-1em}{\begin{tikzpicture}[scale=0.75]
		\coordinate (X1) at (0,0);
		\coordinate (X2) at (1.5,0);
        \coordinate (X3) at (3,0);
            \coordinate (X4) at (4.5,0);
		\node[below] at (X1) {\small{$1$}};
		\node[below] at (X2) {\small{$2$}};
        \node[below] at (X4) {\small{$N$}};
        \node[below] at (X3) {\small{$N{-}1$}};
		\draw[line width=1.5pt,->] (X1)--(0.9,0);
        \draw[line width=1.5pt] (0.75,0)--(X2);
        \draw[line width=1.5pt,dashed] (X2)--(X3);
        \draw[line width=1.5pt,->] (X3)--(3.9,0);
        \draw[line width=1.5pt] (3.75,0)--(X4);
		\path[fill=black] (X1) circle[radius=0.11];
		\path[fill=black] (X2) circle[radius=0.11];
        \path[fill=black] (X3) circle[radius=0.11];
        \path[fill=black] (X4) circle[radius=0.11];
\end{tikzpicture}}
\end{equation}
where $(1\cdots N)$ denote the simple time structure $-\infty<\tau_1<\cdots<\tau_N<0$, and whose  Euler-Mellin representation reads
\begin{equation}\label{eq:chain1}
   (-\text{i})^{N+\tilde{q}_{1}}\Gamma(\tilde{q}_{1})\int_0^\infty \prod_{i=2}^N\left({\rm d}s_i s_i^{\tilde{q}_i{-}1}(1{+}s_i)^{\tilde{q}_{1}{-}\tilde{q}_i{-}1}\right)\ \left(\sum_{i=1}^N\omega_i\prod_{j=2}^i s_j\prod_{j=i{+}1}^N(1{+}s_j)\right)^{-\tilde{q}_{1}}
\end{equation}
where the graph polynomials are given by
\[G_N=\omega_1+\sum_{i=2}^N\left(\omega_i\prod_{j=2}^i\alpha_j\right),\]
according to \eqref{eq:Mellinmassless}, and then perform the change of variables $\alpha_i\to\frac{s_i}{1{+}s_i}$. These are the simplest family tree graphs discussed in \cite{Xianyu:2023ytd,Fan:2024iek}. Following the MoB algorithm, we apply the bracket expansion \eqref{eq:bracketrule2} for each of the factors in \eqref{eq:chain1}. Firstly,we get the indices $n_1,\cdots,n_N$ for the $N$ terms in the graph polynomial, we have 
\begin{align}\label{eq:brkchain1}
\left(\sum_{i=1}^N\omega_i\prod_{j=2}^i s_j\prod_{j=i{+}1}^N(1{+}s_j)\right)^{-\tilde{q}_{1}}{=}&\frac1{\Gamma(\tilde{q}_1)}\sum_{n_1\cdots n_N}\phi_{n_1\cdots n_N}\prod_{i=1}^N\omega_i^{n_i}  \prod_{i=2}^N\left(s_i^{\tilde{n}_i}(1{+}s_i)^{\tilde{n}_1-\tilde{n}_i}\right)\nonumber\\
&\times\langle \tilde{q}_{1}+\tilde{n}_1\rangle \quad,
\end{align}
where, we also introduce the notation $\tilde{n}_i:=\sum_{j=i}^Nn_j$. Furthermore, for each of the factors $(1{+}s_i)$, we collect powers, introduce two more indices and expand it by \eqref{eq:bracketrule2}  as
\begin{equation}\label{eq:brkchain2}
    (1+s_i)^{A_i}=\frac1{\Gamma(-A_i)}\sum_{m_{i,1},m_{i,2}}\phi_{m_{i,1},m_{i,2}}s_i^{m_{i,2}}\langle -A_i{+}m_{i,1}{+}m_{i,2}\rangle \quad ,
\end{equation}
with $A_i=\tilde{n}_1{-}\tilde{n}_i{+}\tilde{q}_1{-}\tilde{q}_i{-}1$. Finally, each $s_i$ factor has power $B_i:=\tilde{n}_i{+}m_{i,2}{+}\tilde{q}_i{-}1$, which gives us $N{-}1$ additional brackets as $\langle B_i{+}1\rangle$, and the summand in \eqref{rmt3} finally reads
\begin{equation}\label{eq:summand2}
    (-\text{i})^{N{+}\tilde{q}_1}\sum_{n_i,m_{i,1},m_{i,2}} \phi_{n_i,m_{i,1},m_{i,2}}\frac{\prod_{i=1}^N\omega_i^{n_i}}{\prod_{i=2}^N\Gamma(-A_i)}\times \langle\tilde{q}_1{+}n_1\rangle\prod_{i=2}^N\langle{-}A_i{+}m_{i,1}{+}m_{i,2}\rangle\langle B_i{+}1\rangle \quad ,
\end{equation}
In total, we have $N{+}2(N{-}1)=3N{-}2$ indices, with $2(N{-}1){+}1=2N{-}1$ brackets as constraints between them. Therefore, we have rank-$(N{-}1)$ bracket system. Different choices for the independent indices will result in series that work in different kinematic regions, as we will show. 
\paragraph{Region $\omega_1>\omega_i$} One of the physical regions, as discussed in \cite{Fan:2024iek}, is
\begin{equation}\label{eq:region1}
    \omega_1>\omega_i\ \ \ \text{for}\ \  i=2,\cdots, N.
\end{equation}
As an illustrative example for this, let us specialize to the two-site chain case, we will then have the summand 
\begin{equation}\label{eq:summand3}  
\begin{aligned}(-\text{i})^{2{+}q_{12}}\sum_{n_1,n_2,m_{2,1},m_{2,2}}&\phi_{n_1,n_2,m_{2,1},m_{2,2}}\frac{\omega_1^{n_1}\omega_2^{n_2}}{\Gamma({-}n_1{-}q_1{+}1)}\times\\
&\langle q_1{+}q_2{+}n_1{+}n_2\rangle\langle {-}n_1{-}q_1{+}1{+}m_{2,1}{+}m_{2,2}\rangle\langle n_2{+}m_{2,2}{+}q_2\rangle
\end{aligned}
\end{equation}
If we are interested in region $\omega_1>\omega_2$, then the only choice for us is solving $\{n_1,m_{2,1},m_{2,2}\}$ in terms of $n_2$, {\it i.e.}, $m^*_{2,1}={-}1,m^*_{2,2}={-}q_2{-}n_2,n^*_1={-}n_2{-}q_1{-}q_2$, such that the argument in power series turns out to be $\left(\frac{\omega_2}{\omega_1}\right)^{n_2}$, which leads to a convergent series. Following the Rule 4 of MoB and \eqref{rmt4}, we then arrive at series solution for two site chain as
\begin{align}\label{eq:twositechainmassless2}
    &(-\text{i})^{2{+}\tilde{q}_{1}}\sum_{n_2}\phi_{n_2}\Gamma(-n_1^*)\Gamma(-m_{2,1}^*)\Gamma(-m_{2,2}^*)\frac{\omega_1^{n_1^*}\omega_2^{n_2}}{\Gamma({-}n_1^*{-}q_1{+}1)}\nonumber\\
    =&(-\text{i})^{2{+}\tilde{q}_{1}}\sum_{n_2}\phi_{n_2}\frac{\Gamma(n_2{+}\tilde{q}_{1})}{n_2{+}q_2}\omega_1^{-n_2{-}\tilde{q}_{1}}\omega_2^{n_2}=\frac{(-\text{i})^2}{(\text{i}\omega_1)^{\tilde{q}_{1}}}\frac{\Gamma(\tilde{q}_{1})}{q_2}\ _2F_1\left(q_2,\tilde{q}_{1},q_2{+}1,-\frac{\omega_2}{\omega_1}\right)
\end{align}
which can be readily seen to yield correct answer for two-site chain, as \eqref{eq:Mellinmassless} is directly the integral representation of the Gauss hypergeometric function ${}_2F_1$.

More generally for the $N$-site chain, using a recursive calculation we can show that, in region \eqref{eq:region1} the only solution of the bracket equations corresponds to choosing $n_2,\cdots, n_N$ as independent indices and solving the others, such that the arguments in the power series are
\begin{equation}
\left\{\left(\frac{\omega_i}{\omega_1}\right)^{n_i}\right\}_{i=2,\cdots,N}.
\end{equation} 
The condition $\omega_1>\omega_i$ for all $i\geq2$ then leads to the convergence of this series. The \emph{unique} solution of bracket equations for $N$-site chain in this region reads
\begin{align}\label{eq:sol1}
    &m_{i,1}^*=-1,\ m_{i,2}^*=-\tilde{q}_i{-}\tilde{n}_{i},\ \ n^*_1=-\tilde{n}_2-\tilde{q}_1.
\end{align}
And the final result for this region is 
\begin{equation}\label{eq:summand1}
    (-\text{i})^{N{+}\tilde{q}_1}\sum_{n_2\cdots n_N}\phi_{n_2\cdots n_N}\Gamma(-n_1^*)\prod_{i=2}^N\Gamma(-m_{i,2}^*)\frac{\omega_1^{n_1^*}\prod_{i=2}^N\omega_i^{n_i}}{\prod_{i=2}^N\Gamma(-A_i)}.
\end{equation}
We see that substituting the solution \eqref{eq:sol1} in \eqref{eq:summand1}, the $\Gamma(-m_{i,2}^*)/\Gamma(-A_i)=1/(\tilde{n}_i{+}\tilde{q}_i)$ cancels, and finally the result reads
\begin{equation}\label{eq:chainres}
    P[(1\cdots N)]=\frac{(-\text{i})^N}{(\text{i}\omega_1)^{\tilde{q}_1}}\sum_{n_2\cdots n_{N}}\phi_{n_2\cdots n_N}\ \Gamma(\tilde{n}_2{+}\tilde{q}_1)\prod_{i=2}^N\frac{1}{(\tilde{n}_i{+}\tilde{q}_i)}\left(\frac{\omega_i}{\omega_1}\right)^{n_i}
\end{equation}
This is exactly the series solution in \cite{Xianyu:2023ytd,Fan:2024iek} for family chain graphs in region $w_1>w_i$ ! 

\paragraph{Region $\omega_j>\omega_i$ for any $j$}
As we have mentioned, one of the advantage for us adopting MoB is that the algorithm generates series representation for all regions automatically. It is then a natural question to move to other physical regions and calculating series expansion of the integral, {\it e.g.}, region with any energy $\omega_j$ being the biggest
\begin{equation}\label{eq:region2}
    \omega_j>\omega_i,\ \ \ {i=1,\cdots,\hat{j},\cdots, N}
\end{equation}
We will have more than one basis series solution in these regions to add up. To illustrate this, we go back to the two site chain again and computing basis series for it in region $\omega_2>\omega_1$ from \eqref{eq:summand3}. Instead of choosing $n_2$ as independent index, to have $\left(\frac{\omega_1}{\omega_2}\right)^k$ as argument in the series, we have two choices now
\begin{align}
   &\text{Series 1}:\ \bar\sigma_1{=}\{m_{2,1},n_1,n_2\}, \ \ m^*_{2,1}=-1,\ n_1^*=m_{2,2}{-}q_1,\ n^*_2={-}m_{2,2}{-}q_2 ,\\ 
   &\text{Series 2}:\ \bar\sigma_1{=}\{m_{2,1},m_{2,2},n_2\}, \ \ m^*_{2,1}=-1,\ m^*_{2,2}=n_1{+}q_1,\ n^*_2={-}n_1{-}q_1{-}q_2,
\end{align}
and correspondingly, the two series read
\begin{align}
    &\text{Series 1}: (-\text{i})^{2{+}q_{12}}\sum_{m_{2,2}}(-1)^{m_{2,2}}\frac{\Gamma(-m_{2,2}{+}q_1)\Gamma(m_{2,2}{+}q_2)}{\omega_1^{q_1}\omega_2^{q_2}}\left(\frac{\omega_1}{\omega_2}\right)^{m_{2,2}}\frac{\sin(\pi m_{2,2})}{\pi m_{2,2}} ,\\
    &\text{Series 2}:-\frac{(-\text{i})^{2}}{(\text{i}\omega_2)^{\tilde{q}_{1}}}\sum_{n_1}\phi_{n_1}\frac{\Gamma(n_1{+}\tilde{q}_{1})}{n_1{+}q_1}\left(\frac{\omega_1}{\omega_2}\right)^{n_1} .
\end{align}
Both of them have argument $\frac{\omega_1}{\omega_2}$ in the summand. Therefore, the result should be a sum over these two series. This result, surprisingly, encodes the nested-time structure from the original definition of the basic family trees.  For Series 1, only term $m_{2,2}=0$ is non-trivial due to the $\sin(x)/x$ factor in the summand, and the series localizes to a single term
\begin{equation}
    (-\text{i})^{2{+}q_{1,2}}\frac{\Gamma(q_1)\Gamma(q_2)}{\omega_{1}^{q_1}\omega_{2}^{q_{2}}}=\prod_{i=1,2}(-\text{i})^{1+q_i}\frac{\Gamma(q_i)}{\omega_i^{q_i}} ,
\end{equation}
which is actually a factorized product of two massless contact functions \eqref{eq:contactmassless}. The Series $2$, on the other hand, is in fact the series expansion \eqref{eq:chainres} for two-site family tree with $2$ as the ancestor node and an extra minus sign. The sum of two series above can be interpreted by the basic relation between family tree integrals as
\begin{equation}\label{eq:masslessid}
    \raisebox{-11pt}{\begin{tikzpicture}
				\coordinate (X1) at (0,0);
				\coordinate (X2) at (1.5,0);
				\node[below] at (X1) {\small{$1$}};
				\node[below] at (X2) {\small{$2$}};
				\draw[line width=1.5pt,->] (X1)--(0.85,0);
				\draw[line width=1.5pt] (0.75,0)--(X2);
				\path[fill=black] (X1) circle[radius=0.1];
				\path[fill=black] (X2) circle[radius=0.1];
		\end{tikzpicture}}=\raisebox{-11pt}{\begin{tikzpicture}
		\coordinate (X1) at (0,0);
		\coordinate (X2) at (1.5,0);
		\node[below] at (X1) {\small{$1$}};
		\node[below] at (X2) {\small{$2$}};
		\draw[line width=1.5pt,dashed] (X1)--(X2);
		\path[fill=black] (X1) circle[radius=0.1];
		\path[fill=black] (X2) circle[radius=0.1];
\end{tikzpicture}}-\raisebox{-11pt}{\begin{tikzpicture}
		\coordinate (X1) at (0,0);
		\coordinate (X2) at (1.5,0);
		\node[below] at (X1) {\small{$1$}};
		\node[below] at (X2) {\small{$2$}};
		\draw[line width=1.5pt] (X1)--(0.85,0);
		\draw[line width=1.5pt,->] (X2)--(0.65,0);
		\path[fill=black] (X1) circle[radius=0.1];
		\path[fill=black] (X2) circle[radius=0.1];
	\end{tikzpicture}}
\end{equation}

This generalizes straightforwardly for higher-site chains. For three-site chain, we have $7$ indices$\{n_1,n_2,n_3,m_{2,1},m_{2,2},m_{3,1},m_{3,2}\}$ satisfying $5$ bracket equations.
\begin{align*}
    n_{123}{+}q_{123}=0&,\ \  1{+}m_{2,1}{+}m_{2,2}{-}n_1{-}q_1=0,\ 1{+}m_{3,1}{+}m_{3,2}{-}n_{12}{-}q_{12}=0\\
    &n_3{+}m_{3,2}{+}q_3=0, n_{23}{+}m_{2,2}{+}q_{23}=0
\end{align*}
Besides $m_{2,1}^*=m_{3,1}^*=-1$ are always fixed, we can choose other three indices to be solved, and in total eight subsets are non-trivial choices. Explicit computation shows that these eight basis series compute eight of the nine possible components in decomposition of three-site-chain cosmological amplitudes, excluding the component shown in red below 
\begin{equation}
	\begin{aligned}
	& \raisebox{-1em}{\begin{tikzpicture}
				\coordinate (X1) at (0,0);
				\coordinate (X2) at (1,0);
				\coordinate (X3) at (2,0);;
				\node[below] at (X1) {\small{$1$}};
				\node[below] at (X2) {\small{$2$}};
				\node[below] at (X3) {\small{$3$}};
				\draw[line width=1.5pt,->] (X1)--(0.6,0);
				\draw[line width=1.5pt] (0.5,0)--(X2);
				\draw[line width=1.5pt,->] (X2)--(1.6,0);
				\draw[line width=1.5pt] (1.5,0)--(X3);
				\path[fill=black] (X1) circle[radius=0.1];
				\path[fill=black] (X2) circle[radius=0.1];
				\path[fill=black] (X3) circle[radius=0.1];
		\end{tikzpicture}} \quad\quad 
		\raisebox{-1em}{\begin{tikzpicture}
		\coordinate (X1) at (0,0);
		\coordinate (X2) at (1,0);
		\coordinate (X3) at (2,0);;
		\node[below] at (X1) {\small{$1$}};
		\node[below] at (X2) {\small{$2$}};
		\node[below] at (X3) {\small{$3$}};
		\draw[line width=1.5pt,->,red] (X1)--(0.6,0);
		\draw[line width=1.5pt,red] (0.5,0)--(X2);
		\draw[line width=1.5pt,->,red] (X3)--(1.4,0);
		\draw[line width=1.5pt,red] (1.5,0)--(X2);
		\path[fill=red] (X1) circle[radius=0.1];
		\path[fill=red] (X2) circle[radius=0.1];
		\path[fill=red] (X3) circle[radius=0.1];
	\end{tikzpicture}} \quad\quad 
	\raisebox{-1em}{\begin{tikzpicture}
	\coordinate (X1) at (0,0);
	\coordinate (X2) at (1,0);
	\coordinate (X3) at (2,0);
	\node[below] at (X1) {\small{$1$}};
	\node[below] at (X2) {\small{$2$}};
	\node[below] at (X3) {\small{$3$}};
	\draw[line width=1.5pt,->] (X1)--(0.6,0);
	\draw[line width=1.5pt] (0.5,0)--(X2);
	\draw[line width=1.5pt,dashed] (X2)--(X3);
	\path[fill=black] (X1) circle[radius=0.1];
	\path[fill=black] (X2) circle[radius=0.1];
	\path[fill=black] (X3) circle[radius=0.1];
\end{tikzpicture}}\\
 & \raisebox{-1em}{\begin{tikzpicture}
		\coordinate (X1) at (0,0);
		\coordinate (X2) at (1,0);
		\coordinate (X3) at (2,0);;
		\node[below] at (X1) {\small{$1$}};
		\node[below] at (X2) {\small{$2$}};
		\node[below] at (X3) {\small{$3$}};
\draw[line width=1.5pt,->] (X2)--(0.4,0);
\draw[line width=1.5pt] (0.5,0)--(X1);
		\draw[line width=1.5pt,->] (X2)--(1.6,0);
		\draw[line width=1.5pt] (1.5,0)--(X3);
		\path[fill=black] (X1) circle[radius=0.1];
		\path[fill=black] (X2) circle[radius=0.1];
		\path[fill=black] (X3) circle[radius=0.1];
\end{tikzpicture}} \quad\quad 
\raisebox{-1em}{\begin{tikzpicture}
		\coordinate (X1) at (0,0);
		\coordinate (X2) at (1,0);
		\coordinate (X3) at (2,0);;
		\node[below] at (X1) {\small{$1$}};
		\node[below] at (X2) {\small{$2$}};
		\node[below] at (X3) {\small{$3$}};
\draw[line width=1.5pt,->] (X2)--(0.4,0);
\draw[line width=1.5pt] (0.5,0)--(X1);
		\draw[line width=1.5pt,->] (X3)--(1.4,0);
		\draw[line width=1.5pt] (1.5,0)--(X2);
		\path[fill=black] (X1) circle[radius=0.1];
		\path[fill=black] (X2) circle[radius=0.1];
		\path[fill=black] (X3) circle[radius=0.1];
\end{tikzpicture}} \quad\quad 
\raisebox{-1em}{\begin{tikzpicture}
		\coordinate (X1) at (0,0);
		\coordinate (X2) at (1,0);
		\coordinate (X3) at (2,0);;
		\node[below] at (X1) {\small{$1$}};
		\node[below] at (X2) {\small{$2$}};
		\node[below] at (X3) {\small{$3$}};
\draw[line width=1.5pt,->] (X2)--(0.4,0);
\draw[line width=1.5pt] (0.5,0)--(X1);
\draw[line width=1.5pt,dashed] (X2)--(X3);
		\path[fill=black] (X1) circle[radius=0.1];
		\path[fill=black] (X2) circle[radius=0.1];
		\path[fill=black] (X3) circle[radius=0.1];
\end{tikzpicture}}\\
 & \raisebox{-1em}{\begin{tikzpicture}
		\coordinate (X1) at (0,0);
		\coordinate (X2) at (1,0);
		\coordinate (X3) at (2,0);;
		\node[below] at (X1) {\small{$1$}};
		\node[below] at (X2) {\small{$2$}};
		\node[below] at (X3) {\small{$3$}};
		\draw[line width=1.5pt,dashed] (X1)--(X2);
		\draw[line width=1.5pt,->] (X2)--(1.6,0);
		\draw[line width=1.5pt] (1.5,0)--(X3);
		\path[fill=black] (X1) circle[radius=0.1];
		\path[fill=black] (X2) circle[radius=0.1];
		\path[fill=black] (X3) circle[radius=0.1];
\end{tikzpicture}} \quad\quad 
\raisebox{-1em}{\begin{tikzpicture}
		\coordinate (X1) at (0,0);
		\coordinate (X2) at (1,0);
		\coordinate (X3) at (2,0);;
		\node[below] at (X1) {\small{$1$}};
		\node[below] at (X2) {\small{$2$}};
		\node[below] at (X3) {\small{$3$}};
		\draw[line width=1.5pt,dashed] (X1)--(X2);
		\draw[line width=1.5pt,dashed] (X2)--(X3);
		\draw[line width=1.5pt,->] (X3)--(1.4,0);
		\draw[line width=1.5pt] (1.5,0)--(X2);
		\path[fill=black] (X1) circle[radius=0.1];
		\path[fill=black] (X2) circle[radius=0.1];
		\path[fill=black] (X3) circle[radius=0.1];
\end{tikzpicture}} \quad\quad 
\raisebox{-1em}{\begin{tikzpicture}
		\coordinate (X1) at (0,0);
		\coordinate (X2) at (1,0);
		\coordinate (X3) at (2,0);
		\node[below] at (X1) {\small{$1$}};
		\node[below] at (X2) {\small{$2$}};
		\node[below] at (X3) {\small{$3$}};
		\draw[line width=1.5pt,dashed] (X2)--(X3);
		\draw[line width=1.5pt,dashed] (X1)--(X2);
		\path[fill=black] (X1) circle[radius=0.1];
		\path[fill=black] (X2) circle[radius=0.1];
		\path[fill=black] (X3) circle[radius=0.1];
\end{tikzpicture}}.
	\end{aligned}
\end{equation}
{\it i.e.} only family tree components in this decomposition. Furthermore, MoB calculation indicates the linear relations between these 8 components. For instance, if we focus on region $\omega_3>\omega_2>\omega_1$, four series with $\bar{\sigma}\in \{\{n_1,n_2,n_3\}$, $\{n_1,m_{3,2},n_3\}$, $\{m_{2,2},n_2,n_3\}$,  $\{m_{2,2},m_{3,2},n_3\}\}$ should be chosen, whose arguments in the series are
\begin{equation}\label{eq:arguments1234}   
\left\{\left(\frac{\omega_2}{\omega_3}\right)^{m_{3,2}}\left(\frac{\omega_1}{\omega_2}\right)^{m_{2,2}},\left(\frac{\omega_2}{\omega_3}\right)^{n_2}\left(\frac{\omega_1}{\omega_3}\right)^{m_{2,2}},\left(\frac{\omega_2}{\omega_3}\right)^{m_{3,2}}\left(\frac{\omega_1}{\omega_2}\right)^{n_1},\left(\frac{\omega_2}{\omega_3}\right)^{n_2}\left(\frac{\omega_1}{\omega_3}\right)^{n_1}\right\}
\end{equation}
respectively. Accordingly, we have four series
\begin{align}
    &\text{Series 1}: (-\text{i})^{3{+}q_{123}}\prod_{i=1}^3\frac{\Gamma(q_i)}{\omega_i^{q_i}},\\
     &\text{Series 2}: -\frac{(-\text{i})^{2}}{(\text{i}\omega_3)^{q_{23}}}\sum_{n_2}\phi_{n_{2}}\frac{\Gamma(n_2{+}q_2{+}q_3)}{(n_2{+}q_2)}\left(\frac{\omega_2}{\omega_3}\right)^{n_2}\times(-\text{i})^{1+q_1}\frac{\Gamma(q_1)}{\omega_1^{q_1}},\\
      &\text{Series 3}: -\frac{(-\text{i})^{2}}{(\text{i}\omega_2)^{q_{12}}}\sum_{n_1}\phi_{n_{1}}\frac{\Gamma(n_1{+}q_1{+}q_2)}{(n_1{+}q_1)}\left(\frac{\omega_1}{\omega_2}\right)^{n_1}\times(-\text{i})^{1+q_3}\frac{\Gamma(q_3)}{\omega_3^{q_3}},\\
    &\text{Series 4}:\frac{(-\text{i})^{3}}{(\text{i}\omega_3)^{q_{123}}}\sum_{n_1,n_2}\phi_{n_1,n_2}\frac{\Gamma(n_1{+}n_2{+}q_{123})}{(n_1{+}q_1)(n_1{+}n_2{+}q_{12})}\left(\frac{\omega_1}{\omega_3}\right)^{n_1}\left(\frac{\omega_2}{\omega_3}\right)^{n_2}.
\end{align}
The sum of the four basis series giving us, the result for the 3-site chain can be interpreted as a consequence of the linear relation
\begin{equation}
    \raisebox{-1em}{\begin{tikzpicture}
				\coordinate (X1) at (0,0);
				\coordinate (X2) at (1,0);
				\coordinate (X3) at (2,0);;
				\node[below] at (X1) {\small{$1$}};
				\node[below] at (X2) {\small{$2$}};
				\node[below] at (X3) {\small{$3$}};
				\draw[line width=1.5pt,->] (X1)--(0.6,0);
				\draw[line width=1.5pt] (0.5,0)--(X2);
				\draw[line width=1.5pt,->] (X2)--(1.6,0);
				\draw[line width=1.5pt] (1.5,0)--(X3);
				\path[fill=black] (X1) circle[radius=0.1];
				\path[fill=black] (X2) circle[radius=0.1];
				\path[fill=black] (X3) circle[radius=0.1];
		\end{tikzpicture}}=\raisebox{-1em}{\begin{tikzpicture}
		\coordinate (X1) at (0,0);
		\coordinate (X2) at (1,0);
		\coordinate (X3) at (2,0);
		\node[below] at (X1) {\small{$1$}};
		\node[below] at (X2) {\small{$2$}};
		\node[below] at (X3) {\small{$3$}};
		\draw[line width=1.5pt,dashed] (X2)--(X3);
		\draw[line width=1.5pt,dashed] (X1)--(X2);
		\path[fill=black] (X1) circle[radius=0.1];
		\path[fill=black] (X2) circle[radius=0.1];
		\path[fill=black] (X3) circle[radius=0.1];
\end{tikzpicture}}-\raisebox{-1em}{\begin{tikzpicture}
		\coordinate (X1) at (0,0);
		\coordinate (X2) at (1,0);
		\coordinate (X3) at (2,0);;
		\node[below] at (X1) {\small{$1$}};
		\node[below] at (X2) {\small{$2$}};
		\node[below] at (X3) {\small{$3$}};
		\draw[line width=1.5pt,dashed] (X1)--(X2);
		\draw[line width=1.5pt,dashed] (X2)--(X3);
		\draw[line width=1.5pt,->] (X3)--(1.4,0);
		\draw[line width=1.5pt] (1.5,0)--(X2);
		\path[fill=black] (X1) circle[radius=0.1];
		\path[fill=black] (X2) circle[radius=0.1];
		\path[fill=black] (X3) circle[radius=0.1];
\end{tikzpicture}}- 
\raisebox{-1em}{\begin{tikzpicture}
		\coordinate (X1) at (0,0);
		\coordinate (X2) at (1,0);
		\coordinate (X3) at (2,0);;
		\node[below] at (X1) {\small{$1$}};
		\node[below] at (X2) {\small{$2$}};
		\node[below] at (X3) {\small{$3$}};
\draw[line width=1.5pt,->] (X2)--(0.4,0);
\draw[line width=1.5pt] (0.5,0)--(X1);
\draw[line width=1.5pt,dashed] (X2)--(X3);
		\path[fill=black] (X1) circle[radius=0.1];
		\path[fill=black] (X2) circle[radius=0.1];
		\path[fill=black] (X3) circle[radius=0.1];
\end{tikzpicture}}+\raisebox{-1em}{\begin{tikzpicture}
		\coordinate (X1) at (0,0);
		\coordinate (X2) at (1,0);
		\coordinate (X3) at (2,0);;
		\node[below] at (X1) {\small{$1$}};
		\node[below] at (X2) {\small{$2$}};
		\node[below] at (X3) {\small{$3$}};
\draw[line width=1.5pt,->] (X2)--(0.4,0);
\draw[line width=1.5pt] (0.5,0)--(X1);
		\draw[line width=1.5pt,->] (X3)--(1.4,0);
		\draw[line width=1.5pt] (1.5,0)--(X2);
		\path[fill=black] (X1) circle[radius=0.1];
		\path[fill=black] (X2) circle[radius=0.1];
		\path[fill=black] (X3) circle[radius=0.1];
\end{tikzpicture}}
\end{equation}
On the other hand, for region $\omega_3>\omega_1>\omega_2$, we have $\bar{\sigma}\in\{\{n_1,m_{3,2},n_3\}$, $\{m_{2,2},m_{3,2},n_3\}$, $\{n_1,m_{2,2},n_3\}\}$. Note that the first two of these solutions are exactly the Series 2 and 4 above. It can be seen from \eqref{eq:arguments1234} that their arguments $\left(\frac{\omega_1}{\omega_3}\right)^{k_1}\left(\frac{\omega_1}{\omega_3}\right)^{k_2}$ imply convergence in both regions. Besides these, we have one more basis series whose argument reads  $\left(\frac{\omega_1}{\omega_3}\right)^{m_{3,2}}\left(\frac{\omega_2}{\omega_1}\right)^{n_2}$ namely
\begin{equation}
    \text{Series 5}: \frac{(-\text{i})^{2}}{(\text{i}\omega_1)^{q_{12}}}\sum_{n_2}\phi_{n_{2}}\frac{\Gamma(n_2{+}q_1{+}q_2)}{(n_2{+}q_2)}\left(\frac{\omega_2}{\omega_1}\right)^{n_2}\times(-\text{i})^{1+q_3}\frac{\Gamma(q_3)}{\omega_3^{q_3}}.
\end{equation}
Summing over the series $2,4,5$ yields the relation
\begin{equation}
    \raisebox{-1em}{\begin{tikzpicture}
				\coordinate (X1) at (0,0);
				\coordinate (X2) at (1,0);
				\coordinate (X3) at (2,0);;
				\node[below] at (X1) {\small{$1$}};
				\node[below] at (X2) {\small{$2$}};
				\node[below] at (X3) {\small{$3$}};
				\draw[line width=1.5pt,->] (X1)--(0.6,0);
				\draw[line width=1.5pt] (0.5,0)--(X2);
				\draw[line width=1.5pt,->] (X2)--(1.6,0);
				\draw[line width=1.5pt] (1.5,0)--(X3);
				\path[fill=black] (X1) circle[radius=0.1];
				\path[fill=black] (X2) circle[radius=0.1];
				\path[fill=black] (X3) circle[radius=0.1];
		\end{tikzpicture}}=\raisebox{-1em}{\begin{tikzpicture}
	\coordinate (X1) at (0,0);
	\coordinate (X2) at (1,0);
	\coordinate (X3) at (2,0);
	\node[below] at (X1) {\small{$1$}};
	\node[below] at (X2) {\small{$2$}};
	\node[below] at (X3) {\small{$3$}};
	\draw[line width=1.5pt,->] (X1)--(0.6,0);
	\draw[line width=1.5pt] (0.5,0)--(X2);
	\draw[line width=1.5pt,dashed] (X2)--(X3);
	\path[fill=black] (X1) circle[radius=0.1];
	\path[fill=black] (X2) circle[radius=0.1];
	\path[fill=black] (X3) circle[radius=0.1];
\end{tikzpicture}}-\raisebox{-1em}{\begin{tikzpicture}
		\coordinate (X1) at (0,0);
		\coordinate (X2) at (1,0);
		\coordinate (X3) at (2,0);;
		\node[below] at (X1) {\small{$1$}};
		\node[below] at (X2) {\small{$2$}};
		\node[below] at (X3) {\small{$3$}};
		\draw[line width=1.5pt,dashed] (X1)--(X2);
		\draw[line width=1.5pt,dashed] (X2)--(X3);
		\draw[line width=1.5pt,->] (X3)--(1.4,0);
		\draw[line width=1.5pt] (1.5,0)--(X2);
		\path[fill=black] (X1) circle[radius=0.1];
		\path[fill=black] (X2) circle[radius=0.1];
		\path[fill=black] (X3) circle[radius=0.1];
\end{tikzpicture}}+\raisebox{-1em}{\begin{tikzpicture}
		\coordinate (X1) at (0,0);
		\coordinate (X2) at (1,0);
		\coordinate (X3) at (2,0);;
		\node[below] at (X1) {\small{$1$}};
		\node[below] at (X2) {\small{$2$}};
		\node[below] at (X3) {\small{$3$}};
\draw[line width=1.5pt,->] (X2)--(0.4,0);
\draw[line width=1.5pt] (0.5,0)--(X1);
		\draw[line width=1.5pt,->] (X3)--(1.4,0);
		\draw[line width=1.5pt] (1.5,0)--(X2);
		\path[fill=black] (X1) circle[radius=0.1];
		\path[fill=black] (X2) circle[radius=0.1];
		\path[fill=black] (X3) circle[radius=0.1];
\end{tikzpicture}}
\end{equation}

\begin{figure}[t]
\centering
\begin{tikzpicture} [every node/.style={draw, circle,fill=black!10, minimum size= 1.22 cm, align=center},
  dot/.style={circle, fill=gray, inner sep=1.5pt}][scale=0.1]

% Left side: G_n
\node (Gn) at (0,0) {$G_N$};

% Equivalence symbol
\node[draw=none,fill=none] (equiv) at (1,0) {$\equiv$};
\node[draw=none,fill=none] (three) at (1.5,0) {\Large 3};

% G_{n-1} node
\node (Gn1) at (3,0) {$G_{N-1}$};

% Vertical line and dot below G_{n-1}
\draw[fill] (3.6,0) circle (2pt);
\draw[line width=1.5pt,->]  (3.6,0) -- (4.8,0);
\draw[line width=1.5pt]  (3.8,0) -- (6,0);
\draw[fill] (6,0) circle (2pt);
% Line connecting to G_{n-2}
%\draw (9,-0.75) -- (4,-1.25);

% G_{n-2} node
\node (Gn2) at (8,0) {$G_{N-2}$};
\node[draw=none,fill=none] (three) at (6.8,0) {${-}$};
% Dot above and below G_{n-2}
\draw[fill] (8.6,0) circle (2pt);
\draw[line width=1.5pt,->]  (8.6,0) -- (9.6,0);
\draw[line width=1.5pt,-_]  (9.6,0) -- (10.6,0);
\draw[fill] (10.6,0) circle (2pt);
\draw[line width=1.5pt,] (10.6,0) -- (11.6,0);
\draw[line width=1.5pt,->] (12.6,0) -- (11.6,0);
\draw[fill] (12.6,0) circle (2pt);
% Y-branch from bottom dot
%\draw (4,-3.25) -- (3.5,-3.75);
%\draw (4,-3.25) -- (4.5,-3.75);
%\draw[fill] (3.5,-3.75) circle (2pt);
%\draw[fill] (4.5,-3.75) circle (2pt);
\end{tikzpicture} 
\caption{Recurrence relation for the number of family chains with $N$ sites}\label{fig:0}
\vspace{2ex}
\end{figure}
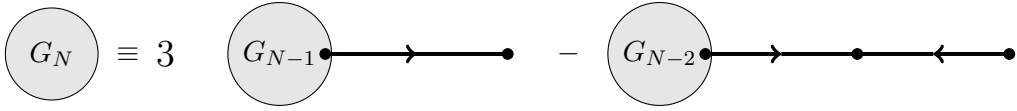

\paragraph{All basis series of family chain from MoB}
We can therefore summarize the physical meanings of all basis series from MoB calculation of an $n$-site chain. For general $N$-site cases, MoB always gives us all family chains, or products of family chains, in the basic time integral decomposition of the full $N$-site chain. More precisely, by choosing different independent indices $\sigma$ from \eqref{eq:summand2}, we have the following different basis series:
\begin{enumerate}
    \item For $\sigma=\{n_2,\cdots,n_N\}$, arguments of the series are $\prod_{i=2}^{N}\left(\frac{\omega_i}{\omega_1}\right)^{n_i}$. This basis series corresponds to the original family chain \eqref{eq:masleschain}.
    \item For $\sigma=\{n_1,n_2,\cdots,n_{i{-}1},n_{i{+}1},\cdots,n_N\}$, arguments of the series are 
    \[\prod_{\substack{j=1\\j\neq i}}^{N}\left(\frac{\omega_j}{\omega_i}\right)^{n_j}\]
    This basis series corresponds to a family chain by reversing the arrows such that node $i$ becomes the ancestor.
    \item For $\sigma=\{n_2,\cdots,n_{i{-}1},m_{i,2},n_{i{+}1},\cdots,n_N\}$, arguments of the series are \[\prod_{j=2}^{i{-}1}\left(\frac{\omega_j}{\omega_1}\right)^{n_j}\prod_{j=i{+}1}^{N}\left(\frac{\omega_j}{\omega_i}\right)^{n_j}\left(\frac{\omega_1}{\omega_i}\right)^{m_{i,2}}.\]  The basis series in this case corresponds to a family chain by replacing edge $(i{-}1i)$ with dashed edge. When multiple $n_i$'s are replaced by $m_{i,2}$'s in $\sigma=\{n_2,\cdots,n_N\}$, we get family chains with more thick edges replaced by dashed edges.
    \item Combining 2 and 3 above, we get a factorized family chain with some arrows reversed. For instance, for $\sigma=\{n_2,\cdots,n_{i{-}1},m_{k,2},n_{i{+}1},\cdots,n_N\}$ with $k<i$, arguments of the series are \[\prod_{j=2}^{k{-}1}\left(\frac{\omega_j}{\omega_1}\right)^{n_j}\prod_{\substack{j={k}\\j\neq i}}^{N}\left(\frac{\omega_j}{\omega_i}\right)^{n_j}\left(\frac{\omega_1}{\omega_i}\right)^{m_{k,2}}.\]  The basis series in this case  corresponds to a family chain obtained by replacing edge $(k{-}1,k)$ with a dashed edge, and reversing the arrows from node $k$ to node $i$\footnote{ $\sigma=\{n_2,\cdots,n_{i{-}1},m_{k,2},n_{i{+}1},\cdots,n_N\}$ with $k>i$ are therefore not allowed by the bracket equations.}. 
\end{enumerate}
The total number of basis series solutions from MoB for $N$-site chain $a_N$, satisfies the recursion relation, and by solving this recursive relation we have 
\begin{equation}
    a_N{=}\frac{1}{\sqrt{5}}\left(\left(\frac{3{+}\sqrt{5}}{2}\right)^N{-}\left(\frac{3{-}\sqrt{5}}{2}\right)^N\right)
\end{equation}
This can be understood as follows, adding an edge from node $N{-}1$ of an $(N{-}1)$-site family chain, naively triples the number of family chains (as this edge can be pointing inward,outward or be dashed). However, an extra edge with arrow pointing from node $N$ to node $N{-}1$ will not be allowed, if in the original chain there is already an arrow pointing out from node $N{-}2$ to $N{-}1$, which therefore results in the counting (Fig.\ref{fig:0}). 

\paragraph{Series representation for general family tree graphs}
Finally, we move on to the most general family trees. In \cite{Xianyu:2023ytd}, a general series formula for general family tree integrals has been presented as 
\begin{equation}\label{eq:treeres}
    P[\mathcal{N}(1\cdots N)]=\frac{(-\text{i})^N}{(\text{i}\omega_1)^{\tilde{q}_1}}\sum_{n_2\cdots n_{N}}\phi_{n_2\cdots n_N}\ \Gamma(\tilde{n}_2{+}\tilde{q}_1)\prod_{i=2}^N\frac{1}{(\tilde{n}_i{+}\tilde{q}_i)}\left(\frac{\omega_i}{\omega_1}\right)^{n_i},
\end{equation}
which is similar to the chain result \eqref{eq:chainres} with the only difference being $\tilde{q}_i$ and $\tilde{n}_i$ here are the sum of $q_j$ and $n_j$ of all descendants of node $i$. The ancestor of the diagram is again chosen as node $1$, and $\mathcal{N}$ here denotes the time ordered structure of the family tree. We can directly derive the series solution by using MoB, using an almost identical calculation as for the family chain graphs. After a change of variables $\alpha_i\to\frac{s_i}{s_{i{+}1}}$ in \eqref{eq:Mellinmassless}, Euler-Mellin representation for general family tree graph reads
\begin{equation}\label{eq:tree1}
   (-\text{i})^{N+\tilde{q}_{1}}\Gamma(\tilde{q}_{1})\int_0^\infty \prod_{i=2}^N\left({\rm d}s_i s_i^{\tilde{q}_i{-}1}(1{+}s_i)^{\tilde{q}_{1}{-}\tilde{q}_i{-}1}\right)\ \left(\sum_{i=1}^N\omega_i\prod_{j=2}^N s_j^{k_{i,j}}\prod_{j=2}^N(1{+}s_j)^{1{-}k_{i,j}}\right)^{-\tilde{q}_{1}},
\end{equation}
where we define the notation $k_{i,j}$ as
\begin{equation}
	k_{i,j}=\left\{
	\begin{aligned}
		&1,\qquad & \text{if $i{=}j$ or  is descendant of $j$}\\
		&0,\qquad & \text{otherwise}\\
	\end{aligned}
\right .
\end{equation}
and we still denote $\tilde{q}_i$ as sum of all twists for node $i$ and its descendants $\tilde{q}_i=\sum_{j=1}^Nk_{j,i}q_j$. As we did in the  MoB computation for family chain graphs, we still get $3N-2$ indices and $2N-1$ brackets equations with the same summand \eqref{eq:summand2}. The only difference now is that we have 
\[A_i=\sum_{j=1}^N(1{-}k_{j,i})n_j{+}\tilde{q}_1-\tilde{q}_i-1,\ B_i=\sum_{j=1}^Nk_{j,i} n_j{+}\tilde{q}_i{-}1.\]
Suppose we again work in region \eqref{eq:region1}, we still need to solve $n_2,\cdots,n_N$ from bracket equations, whose solution will be
\begin{align}
    &m_{i,1}=-1,\ m_{i,2}=-\tilde{q}_{i}{-}\sum_{j=2}^Nk_{i,j}n_{j}=-\tilde{q}_i{-}\tilde{n}_i,\ \ n_1=-\sum_{i=2}^Nn_i-\tilde{q}_1.
\end{align}
Starting with the same summand \eqref{eq:summand1} and noticing that $-A_i={-}n_1{-}\sum_{j=2}^N(1{-}k_{j,i})n_j{-}\tilde{q}_1{+}\tilde{q}_i+1=-m_{i,2}{+}1$, we get the result quoted in \eqref{eq:treeres}.

\section{Cosmological amplitudes with massive exchanges}
\label{sec:massive correlators}
After talking about the massless cases as a warm-up, now we move on to cosmological amplitudes with massive exchanges. We will apply MoB to cosmological integrals with a general $\nu$ in the propagators, satisfying $|\Re(-\text{i}\nu)|<\frac12$, which are again multivariable generalized hypergeometric functions. We begin with defining {\it massive family trees}, which form a basis for all tree-level massive amplitudes.  We then present general basis series solutions for tree-level cosmological amplitudes by a systematic computation using MoB. For an $N$-site massive family tree, its result is composed of $2^{2(N{-}1)}$ basis series, which can be organized into one basic building block called primary series, together with other $2^{2(N{-}1)}{-}1$ basis series which are related to the primary series by shifting parameters. 
For the summands of these basis series, we obtain Feynman-like rules, and we see a recursive structure generated from nested-time structure between the series solutions. Finally, to demonstrate features of the solutions, we will consider four point cosmological amplitudes in more detail and  discuss their result in different physical regions, which MoB naturally provides.

It is crucial to mention at the outset that, since we have the relation \eqref{eq:1} for the Hankel functions and also its conjugate $\text{H}_\mu^{(1)}(x)$, the cosmological amplitudes with massive exchanges can be related to massless cases by extra folds of integration. Suppose that we are dealing with a tree-level integral with a massive exchange between nodes $i$ and $i{+}1$ 
\begin{equation}\label{eq:4.1}
    \int_{-\infty}^0{\rm d}\tau_i{\rm d}\tau_{i{+}1}\mathcal{I}_{L}(\tau_{i})D_{a_1a_2}(k,\tau_i,\tau_{i{+}1})\mathcal{I}_{R}(\tau_{i{+}1})
\end{equation}
According to \eqref{eq:btobulk1} and \eqref{eq:btobulk2}, all terms from \eqref{eq:4.1} are generally of the form 
\begin{equation}
    \int_{-\infty}^0{\rm d}\tau_i{\rm d}\tau_{i{+}1}\mathcal{I}_{L}(\tau_{i})\text{H}_{-\text{i}\nu}^{(2)}(-k\tau_i)\text{H}_{\text{i}\nu}^{(1)}(-k\tau_{i{+}1})\mathcal{I}_{R}(\tau_{i{+}1})\theta_{i,i{+}1},
\end{equation}
its complex conjugate, or without the nested-time factor $\theta_{i,i{+}1}$. For any of these cases, following the relation \eqref{eq:1}, we have
\begin{align}\label{1}
    {\text H}_{-\text{i}\nu}^{(2)}(-k\tau_1){\text H}_{\text{i}\nu}^{(1)}(-k\tau_2){=}&\frac{4{\rm e}^{2\pi \nu}{\rm cosh}(\pi \nu)}{\pi^2}\left(\frac{\tau_2}{\tau_1}\right)^{\text{i}\nu}{\rm e}^{-\text{i}k(\tau_2{-}\tau_1)}\\
    &\int_0^{\infty}{\rm d}s_1{\rm d}s_2{\rm e}^{-2\text{i}k\tau_2 s_2+2\text{i}k\tau_1s_1}(s_1(1{+}s_1))^{-\text{i}\nu-1/2}(s_2(1{+}s_2))^{\text{i}\nu-1/2}, \nonumber
\end{align}
where we have used
\begin{equation*}
    \frac{1}{\Gamma(\frac12{-}\text{i}\nu)\Gamma(\frac12{+}\text{i}\nu)}=\frac{\cosh(\pi \nu)}{\pi}.
\end{equation*}
In this section, we will often use the shorthand notations
\begin{equation}\label{eq:BC}
    B:=\frac{\cosh(\pi \nu)}{\pi},\ \ C:=\frac{4{\rm e}^{2\pi \nu}{\rm cosh}(\pi \nu)}{\pi^2}
\end{equation}
for simplicity. We therefore obtain
\begin{equation}\label{eq:2}
    \frac{4{\rm e}^{2\pi \nu}{\rm cosh}(\pi \nu)}{\pi^2}\int_0^{\infty}{\rm d}s_1{\rm d}s_2(s_1(1{+}s_1))^{-\text{i}\nu-1/2}(s_2(1{+}s_2))^{\text{i}\nu-1/2}\hat{\mathcal{I}}(s_1,s_2)
\end{equation}
with $\hat{\mathcal{I}}(s_1,s_2)$ being the deformed cosmological integral with massless exchange between $\tau_i$ and $\tau_{i{+}1}$ as
\begin{equation}
    \hat{\mathcal{I}}(s_1,s_2)=\int_{-\infty}^0{\rm d}\tau_i{\rm d}\tau_{i{+}1}\mathcal{I}_{L}(\tau_i)(-\tau_i)^{-\text{i}\nu}(-
    \tau_{i{+}1})^{\text{i}\nu}{\rm e}^{-\text{i}k\tau_{i{+}1}(2s_2{+}1)}{\rm e}^{\text{i}k\tau_{i}(2s_1{+}1)}\mathcal{I}_R(\tau_{i{+}1})\theta_{i,i{+}1}.
\end{equation}
The twist for node $i$ and $i{+}1$ are shifted by $-\text{i}\nu$ and $\text{i}\nu$ respectively, and the energies from the bulk-to-boundary propagators of the two nodes are shifted by ${+}k(2s_1{+}1)$ and ${-}k(2{s_2{+}1})$ respectively. Using the Euler-Mellin representation for massless cosmological integrals \eqref{eq:Mellinmassless}, calculation for integrals with massive exchanges will therefore be straightforward with MoB. We will therefore begin with \eqref{eq:2}, recast any massive $N$-site cosmological integral, as $2(N-1)$-fold of integration over the corresponding massless integrals, and  present compact solutions for general massive cosmological amplitudes by using MoB.\footnote{To make the mode functions for massless and massive scalar fields to decay at $\tau\to-\infty$, for a node $v$ with $a_v=\pm$,  it is necessary to assign a small imaginary part to $\tau_v$ as $\tau_v\to \tau_v(1\mp\text{i}0^+)$. The integrand of cosmological amplitudes \eqref{1} are then suppressed at $\tau=-\infty$ by the condition $k,s>0$ and the exponential function. 
}

\subsection{Decomposition into massive family trees}
Similar to the massless case, we first introduce a convenient basis and consider the decomposition of a general massive cosmological amplitude into this basis, which will be our starting point for applying MoB. Compared to the massless cases \eqref{eq:correprop}, apart from the nested-time structure, the massive propagators \eqref{eq:btobulk1} and \eqref{eq:btobulk2} enjoy an additional structure due to the location of different Hankel functions. Again with the help of $\theta_{i,j}=1{-}\theta_{i,j}$, we can generally expand any massive cosmological amplitude in terms of  tree integrals or their products with partial-ordered nested-time structure. However, each edge of the tree integral now belongs to one of the three types in the following (Taking two-site basis as an example)\footnote{Here we slightly abuse the diagrammatic notation, since in this section we use the black thick arrow again to represent one kind of basic time integral with Hankel functions. This should not be confused with the notation in the last section, where we used black thick arrow to represent conformally-coupled basic time integrals.}: 
\begin{align} 
    \mathcal{I}_2{:=}\raisebox{-1em}{\begin{tikzpicture}[scale=0.75]
		\coordinate (X1) at (0,0);
		\coordinate (X2) at (1.5,0);
		\node[below] at (X1) {\small{$1$}};
		\node[below] at (X2) {\small{$2$}};
		\draw[line width=1.5pt,->] (X1)--(0.9,0);
        \draw[line width=1.5pt] (0.75,0)--(X2);
		\path[fill=black] (X1) circle[radius=0.11];
		\path[fill=black] (X2) circle[radius=0.11];
\end{tikzpicture}}&{=}(-\text{i})^2\int_{-\infty}^0(-\tau_1)^{p_1-1}(-\tau_2)^{p_2-1}{\rm e}^{\text{i}w_i\tau_1+\text{i}w_2\tau_2}\text{H}_{-\text{i}\nu}^{(2)}(-k\tau_1)\text{H}_{\text{i}\nu}^{(1)}(-k\tau_2)\theta_{2,1},\label{eq:twositechain}  \\
\tilde{\mathcal{I}}_2{:=}\raisebox{-1em}{\begin{tikzpicture}[scale=0.75]
		\coordinate (X1) at (0,0);
		\coordinate (X2) at (1.5,0);
		\node[below] at (X1) {\small{$1$}};
		\node[below] at (X2) {\small{$2$}};
		\draw[line width=1.5pt,->,black!30] (X1)--(0.9,0);
        \draw[line width=1.5pt,black!30] (0.75,0)--(X2);
		\path[fill=black] (X1) circle[radius=0.11];
		\path[fill=black] (X2) circle[radius=0.11];
\end{tikzpicture}}&{=}(-\text{i})^2\int_{-\infty}^0(-\tau_1)^{p_1-1}(-\tau_2)^{p_2-1}{\rm e}^{\text{i}w_i\tau_1+\text{i}w_2\tau_2}\text{H}_{-\text{i}\nu}^{(1)}(-k\tau_1)\text{H}_{\text{i}\nu}^{(2)}(-k\tau_2)\theta_{2,1},\label{eq:tI2}\\
\mathcal{I}_{1,1}{:=}\raisebox{-1em}{\begin{tikzpicture}[scale=0.75]
		\coordinate (X1) at (0,0);
		\coordinate (X2) at (1.5,0);
		\node[below] at (X1) {\small{$1$}};
		\node[below] at (X2) {\small{$2$}};
		\draw[line width=1.5pt,dashed,->] (X1)--(0.9,0);
        \draw[line width=1.5pt,dashed] (0.75,0)--(X2);
		\path[fill=black] (X1) circle[radius=0.11];
		\path[fill=black] (X2) circle[radius=0.11];
\end{tikzpicture}}&{=}(-\text{i})^2\int_{-\infty}^0(-\tau_1)^{p_1-1}(-\tau_2)^{p_2-1}{\rm e}^{\text{i}w_i\tau_1+\text{i}w_2\tau_2}\text{H}_{-\text{i}\nu}^{(2)}(-k\tau_1)\text{H}_{\text{i}\nu}^{(1)}(-k\tau_2).\label{eq:I11}
\end{align}
Furthermore,  for wavefunction coefficients we need another piece namely,
\begin{equation} \label{mft4}
   \tilde{\mathcal{I}}_{1,1}{:=}\raisebox{-1em}{\begin{tikzpicture}[scale=0.75]
		\coordinate (X1) at (0,0);
		\coordinate (X2) at (1.5,0);
		\node[below] at (X1) {\small{$1$}};
		\node[below] at (X2) {\small{$2$}};
		\draw[line width=1.5pt,dashed] (X1)--(X2);
		\path[fill=black] (X1) circle[radius=0.11];
		\path[fill=black] (X2) circle[radius=0.11];
\end{tikzpicture}}{=}(-\text{i})^2\int_{-\infty}^0(-\tau_1)^{p_1-1}(-\tau_2)^{p_2-1}{\rm e}^{\text{i}w_i\tau_1+\text{i}w_2\tau_2}\text{H}_{-\text{i}\nu}^{(2)}(-k\tau_1)\text{H}_{{-}\text{i}\nu}^{(2)}(-k\tau_2). 
\end{equation}
Note that in these expression and also in this section, because of the extra factor $(\tau_1\tau_2)^{\frac12}$ in the propagators \eqref{eq:btobulk0} and \eqref{eq:btobulk1}, we introduce the notation
\begin{equation}\label{eq:shiftq}
    p_i=q_i{+}\frac k2,\ \ k=\text{number of Hankel functions connected with node $i$}
\end{equation}
for simplicity. This decomposition, similar to family tree decomposition for massless amplitudes, is again not unique. For instance, two-site chain function $\mathcal{I}_{++}$ can be decomposed as
\begin{align}
    \mathcal{I}_{++}&=(-\text{i})^2\int_{-\infty}^0{\rm d}\tau_1{\rm d}\tau_1(-\tau_1)^{q_1-1}(-\tau_2)^{q_2-1}{\rm e}^{\text{i}w_1\tau_1+\text{i}w_2\tau_2}D_{++}(\tau_1,\tau_2,k)\nonumber\\
    &=\frac\pi4{\rm e}^{{-}\pi \nu}\left(\raisebox{-1em}{\begin{tikzpicture}[scale=0.75]
		\coordinate (X1) at (0,0);
		\coordinate (X2) at (1.5,0);
		\node[below] at (X1) {\small{$1$}};
		\node[below] at (X2) {\small{$2$}};
		\draw[line width=1.5pt,->] (X1)--(0.9,0);
        \draw[line width=1.5pt] (0.75,0)--(X2);
		\path[fill=black] (X1) circle[radius=0.11];
		\path[fill=black] (X2) circle[radius=0.11];
\end{tikzpicture}}{+}\raisebox{-1em}{\begin{tikzpicture}[scale=0.75]
		\coordinate (X1) at (0,0);
		\coordinate (X2) at (1.5,0);
		\node[below] at (X1) {\small{$2$}};
		\node[below] at (X2) {\small{$1$}};
		\draw[line width=1.5pt,->] (X1)--(0.9,0);
        \draw[line width=1.5pt] (0.75,0)--(X2);
		\path[fill=black] (X1) circle[radius=0.11];
		\path[fill=black] (X2) circle[radius=0.11];
\end{tikzpicture}}\right)\nonumber\\
    &=\frac\pi4{\rm e}^{{-}\pi \nu}\left(\raisebox{-1em}{\begin{tikzpicture}[scale=0.75]
		\coordinate (X1) at (0,0);
		\coordinate (X2) at (1.5,0);
		\node[below] at (X1) {\small{$1$}};
		\node[below] at (X2) {\small{$2$}};
		\draw[line width=1.5pt,->] (X1)--(0.9,0);
        \draw[line width=1.5pt] (0.75,0)--(X2);
		\path[fill=black] (X1) circle[radius=0.11];
		\path[fill=black] (X2) circle[radius=0.11];
\end{tikzpicture}}-\raisebox{-1em}{\begin{tikzpicture}[scale=0.75]
		\coordinate (X1) at (0,0);
		\coordinate (X2) at (1.5,0);
		\node[below] at (X1) {\small{$1$}};
		\node[below] at (X2) {\small{$2$}};
		\draw[line width=1.5pt,->,black!30] (X1)--(0.9,0);
        \draw[line width=1.5pt,black!30] (0.75,0)--(X2);
		\path[fill=black] (X1) circle[radius=0.11];
		\path[fill=black] (X2) circle[radius=0.11];
\end{tikzpicture}}+\raisebox{-1em}{\begin{tikzpicture}[scale=0.75]
		\coordinate (X1) at (0,0);
		\coordinate (X2) at (1.5,0);
		\node[below] at (X1) {\small{$2$}};
		\node[below] at (X2) {\small{$1$}};
		\draw[line width=1.5pt,dashed,->] (X1)--(0.9,0);
        \draw[line width=1.5pt,dashed] (0.75,0)--(X2);
		\path[fill=black] (X1) circle[radius=0.11];
		\path[fill=black] (X2) circle[radius=0.11];
\end{tikzpicture}}\right),
\end{align}
leveraging the relation
\begin{align}
D_{++}(\tau_1,\tau_2,k){=}\frac\pi4{\rm e}^{{-}\pi \nu}(\tau_1\tau_2)^{\frac12}\left( {\text H}_{\text{i}\nu}^{(1)}\right.&(-k\tau_1){\text H}_{-\text{i}\nu}^{(2)}(-k\tau_2){+}\theta(\tau_2{-}\tau_1)\times \\
&\left. [{\text H}_{-\text{i}\nu}^{(2)}(-k\tau_1){\text H}_{\text{i}\nu}^{(1)}(-k\tau_2){-}{\text H}_{\text{i}\nu}^{(1)}(-k\tau_1){\text H}_{-\text{i}\nu}^{(2)}(-k\tau_2)]\right).\nonumber
\end{align}
Note that, since piece $\mathcal{I}_{1,1}$ does not involve $\theta_{i,j}$, tree diagrams with one dashed-line propagator are indeed products of two non-overlapping integrals. However, due to the existence of two Hankel functions in \eqref{eq:I11}, it is still not equivalent to a product of two lower massive family tree diagrams. Arrow on the dashed line denotes the location of $\text{H}_{-\text{i}\nu}^{(2)}$ or $\text{H}_{\text{i}\nu}^{(1)}$. Each of the pieces in a tree with dashed-propagators will then be massive nested-time integrals with extra Hankel function factors on the leaves. For instance, massive integral
\begin{equation}
    \raisebox{-1em}{\begin{tikzpicture}[scale=0.75]
		\coordinate (X1) at (0,0);
		\coordinate (X2) at (1.5,0);
        \coordinate (X3) at (3,0);
            \coordinate (X4) at (4.5,0);
		\node[below] at (X1) {\small{$1$}};
		\node[below] at (X2) {\small{$2$}};
        \node[below] at (X4) {\small{$4$}};
        \node[below] at (X3) {\small{$3$}};
		\draw[line width=1.5pt,->] (X1)--(0.9,0);
        \draw[line width=1.5pt] (0.75,0)--(X2);
        \draw[line width=1.5pt,dashed,->] (X2)--(2.4,0);
        \draw[line width=1.5pt,dashed] (2.25,0)--(X3);
        \draw[line width=1.5pt,->] (X3)--(3.9,0);
        \draw[line width=1.5pt] (3.75,0)--(X4);
		\path[fill=black] (X1) circle[radius=0.11];
		\path[fill=black] (X2) circle[radius=0.11];
        \path[fill=black] (X3) circle[radius=0.11];
        \path[fill=black] (X4) circle[radius=0.11];
\end{tikzpicture}}
\end{equation}
involves two pieces which are deformed two-site nested-time integrals
\begin{align}\label{eq:twositechain2p}
    \mathcal{I}_{2p}= \raisebox{-1em}{\begin{tikzpicture}[scale=0.75]
		\coordinate (X1) at (0,0);
		\coordinate (X2) at (1.5,0);
		\node[below] at (X1) {\small{$1$}};
		\node[below] at (X2) {\small{$2_{2,k_{23}}$}};
		\draw[line width=1.5pt,->] (X1)--(0.9,0);
        \draw[line width=1.5pt] (0.75,0)--(X2);
		\path[fill=black] (X1) circle[radius=0.11];
		\path[fill=black] (X2) circle[radius=0.11];
\end{tikzpicture}}:=(-\text{i})^2\int_{-\infty}^{0}{\rm d}\tau_1{\rm d}\tau_2(-\tau_1)^{p_1{-}1}&(-\tau_2)^{p_2{-}1}{\rm e}^{\text{i}w_1\tau_1+\text{i}w_2\tau_2}\\&\text{H}_{-\text{i}\nu}^{(2)}(-k_{12}\tau_1)\text{H}_{\text{i}\nu}^{(1)}(-k_{12}\tau_2)\theta_{2,1}\text{H}_{-\text{i}\nu}^{(2)}(-k_{23}\tau_2),\nonumber
\end{align}
which is a generalization of \eqref{eq:twositechain} with an extra Hankel function on the node $2$, and in the diagram we use $2_{2,k}$ to indicate that Hankel function $\text{H}^{(2)}_{-\text{i}\nu}(-k\tau_2)$ on node $2$. Therefore, to form the most general basis for all massive cosmological amplitudes, we should not only consider family trees with arbitrary black and gray edges as above, but also consider their deformation by adding arbitrarily many factors $\prod_b\text{H}_{\pm \text{i}\nu}^{s_b}(-k_{ab}\tau_a)$ on each of their leaves. 

In the following, we will call a massive family tree with only black and gray edges and no extra Hankel functions on its leaves as {\it basic massive family trees}, and call it {\it general massive family trees} if it has more Hankel kernels on leaves. Thanks to the relation \eqref{eq:1}, both of the cases can always be converted into Euler-Mellin integrals, and MoB is applicable. General massive family trees involve additional integrations over corresponding basic massive family trees. In what follows, we will mainly focus on the case of basic massive family tree, and offer some comments on general massive family trees towards the end of this section.

\subsection{MoB series solutions for massive family trees}
\label{sec:massive solution}
In this section, we provide basis series solutions for massive family trees. Following the decomposition in the last subsection, we will  consider basic massive family trees first. We begin by presenting results of $N$-site basic massive family trees in basis series solutions, and then provide some specific examples. Then we briefly discuss general massive family trees. The reader interested in the explicit details can refer to Appendix.\ref{sec:basic massive solution derivation}. 

\subsubsection{General solutions for basic massive family trees}
We consider an $N$-site basic massive family tree with node $1$ as the ancestor and only one edge $(1,2)$ from ancestor $1$. $w_i$ is the energies flowing through the bulk-to-boundary propagators on node $i$, $k_{i,j}$ is the energy through bulk-to-bulk propagators from node $i$ to node $j$, and $p_i$ is the twist \eqref{eq:shiftq} on node $i$.

\paragraph{``Feynman rules" and notations for basic family trees} To organize the result, we first introduce basic ingredients that appear as building blocks in our series solutions, which can be treated as ``Feynman rules" of our series result.  The series are always expanded around $w_1>w_i, k_{i,j}$. 
\begin{enumerate}   
\item For every edge $(i,j)$, either black or gray, we introduce two indices $\{n_{i,j}^{(1)},n_{i,j}^{(2)}\}$, in total $2(N{-}1)$ indices, and assign a factor 
    \begin{equation}\label{eq:E}     
    \mathcal{E}_{i,j}:{=}(-1)^{n_{i,j}^{(1)}}(2k_{i,j})^{n_{i,j}^{(1)}{+}n_{i,j}^{(2)}}f(n_{i,j}^{(1)})g(n_{i,j}^{(2)}) 
    \end{equation}
    to the edge. The shorthand notations $f$, $g$ are defined as 
 \begin{equation}\label{eq:fg}
f(x):=\Gamma\left(\frac12{+}x{+}\text{i}\nu\right)\Gamma\left({-}x{-}2\text{i}\nu\right),\ g(x):=\Gamma\left(\frac12{+}x{-}\text{i}\nu\right)\Gamma({-}x{+}2\text{i}\nu).
\end{equation}
    \item For every node $j$ except the ancestor, we introduce an index $m_j$, in total $N{-}1$ indices, and assign a factor
    \begin{equation}\label{eq:deformw}
        \raisebox{-5em}{\begin{tikzpicture}[scale=1.2]
		\draw[line width=1.5pt,->] (0,0)--(0.65*0.819152, 0.65*0.573576);
		\draw[line width=1.5pt] (0,0)--(0.819152, 0.573576);
		\draw[fill=black!10] (0.819152*1.3, 0.573576*1.3) circle [radius=0.3];
		\draw[line width=1.5pt,->] (0,0)--(-0.34202*0.65, 0.939693*0.65);
		\draw[line width=1.5pt] (0,0)--(-0.34202, 0.939693);
		\draw[fill=black!10] (-0.34202*1.3, 0.939693*1.3) circle [radius=0.3];
		\draw[line width=1.5pt,->] (-1.2,0)--(-1.2*0.45, 0);
		\draw[line width=1.5pt] (0,0)--(-1.2,0);
		\draw[fill=black!10] (-1.2*1.3,0) circle [radius=0.3];
		\draw[line width=1.5pt,->,gray] (0,0)--(0.65*0.819152, -0.65*0.573576);
		\draw[line width=1.5pt,gray] (0.819152, -0.573576)--(0.45*0.819152, -0.45*0.573576);
		\draw[fill=black!10] (0.819152*1.3, -0.573576*1.3) circle [radius=0.3];
		\draw[line width=1.5pt,gray] (-0.34202, -0.939693)--(-0.34202*0.45, -0.939693*0.45);
		\draw[line width=1.5pt,->,gray] (0,0)--(-0.34202*0.65, -0.939693*0.65);
		\draw[fill=black!10] (-0.34202*1.3, -0.939693*1.3) circle [radius=0.3];
		\path[fill=black] (0,0) circle[radius=0.075];
		\path[fill=black] (0.59131, 0.806445) circle[radius=0.025];
		\path[fill=black] (0.300706, 0.9537170,0) circle[radius=0.025];
		\path[fill=black] (-0.0218149, 0.999762) circle[radius=0.025];
		\path[fill=black] (0.59131, -0.806445) circle[radius=0.025];
		\path[fill=black] (0.300706,-0.9537170) circle[radius=0.025];
		\path[fill=black] (-0.0218149, -0.999762) circle[radius=0.025];
		\path[fill=black] (0.819152, 0.573576) circle[radius=0.075];
		\path[fill=black] (-0.34202, 0.939693) circle[radius=0.075];
		\path[fill=black] (0.819152, -0.573576) circle[radius=0.075];
		\path[fill=black] (-0.34202, -0.939693) circle[radius=0.075];
		\path[fill=black] (-1.2, -0) circle[radius=0.075];
		\node at (0.819152*1.3, 0.573576*1.3) {\tiny{$G_u$}};
		\node at (-0.34202*1.3, 0.939693*1.3) {\tiny{$G_2$}};
		\node at (-1.2*1.3, 0) {\tiny{$G_1$}};
		\node at (0.819152*1.3, -0.573576*1.3) {\tiny{$G_{u+1}$}};
		\node at (-0.34202*1.3, -0.939693*1.3) {\tiny{$G_{v}$}};
		\node[right] at (-0.07,0.25) {\small{$j$}};
	\end{tikzpicture}}\sim \mathcal{V}_j:=\left(w_j+b_{1,j}k_{1,j}+\sum_{i=2}^{u}k_{j,i}-\sum_{i=u+1}^{v}k_{j,i}\right)^{m_j},
    \end{equation}
    to the node, where $b_{1,j}=1$ if edge $(1,j)$ is gray, and  $b_{1,j}=-1$ if edge $(1,j)$ is black.     
    \item For the ancestor node $1$, we assign one more factor
    \begin{equation}
        \raisebox{-1.2em}{\begin{tikzpicture}
		\draw[fill=black!10] (0,0) circle [radius=0.5];
		\path[fill=black] (-0.5,0) circle[radius=0.1];
		\draw[line width=1.5pt] (-0.5,0)--(-1.1,0);
		\draw[line width=1.5pt,->] (-1.5,0)--(-0.9,0);
		\path[fill=black] (-1.5,0) circle[radius=0.1];
		\node at (0,0) {\small{$G_2$}};
		\node[below] at (-1.5,0) {\small{$1$}};
	\end{tikzpicture}}\sim \mathcal{V}_1:=(w_1{+}b_{1,2}k_{1,2})^{-\sum_i m_i-\sum_{i,j}(n_{i,j}^{(1)}+n_{i,j}^{(2)})-\tilde{p}_1}.
    \end{equation}
    where $b_{1,2}=-1$ if edge $(1,2)$ is gray, and  $b_{1,2}=1$ if edge $(1,2)$ is black.

\end{enumerate}

Before we give the general result, we introduce the following shorthand notations for ease of presentation
\begin{enumerate}
    \item For each of the node, we introduce its deformed twist as 
    \begin{equation}\label{eq:deformtwist}
    \begin{aligned}
        &\hat{p}_k:=p_k{+}(b_{in}{-}b_{out}{-}g_{in}{+}g_{out})\text{i}\nu,\\\ \ &b_{in},b_{out},g_{in},g_{out}: \ \text{number of black \& gray edges attached to node k}
        \end{aligned}
    \end{equation}
    and we will use $\tilde{\mathrm{p}}_i=\sum_{k}\hat{p}_k,\ \  k\in\{\text{$i$ and its descendants}\}$ to denote the sum of all deformed twists of node $i$ and all its descendants. 
    \item We also use 
    \begin{equation}\label{eq:deformn}
    \begin{aligned}
        &\hat{n}_i=\sum_k m_k+\sum_{k,l} n_{k,l}^{(1)}+n_{k,l}^{(2)}+n^\prime, \text{$k$ or $l$ is descendant of $i$ and } \\&n^\prime=\left\{
	\begin{aligned}
		&n_{i^\prime,i}^{(1)},\qquad & \text{if $i^\prime$ is the parent of $i$ and $(i^\prime,i)$ is black}\\
		&n_{i^\prime,i}^{(2)},\qquad & \text{if $i^\prime$ is the parent of $i$ and $(i^\prime,i)$ is gray}\\
	\end{aligned}
\right .
\end{aligned}
    \end{equation}
    This is in fact a sum over all indices adjacent to node $i$ and all its descendants.
    \end{enumerate}

\paragraph{Result for a basic massive family tree}After these ingredients are introduced, we are ready to present our most general result.

\begin{enumerate}

    \item A basic massive family tree with nested-time structure $\mathcal{N}$ is given by a sum of $2^{2(N{-}1)}$ basis series
    \begin{equation}\label{eq:resultmass}
        \mathcal{I}[\mathcal{N}(12\cdots N)]=(-\text{i})^{N+\tilde{p}_1}  (BC)^{N{-}1}\times\sum_{e_{i,j}^{(s)}\in\{0,1\}}\sum_{n_{i,j}^{(1)},n_{i,j}^{(2)},m_i}\mathbf{T}_{e_{i,j}^{(s)}}.
    \end{equation}
    Labels $e_{i,j}^{(1)}$ and $e_{i,j}^{(2)}$, in total $2(N{-}1)$ of them, correspond to the edge $(i,j)\in\mathcal{E}$ from node $i$ to node $j$, and $B$ and $C$ are defined as \eqref{eq:BC}.  Each of the summand $\mathbf{T}_{e_{i,j}^{(s)}}$ has  $3(N{-}1)$ free indices $\{n_{i,j}^{(1)},n_{i,j}^{(2)}\}_{(i,j)\in\mathcal{E}}\cup\{m_k\}_{k=2,\cdots,N}$ to be summed over.
    
 \item    The case with all labels $e_{i,j}^{(s)}=0$, which we call the {\it primary series}, is given by 
   \begin{align}\label{eq:T000}
&\mathbf{T}_{0,\cdots,0}=\phi_{n_{i,j}^{(1)},n_{i,j}^{(2)},m_i}\Gamma(\tilde{p}_1{+}\hat{n}_1)\prod_{i=2}^N\frac{1}{\hat{n}_2{+}\tilde{\mathrm{p}}_i}\prod_{(i,j)}\mathcal{E}_{i,j}\prod_{i=1}^N\mathcal{V}_i\quad.
    \end{align}   
One can see that this series looks similar to the massless family trees \eqref{eq:treeres}. 
    \item Other $2^{2(N{-}1)}-1$ series, which we call the {\it descendant series}, can be generated from the primary series by two types of simple shifting operations namely
    \begin{align}\label{eq:T111}
        \mathbf{T}_{\cdots,e_{i,j}^{(1)}= 1,0,\cdots}&:=\frac{(-1)^{2\text{i}\nu}\ \Gamma({-}n_{i,j}^{(1)}{+}2\text{i}\nu)\Gamma(n_{i,j}^{(1)}{-}2\text{i}\nu{+}1)}{\Gamma(n_{i,j}^{(1)}{+}1)\Gamma(-n_{i,j}^{(1)})}\mathbf{T}_{0,\cdots,0}\left(n_{i,j}^{(1)}\to n_{i,j}^{(1)}{-}2\text{i}\nu\right),\nonumber\\
        \mathbf{T}_{\cdots,0,e_{i,j}^{(2)}= 1,\cdots}&:=\frac{(-1)^{-2\text{i}\nu}\ \Gamma({-}n_{i,j}^{(2)}{-}2\text{i}\nu)\Gamma(n_{i,j}^{(1)}{+}2\text{i}\nu{+}1)}{\Gamma(n_{i,j}^{(2)}{+}1)\Gamma(-n_{i,j}^{(2)})}\mathbf{T}_{0,\cdots,0}\left(n_{i,j}^{(2)}\to n_{i,j}^{(2)}{+}2\text{i}\nu\right),
    \end{align}
    and similarly for the other series with more than one  $e_{i,j}^{(s)}{=}1$. Note that the divergence of the prefactor $\frac1{\Gamma(n_{i,j}^{(s)}{+}1)\Gamma(-n_{i,j}^{(s)})}$ is always canceled with the divergent factor obtained by shifting $n_{i,j}^{(s)}$ in $\Gamma(-n_{i,j}^{(s)}+(-1)^{s}\ 2\text{i}\nu)$ of \eqref{eq:fg}.
     \end{enumerate}   

Let us present some explicit examples to illustrate this general solution.

\subsubsection{Explicit examples of basic massive family trees} 
Before we go to non-trivial basic massive family trees, it is already worth mentioning that this primary-descendant structure also applies to the massive contact function \eqref{eq:massivecontact2}. By implementing \eqref{eq:1}, we can express the massive contact function as an one-fold integration over the massless contact function as
\begin{equation}
  \int_{-\infty}^0{\rm d}\tau\ \text{e}^{\text{i}w\tau}(-\tau)^{q{-}1}\text{H}^{(2)}_{-\text{i}\nu}(-k\tau){=}Ak^{-\text{i}\nu}\int_0^{\infty}{\rm d}s(s(1{+}s))^{-\frac12-\text{i}\nu}\times\frac{\Gamma(q{-}\text{i}\nu)}{(i(w{+}k(2s{+}1)))^{q{-}\text{i}\nu}},
\end{equation}
with 
\begin{equation}\label{eq:A}
    A:=\frac{\text{i}}{\sqrt{\pi}}\frac{2^{-\text{i}\nu{+}1}{\rm e}^{\nu\pi}}{\Gamma[-\text{i}\nu{+}1/2]}.
\end{equation}
Therefore, we can apply MoB to the $s$-integration and finally arrive at
\begin{equation}\label{eq:massivecontact3}
    Ak^{-\text{i}\nu}\times (2\text{i}k)^{-q{+}\text{i}\nu}\sum_{n=0}^{\infty}\left(\mathbf{T}_0{+}\mathbf{T}_1\right),
\end{equation}
where, the two series read
\begin{align}    &\mathbf{T}_0=\frac{\phi_n\Gamma(n{+}q{-}\text{i}\nu)g(n)}{\Gamma(\text{i}\nu{+}\frac12)}\left(\frac{2k}{w{+}k}\right)^{n{+}q{-}\text{i}\nu},\ \nonumber\\
&\mathbf{T}_1=\frac{(-1)^{-2\text{i}\nu}\ \Gamma({-}n{-}2\text{i}\nu)\Gamma(n{+}2\text{i}\nu{+}1)}{\Gamma(n{+}1)\Gamma(-n)}\mathbf{T}_0(n\to n{+}2\text{i}\nu).
\end{align}
It can be easily checked that by summing over $n$ in \eqref{eq:massivecontact3}, it yields the same result $\mathcal{F}^{(2)}(p,w,k,\nu)$ as \eqref{eq:massivecontact2}. A similar series result can be deduced for $\mathcal{F}^{(1)}(p,w,k,\nu)$ as well.

Now we move on to basic massive family trees. We first present explicit result for two-site chain cosmological integral \eqref{eq:twositechain}. Following our general discussion, its result is a sum over four series $\mathbf{T}_{e_{1,2}^{(1)},e_{1,2}^{(2)}}$, each of which is a three-fold sum
\begin{equation}\label{eq:twositefirst}
    \mathcal{I}_2=(-\text{i})^{2+\tilde{p}_1}(BC)\times\sum_{m_2,n_{1,2}^{(1)},n_{1,2}^{(2)}}\left(\mathbf{T}_{0,0}+\mathbf{T}_{0,1}+\mathbf{T}_{1,0}+\mathbf{T}_{1,1}\right),
\end{equation}
with the four summands
{\small
\begin{equation}\label{eq:twositefirst2}
\begin{aligned}    &\mathbf{T}_{0,0}{=}\frac{\phi_{m_2,n_{1,2}^{(1)},n_{1,2}^{(2)}}\Gamma(m_2{+}n_{1,2}^{(1)}{+}n_{1,2}^{(2)}{+}\tilde{p}_1)}{(2k)^{\tilde{p}_1}(m_2{+}n_{1,2}^{(1)}{+}{p}_2{+}\text{i}\nu)}({-})^{n_{1,2}^{(1)}}f(n_{1,2}^{(1)})g(n_{1,2}^{(2)})\left(\frac{w_2{-}k}{w_1{+}k}\right)^{m_2}\left(\frac{2k}{w_1{+}k}\right)^{n_{1,2}^{(1)}{+}n_{1,2}^{(2)}{+}\tilde{p}_1},\\    
&\mathbf{T}_{1,0}{=}\frac{\phi_{m_2,n_{1,2}^{(1)},n_{1,2}^{(2)}}\Gamma(m_2{+}n_{1,2}^{(1)}{+}n_{1,2}^{(2)}{+}\tilde{p}_1{-}2\text{i}\nu)(-1)^{-2\text{i}\nu}}{(2k)^{\tilde{p}_1}(m_2{+}n_{1,2}^{(1)}{+}{p}_2{-}\text{i}\nu)}(-)^{n_{1,2}^{(1)}}g(n_{1,2}^{(1)})g(n_{1,2}^{(2)})\left(\frac{w_2{-}k}{w_1{+}k}\right)^{m_2}\left(\frac{2k}{w_1{+}k}\right)^{n_{1,2}^{(1)}{+}n_{1,2}^{(2)}{+}\tilde{p}_1{-}2\text{i}\nu},\\    &\mathbf{T}_{0,1}{=}\frac{\phi_{m_2,n_{1,2}^{(1)},n_{1,2}^{(2)}}\Gamma(m_2{+}n_{1,2}^{(1)}{+}n_{1,2}^{(2)}{+}\tilde{p}_1{+}2\text{i}\nu)}{(2k)^{\tilde{p}_1}(m_2{+}n_{1,2}^{(1)}{+}{p}_2{+}\text{i}\nu)}(-)^{n_{1,2}^{(1)}}f(n_{1,2}^{(1)})f(n_{1,2}^{(2)})\left(\frac{w_2{-}k}{w_1{+}k}\right)^{m_2}\left(\frac{2k}{w_1{+}k}\right)^{n_{1,2}^{(1)}{+}n_{1,2}^{(2)}{+}\tilde{p}_1{+}2\text{i}\nu},\\    &\mathbf{T}_{1,1}{=}\frac{\phi_{m_2,n_{1,2}^{(1)},n_{1,2}^{(2)}}\Gamma(m_2{+}n_{1,2}^{(1)}{+}n_{1,2}^{(2)}{+}\tilde{p}_1) (-1)^{-2\text{i}\nu}}{(2k)^{\tilde{p}_1}(m_2{+}n_{1,2}^{(1)}{+}{p}_2{-}\text{i}\nu)}(-)^{n_{1,2}^{(1)}}g(n_{1,2}^{(1)})f(n_{1,2}^{(2)})\left(\frac{w_2{-}k}{w_1{+}k}\right)^{m_2}\left(\frac{2k}{w_1{+}k}\right)^{n_{1,2}^{(1)}{+}n_{1,2}^{(2)}{+}\tilde{p}_1}.
    \end{aligned}
\end{equation}}
\noindent The correctness can be checked by a direct numerical comparison\footnote{We caution the reader that phase-factor components of our series solutions, e.g., the function $(-1)^{-2\text{i}\nu}=\text{exp}({-}2\nu(-\pi{+}2n\pi))$ for $n\in\mathbb{Z}$, are multivalued functions in the complex plane. When numerically verifying these results, special care must be taken regarding branch choices. A naive comparison, for instance, direct numerical evaluation via {\tt Mathematica} against the original integral, may yield mismatches due to phase differences induced by branch selection.}. We will offer a more elaborate discussion for this result in section \ref{sec:two site massive}. 

One important advantage of our result \eqref{eq:twositefirst} (as well as general result \eqref{eq:resultmass}) is that it is easy to take the massless limit $\hat{m}\to0$ or $\nu\to\frac{\text{i}}2$ smoothly. This point is based on smooth massless limit for the two combinations 
\begin{equation}
\lim_{\nu\to\frac{\text{i}}2}\frac{f(n)}{\Gamma(\frac12+\text{i}\nu)}=\left\{\begin{aligned}        &\phantom{aa}1\quad\quad\quad\quad\text{if}\  n=0\\&\frac{({-}1)^n}{2}\phantom{a}\quad\quad\text{if}\  n\geq1
    \end{aligned} \right., \ \ \  \lim_{\nu\to\frac{\text{i}}2}\frac{g(n)}{\Gamma(\frac12+\text{i}\nu)}=\frac{(-1)^{n{+}1}}{2(n{+}1)},
\end{equation}
where the factors $1/\Gamma(\frac12{+}\text{i}\nu)$, come from the $\cosh(\pi \nu)$ prefactors contributed by $(BC)$ in \eqref{eq:twositefirst}. For other factors in \eqref{eq:twositefirst2}, we can simply set $\nu\to\frac{\text{i}}2$ freely without taking any limits. Finally, by taking sum over the three indices, each of them yields a finite result. Summing over the four terms, we can get the correct result for massless two-site chain \eqref{eq:twositechainmassless2} .

The second non-trivial example we present here is a three-site integral, firstly computed in \cite{Aoki:2024uyi}, whose definition as a nested-time integral is 
\begin{align}\label{eq:3sitemassivechain1}
\mathcal{I}_3=&\raisebox{-1em}{\begin{tikzpicture}
				\coordinate (X1) at (0,0);
				\coordinate (X2) at (1,0);
				\coordinate (X3) at (2,0);;
				\node[below] at (X1) {\small{$1$}};
				\node[below] at (X2) {\small{$2$}};
				\node[below] at (X3) {\small{$3$}};
				\draw[line width=1.5pt,->] (X1)--(0.6,0);
				\draw[line width=1.5pt] (0.5,0)--(X2);
				\draw[line width=1.5pt,->] (X2)--(1.6,0);
				\draw[line width=1.5pt] (1.5,0)--(X3);
				\path[fill=black] (X1) circle[radius=0.1];
				\path[fill=black] (X2) circle[radius=0.1];
				\path[fill=black] (X3) circle[radius=0.1];
		\end{tikzpicture}}\nonumber\\
        =&(-\text{i})^3\int_{-\infty}^0\prod_{i=1}^3\left({\rm d}\tau_i(-\tau_i)^{p_i{-}1}{\rm e}^{\text{i}w_i\tau_i}\right){\text H}_{-\text{i}\nu}^{(2)}(-k_{1,2}\tau_1){\text H}_{\text{i}\nu}^{(1)}(-k_{1,2}\tau_2)\theta_{2,1}\\
&\phantom{aaaaaaaaaaaaaaaaaaaaaaaaaaaaaaaaaa}\times{\text H}_{-\text{i}\nu}^{(2)}(-k_{2,3}\tau_2){\text H}_{\text{i}\nu}^{(1)}(-k_{2,3}\tau_3)\theta_{3,2},\nonumber
\end{align}
and admits an Euler-Mellin representation
\begin{equation}\label{eq:I3Mellin}  
\begin{aligned}
\mathcal{I}_3=\int_0^{\infty}\prod_{\substack{i=1,2\\t=1,2}}[{\rm d}s_{i,i+1}^{(t)}(s_{i,i{+}1}^{(t)}(1{+}s_{i,i{+}1}^{(t)}))^{(-1)^{t{+}1}\text{i}\nu{-}\frac12}]\int_0^{\infty}\prod_{i=2,3}{\rm d}r_i r_i^{\tilde{p}_i{+}\text{i}\nu{-}1}(1{+}r_i)^{\tilde{p}_1-\tilde{p}_i{-}\text{i}\nu{-}1}\\
\times(-\text{i})^{3{+}\tilde{p}_1}\Gamma(\tilde{p}_1)[\tilde{w}_1(1{+}r_2)(1{+}r_3){+}\tilde{w}_2r_2(1{+}r_3){+}\tilde{w}_3r_2r_3]^{-\tilde{p_1}},
\end{aligned}
\end{equation}
with $\tilde{w}_1{=}w_1{+}k_{1,2}(1{+}2s_{1,2}^{(2)})$, $\tilde{w}_2{=}w_2{-}k_{1,2}(1{+}2s_{1,2}^{(1)}){+}k_{2,3}(1{+}2s_{2,3}^{(2)})$ and $\tilde{w}_3{=}w_3{-}k_{2,3}(1{+}2s_{2,3}^{(1)})$. Following our discussion, its result can be expressed by a sum of $16$ series as
\begin{equation}    
\mathcal{I}_3=(-\text{i})^{3{+}\tilde{p}_1}(BC)^2\times\sum_{e_{1,2}^{(1)},e_{1,2}^{(2)},e_{2,3}^{(1)},e_{2,3}^{(2)}\in\{0,1\}}\sum_{n_{i,i{+}1}^{(s)},m_i}\mathbf{T}_{e_{1,2}^{(1)},e_{1,2}^{(2)},e_{2,3}^{(1)},e_{2,3}^{(2)}}.
\end{equation}
Each of the series has six folds of summation. Indices of the summation are denoted as
\begin{equation}
    \{m_2,m_3,n_{1,2}^{(1)},n_{1,2}^{(2)},n_{2,3}^{(1)},n_{2,3}^{(2)}\}.
\end{equation}
The primary series is then (denote $\tilde{m}_i=\sum_{k}m_k$ for node $i$ and all its descendants as well)
\begin{equation}
    \mathbf{T}_{0,0,0,0}=\frac{\phi_{n_{i,i{+}1}^{(s)},m_i}\Gamma(\tilde{m}_2{+}\sum_{i=1}^2(n_{i,i{+}1}^{(1)}{+}n_{i,i{+}1}^{(2)}){+}\tilde{p}_1)}{(\tilde{m}_2{+}n_{1,2}^{(1)}{+}n_{2,3}^{(1)}{+}n_{2,3}^{(2)}{+}p_2{+}p_3{+}\text{i}\nu)(m_3{+}n_{2,3}^{(1)}{+}p_3{+}\text{i}\nu)}\mathcal{E}_{1,2}\mathcal{E}_{2,3}\mathcal{V}_1\mathcal{V}_2\mathcal{V}_3,
\end{equation}
with the factors $\mathcal{E}_{i,j}$ from \eqref{eq:E} and $\mathcal{V}_j$ from nodes \eqref{eq:deformw} as
\begin{equation}
    \mathcal{V}_1{=}(w_1{+}k_{1,2})^{-m_2{-}m_3{-}n_{1,2}^{(1)}{-}n_{1,2}^{(2)}{-}n_{2,3}^{(1)}{-}n_{2,3}^{(2)}-\tilde{p}_1}, \ \mathcal{V}_2{=}(w_2{-}k_{1,2}{+}k_{2,3})^{m_2},\ \mathcal{V}_3{=}(w_3{-}k_{2,3})^{m_3},
\end{equation}
and we have correspondingly $15$ more descendant series,obtained by the shifts \eqref{eq:T111}. For instance, one of the series we need to sum over is
\begin{equation}   
\begin{aligned}
\mathbf{T}_{0,0,0,1}{=}\frac{\phi_{n_{i,i{+}1}^{(s)},m_i}\Gamma(\tilde{m}_2{+}\sum_{i=1}^2(n_{i,i{+}1}^{(1)}{+}n_{i,i{+}1}^{(2)}){+}\tilde{p}_1{+}2\text{i}\nu)(-)^{n_{1,2}^{(1)}+n_{2,3}^{(1)}}}{(\tilde{m}_2{+}n_{1,2}^{(1)}{+}n_{2,3}^{(1)}{+}n_{2,3}^{(2)}{+}p_2{+}p_3{+}3\text{i}\nu)(m_3{+}n_{2,3}^{(1)}{+}p_3{+}\text{i}\nu)}\left(\frac{2k_{2,3}}{w_1{+}k_{1,2}}\right)^{2\text{i}\nu}\phantom{aaaaaaaaaa}\\\times f(n_{1,2}^{(1)})g(n_{1,2}^{(2)})f(n_{2,3}^{(1)})f(n_{2,3}^{(2)})\mathcal{V}_1\mathcal{V}_2\mathcal{V}_3,
\end{aligned}
\end{equation}
with index $e_{2,3}^{(2)}$ shifted from $0$ to $1$. We also check its correctness by a direct numerical evaluation of the Euler-Mellin representation of $\mathcal{I}_3$ \eqref{eq:I3Mellin}. 

In Appendix \ref{app:results}, the reader can find more explicit examples, which not only serve as illustration for our general result, but also play a role as the necessary building blocks of results for the full {\it in-in} correlators and wavefunction coefficients.

\subsubsection{Generalization to general massive family trees}
\label{sec:generl massive trees}
Finally, we offer some comments over general massive family trees, {\it i.e.,} massive family trees with extra Hankel functions on its leaves like \eqref{eq:twositechain2p}. A general calculation for these integrals is almost the same as derivation for basic massive trees, except that we have to introduce one more fold of integration for each extra Hankel function, which finally results in one more fold of series.  For illustration, we present two special examples for these generalized cases without presenting details of derivation, whose results can be checked via numerical comparison. The simplest case of general massive family tree is the general massive contact functions which can be defined as
\begin{equation}
    (-\text{i})\int_{-\infty}^0{\rm d}\tau(-\tau)^{q{-}1}\text{e}^{\text{i}w\tau}\prod_{i=1}^{n_1}\text{H}^{(2)}_{{-}\text{i}\nu}(-k_i\tau)\prod_{i=n_1{+}1}^{n}\text{H}^{(1)}_{{}\text{i}\nu}(-k_i\tau),
\end{equation}
and will be one of the building blocks when discussing higher-site tree graphs in Appendix.\ref{app:results}. We present a special case with valency-two, 
\begin{equation}
    \mathcal{F}^{(2,2)}(p,w,k_1,k_2,\nu):=(-\text{i})\int_{-\infty}^0{\rm d}\tau\ (-\tau)^{p{-}1}\text{e}^{\text{i}w\tau}\ \text{H}_{-\text{i}\nu}^{(2)}(-k_1\tau)\text{H}_{-\text{i}\nu}^{(2)}(-k_2\tau),
\end{equation}
which we will also see later as one of the building block of the three-site-chain graph. Its result is a sum over four series
\begin{equation}
    \mathcal{F}^{(2,2)}(p,w,k_1,k_2,\nu)=(-\text{i})^{1{+}p{-}2\text{i}\nu}A^2 (k_1k_2)^{-\text{i}\nu}\times\sum_{e_{1}^{(2)},e_2^{(2)}\in\{0,1\}}^{\infty}\sum_{n_{1}^{(2)},n_2^{(2)}}\mathbf{T}_{e_{1}^{(2)},e_2^{(2)}},
\end{equation}
with $A$ defined as \eqref{eq:A} and the primary series
\begin{equation}    \mathbf{T}_{0,0}=\frac{\phi_{n_{1}^{(2)},n_{2}^{(2)}}\Gamma(n_{1}^{(2)}{+}n_{2}^{(2)}{+}p{-}2\text{i}\nu)}{\Gamma(\text{i}\nu{+}\frac12)^2} g(n_1^{(2)})g(n_2^{(2)})(2k_1)^{n_1^{(2)}}(2k_2)^{n_2^{(2)}}(w{+}k_1{+k}_2)^{{-}n_{1}^{(2)}{-}n_{2}^{(2)}{-}p{+}2\text{i}\nu},
\end{equation}
and the other three series are related with this by the second type of shift in \eqref{eq:T111}.

Secondly, an explicit derivation shows that \eqref{eq:twositechain2p} can be expressed by a sum over $8$ series
\begin{equation}    \mathcal{I}_{2p}=(-\text{i})^{2{+}\tilde{p}_1{-}\text{i}\nu}(ABC)k_{2,3}^{-\text{i}\nu}\times\sum_{e_{1,2}^{(1)},e_{1,2}^{(2)},e_{2,3}^{2}=\{0,1\}}\sum_{n_{1,2}^{(1)},n_{1,2}^{(2)},n_{2,3}^{(2)},m_2}\mathbf{T}_{e_{1,2}^{(1)},e_{1,2}^{(2)},e_{2,3}^{(2)}},
\end{equation}
and its primary solution reads
\begin{align}
    \mathbf{T}_{0,0,0}{=}\frac{\phi_{n,m}(2k_{2,3})^{n_{2,3}^{(2)}}\Gamma(m_2{+}n_{1,2}^{(1)}{+}n_{1,2}^{(2)}{+}n_{2,3}^{(2)}{+}\tilde{p}_1{-}\text{i}\nu) }{(m_2{+}n_{1,2}^{(1)}{+}n_{2,3}^{(2)}{+}p_2)\Gamma(\text{i}\nu{+}\frac12)}g(n_{2,3}^{(2)}) \mathcal{E}_{1,2}\hat{\mathcal{V}}_1\hat{\mathcal{V}}_2,
\end{align}
with $A$ from \eqref{eq:A}, and the deformed vertex factors
\begin{equation}
\hat{\mathcal{V}}_1=(w_1{+}k_{1,2})^{-m_2{-}n_{1,2}^{(1)}{-}n_{1,2}^{(2)}{-}n_{2,3}^{(2)}{-}\tilde{p}_{1}{+}\text{i}\nu},\ \hat{\mathcal{V}}_2=(w_2{-}k_{1,2}{+}k_{2,3})^{m_2}.
\end{equation}
Other seven series can be generated still following the rule \eqref{eq:T111}.

We end this section by presenting basis series solution of generalized massive family tree integrals without a derivation.
\begin{enumerate}
    \item Each extra Hankel function factor $\text{H}_{(-1)^{s-1} \text{i}\nu}^{(s)}(-k_{a,b}\tau)$ on any of the leaves (node $a$) introduces an additional subscript $e_{a,b}^{(s)}\in\{0,1\}$ for $\mathbf{T}_{e_{i,j}^{(s)}}$.  
    \item Each extra factor of $\text{H}_{-\text{i}\nu}^{(2)}(-k_{a,b}\tau_a)$ on node $a$, introduces one more fold of summation over an extra index $n_{a,b}^{(2)}$. To obtain its primary solution, we firstly deform
    \begin{equation} \label{gr1}
        \hat{n}_a\to \hat{n}_a{+}n_{a,b}^{(2)},\ \ \hat{p}_a\to \hat{p}_a{-}\text{i}\nu,\ \ w_a\to w_a{+}k_{a,b}
    \end{equation}
    in addition to  \eqref{eq:deformw}, \eqref{eq:deformtwist} and \eqref{eq:deformn}, and then add an extra factor to obtain
    \begin{equation}  \label{gr2}
\mathbf{T}_{0,\cdots,0,0}=\frac{\phi_{n_{a,b}^{(2)}} (2k_{a,b})^{n_{a,b}^{(2)}}g(n_{a,b}^{(2)})}{\Gamma(\text{i}\nu{+}\frac12)}\times \mathbf{T}_{0\cdots,0}.
    \end{equation}
    \item Each additional factor of  $\text{H}_{\text{i}\nu}^{(1)}(-k_{a,b}\tau_a)$ on node $a$,  introduces one more fold of summation over an extra index $n_{a,b}^{(1)}$. To obtain its primary solution, we firstly deform
    \begin{equation} \label{gr3}
        \hat{n}_a\to \hat{n}_a{+}n_{a,b}^{(1)},\ \ \hat{p}_a\to \hat{p}_a{+}\text{i}\nu,\ \ w_a\to w_a{-}k_{a,b}
    \end{equation}
    in addition to  \eqref{eq:deformw}, \eqref{eq:deformtwist} and \eqref{eq:deformn}, and then add an extra factor to obtain
    \begin{equation}  \label{gr4}
\mathbf{T}_{0,\cdots,0,0}=\frac{\phi_{n_{a,b}^{(1)}} (-2k_{a,b})^{n_{a,b}^{(1)}}f(n_{a,b}^{(1)})}{{\Gamma({-}\text{i}\nu{+}\frac12)}}\times \mathbf{T}_{0\cdots,0}.
    \end{equation}
    \item Finally, descendant series can be obtained by rule \eqref{eq:T111}, after treating $n_{a,b}^{(s)}$ as the same as other $n_{i,j}^{(s)}$ indices from the thick edges. An extra prefactor 
    \begin{equation}(-\text{i})^{(l_1{-}l_2)  \label{gr4}
    \text{i}\nu} \left(\bar{A}k_{a,b}^{\text{i}\nu} \right)^{l_1} \left(Ak_{a,b}^{-\text{i}\nu} \right)^{l_2},\end{equation} 
    should be included in addition to \eqref{eq:resultmass}, where $l_i$ is the number of extra Hankel functions $\text{H}^{(i)}$.
\end{enumerate}

\subsection{Massive four-point cosmological amplitude and their MoB solution}
\label{sec:two site massive}
\begin{figure}[t]
    \centering
        \begin{tikzpicture}
        \draw[line width=2pt] (0,0)--(10,0);
        \draw[line width=2pt,dashed] (0,-5)--(10,-5);
        \draw[line width=1pt] (2,0)--(3,-4)--(4,0);
        \draw[line width=1pt] (8,0)--(7,-3)--(6,0);
        \draw[decorate, decoration=snake, segment length=12pt, segment amplitude=2pt, black,line width=1.5pt] (3,-4)--(7,-3);
         \filldraw[black] (2.5,-1.5) node[anchor=east] {{$\mathbf{k}_1$}};
          \filldraw[black] (3.6,-1.5) node[anchor=west] {{$\mathbf{k}_2$}};
        \filldraw[black] (7.5,-2) node[anchor=west] {{$\mathbf{k}_3$}};
        \filldraw[black] (6.5,-2) node[anchor=east] {{$\mathbf{k}_4$}};
        \filldraw[black] (5,-3.75) node[anchor=north] {{$\sum_{i=1,2}\mathbf{k}_i$,\ \ $\nu$}};
        \filldraw[black] (0,0) node[anchor=east] {{$\tau=0$}};
        \filldraw[black] (0,-5) node[anchor=east] {{$\tau=-\infty$}};
        \filldraw[black] (3,-4) node[anchor=east] {{$\tau_1$}};
         \filldraw[black] (7,-3) node[anchor=west] {{$\tau_2$}};
    \end{tikzpicture}
    \caption{Two-site correlation function with massive exchange}
    \label{fig:twositechain}
    \vspace{3ex}
\end{figure}
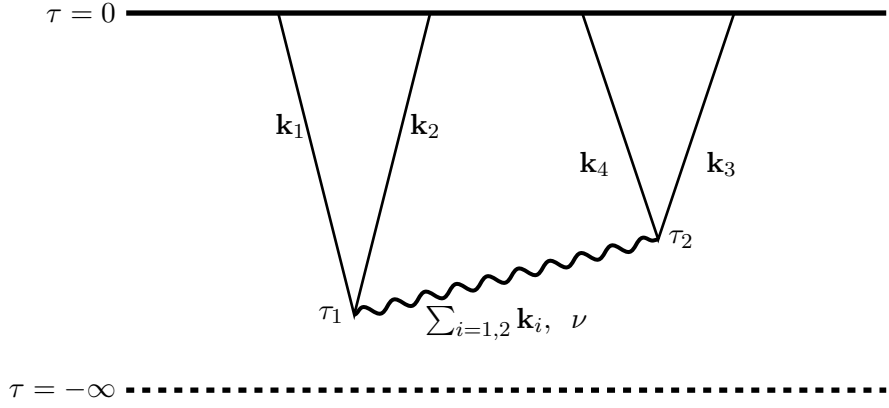

Finally, we offer a special discussion for the two-site chain cosmological integrals with a massive exchange, which have been discussed in various previous works \cite{Arkani-Hamed:2018kmz,Werth:2023pfl,Liu:2024str}. Without loss of generality, we assume there are two $\phi^3$ interactions in the procedure (Fig.\ref{fig:twositechain}). Therefore the kinematical input reads 
\begin{equation}\label{eq:kinematicsI2}
\omega_1=|\mathbf{k}_1|+|\mathbf{k}_2|, \ \omega_2=|\mathbf{k}_3|+|\mathbf{k}_4|, \ \  k=|\mathbf{k}_1+\mathbf{k}_2|=|\mathbf{k}_3+\mathbf{k}_4|
\end{equation} 
in \eqref{eq:twositechain}. Furthermore, at $d=3$, for FRW models of physical interest $-1<\rho<0$, which translates to $0<q_v<1$ for all $q_v=(4{-}n)\rho{+}1$ with $n{=}3$. We therefore still consider $q_v$ as general parameter, imposing only the requirement that $0<q_v<1$ in all our calculations. We will firstly focus on one of its massive family tree basis  elements $\mathcal{I}_2$ (eq.\eqref{eq:twositechain}). We have already solved this integral in the region $\{w_1>w_2,k\}$ as a sum of four series \eqref{eq:twositefirst}. In this section we will firstly present more details of this calculation, and solve this case in all the physical regions. We will see that similar to the result for massless model, MoB series solutions of this integral in different regions also encode nested-time structures. Finally, we add up MoB series solutions for all building blocks to get the full result for four-point cosmological amplitudes.

According to the general discussion, after the Hankel functions have been converted to Euler-Mellin integrals by \eqref{eq:1}, massive family tree $\mathcal{I}_2$ is a two-fold integral given by
\begin{equation} \label{2site1}   \mathcal{I}_2=C\int_{0}^{\infty}\int_{0}^{\infty} ds_1 ds_2 s_1^{-\text{i}\nu-1/2}s_2^{\text{i}\nu-1/2}(1+s_1)^{-\text{i}\nu-1/2}(1+s_2)^{\text{i}\nu-1/2}\left[P(\hat {12})\right], 
\end{equation}
with the deformed massless two-site family tree integral 
\begin{equation}
    P[(\hat{12})]=(-\text{i})^2\int_{-\infty}^0{\rm d}\tau_1{\rm d}\tau_2(-\tau_1)^{p_1{-}\text{i}\nu}(-\tau_2)^{p_2{+}\text{i}\nu}{\rm e}^{\text{i}(w_1{+}k(1+2s_1))\tau_1}{\rm e}^{\text{i}(w_2{-}k(1+2s_2))\tau_2}\theta_{2,1},
\end{equation}
and $C=\frac{4{\rm e}^{2\pi \nu}{\rm cosh}(\pi \nu)}{\pi^2}$. To proceed by MoB, we rewrite the nested-time integral into Euler-Mellin integral by \eqref{eq:Mellinmassless}, and change variable $\alpha_2\to \frac{r_2}{1{+}r_2}$ again. Finally, $\mathcal{I}_2$ is a three-fold Euler-Mellin integral 
\begin{align}\label{eq:twositeMellin}
    \mathcal{I}_2={-}C\frac{\ \Gamma{(\tilde{p}_1)}}{(2\text{i}k)^{\tilde{p}_1}}\int_0^\infty ds_1ds_2dr_2 &s_1^{{-}\text{i}\nu{-}1/2}s_2^{\text{i}\nu{-}1/2}(1{+}s_1)^{{-}\text{i}\nu{-}1/2}(1{+}s_2)^{\text{i}\nu{-}1/2}\nonumber\\
    &r_2^{p_2 + i \nu-1}(1{+}r_2)^{p_1{-}\text{i}\nu{-}1}
    {\left((1{+}r_2)(X{+}s_1){-}r_2(Y{-}s_2)\right)^{-\tilde{p}_1}},
\end{align}
where we  define 
\[X:=\frac{w_1{+}k}{2k},\ Y:=\frac{w_2{-}k}{2k}\]
to be new variables for simplicity in this subsection. In this work we will focus on the regions where  physical vectors satisfy the triangle-inequality $|\mathbf{k}_1{+}\mathbf{k}_2|<|\mathbf{k}_1|{+}|\mathbf{k}_2|$. Basic definitions of the kinematic variables \eqref{eq:kinematicsI2} give the ranges of $X, Y$ to be $X>1$ and  $Y>0$\footnote{Region $X<1$ corresponds to the case that external legs have a reduced sound speed \cite{Jazayeri:2023xcj,Qin:2025xct}. MoB also gives us series solutions valid in this region but we do not consider them in this work.}

Eq. \eqref{eq:twositeMellin} is then the appropriate beginning point for us to apply MoB. This is a special case of the general derivation in Appendix. \ref{sec:basic massive solution derivation}, but this time we will go through all MoB series and get series result for all physical regions. We put some details of this derivation in Appendix.\ref{sec:basic massive solution derivation} as well, and only mention here that we can get $\mathbf{40}$ series contributing to various regions in $X$ and $Y$.  As emphasized before, the regions we focus on in this work are $R_1=\{0<Y<X\}$ and $R_2=\{ 1<X<Y\}$. By looking at the arguments of the series, it can be argued that out of the {\bf 40} series only {\bf 4} and {\bf 8} series need to be added up to get the results for the regions $R_1$ and $R_2$ respectively.

\begin{figure*}[t!]
    \centering
    \begin{subfigure}[b]{0.5\textwidth}
    \includegraphics[width=0.95\linewidth]{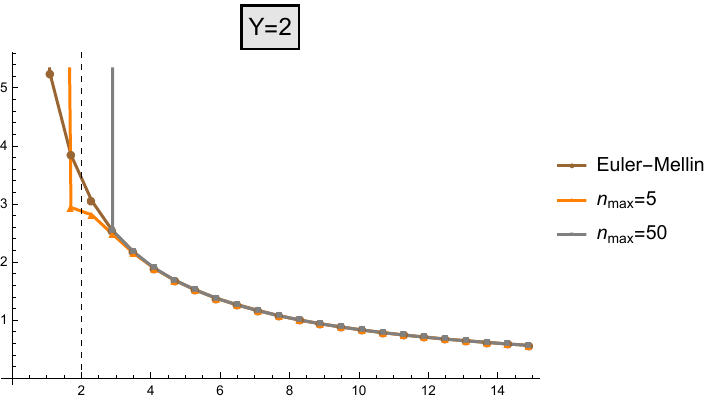}
    \caption{{\scriptsize Region $R_1 =\{0<Y<X \}$ with $Y=2$ and $\nu=\frac{\text{i}}{10}$.}}
     \vspace{5ex}
      \end{subfigure}%
        ~
    \begin{subfigure}[b]{0.5\textwidth}
    \includegraphics[width=0.95\linewidth]{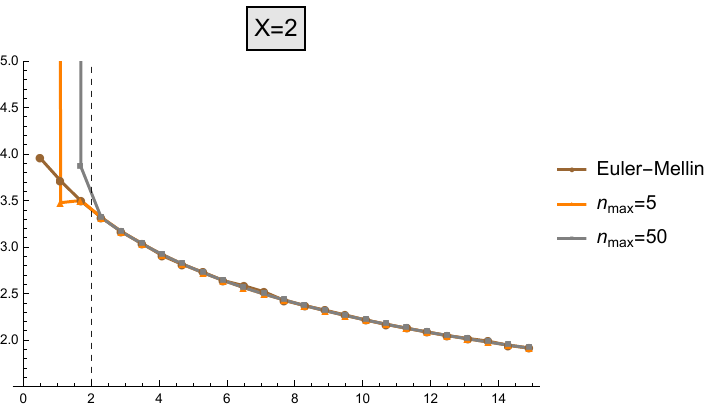}
    \caption{{\scriptsize Region $R_2 =\{1<X<Y \}$ with $X=2$ and $\nu=\frac{\text{i}}{10}$.}}
     \vspace{5ex}
      \end{subfigure} \\
      
      \begin{subfigure}[b]{0.5\textwidth}
    \includegraphics[width=0.95\linewidth]{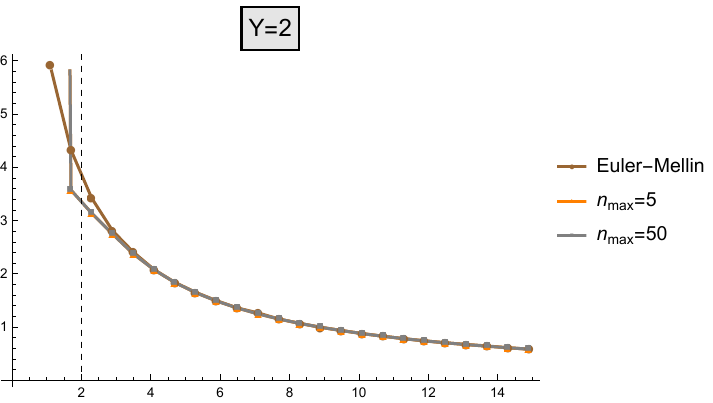}
    \caption{{\scriptsize Region $R_1 =\{0<Y<X \}$ with $Y=2$ and $\nu=\frac{1}{10}$.}}
     \vspace{5ex}
      \end{subfigure}%
        ~
    \begin{subfigure}[b]{0.5\textwidth}
    \includegraphics[width=0.95\linewidth]{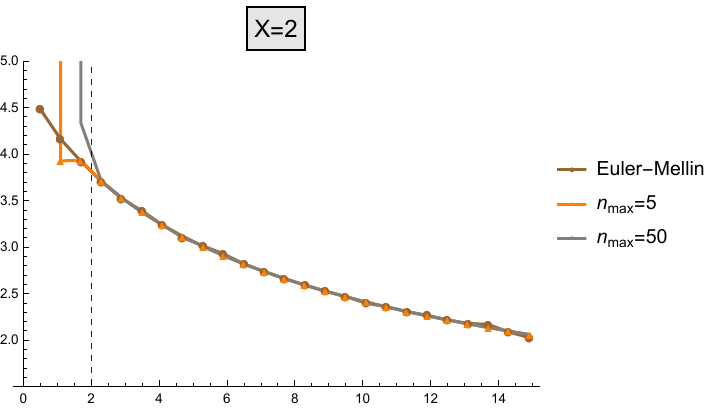}
    \caption{{\scriptsize Region $R_2 =\{1<X<Y \}$ with $X=2$ and $\nu=\frac{1}{10}$.}}
     \vspace{5ex}
      \end{subfigure}% 
    \caption{The figure above shows convergence of the series solutions of the 2-site basic massive family tree $\mathcal{I}_2$. We choose $p_1=\frac34$, $p_2=\frac57$, fix $Y=2$ or $X=2$ and plot the value of $|\mathcal{I}_2|$ from \eqref{eq:twositeMellin} and comparing with \eqref{eq:R1region} or \eqref{eq:R2region} respectively. The plots in the two rows correspond to the cases where the mass of the scalar is larger ($m >H$, i.e., $\nu \in  \mathbb{R}$) and smaller ($m< H$, i.e., $\nu \in \text{i}\mathbb{R}$) than the Hubble scale  respectively. We choose $\nu=\frac{\text{i}}{10}$ and  $\nu=\frac{1}{10}$ to generate the plots.
    The series converge in the region to the right of the dotted line and here we see that if we truncate to 5 terms then we already get excellent convergence.}
    \label{fig:R2region}
    \vspace{5ex}
\end{figure*}

\paragraph{$R_1$-region} 
In region $R_1$ our final result reads
\begin{equation}  \label{eq:R1region}
\mathcal{I}_2=(2\text{i}k)^{-\tilde{p}_1}C\times\sum_{i=1}^4\sum_{n=0}^{\infty}\mathbf{T}_i,
\end{equation}
with 
\begin{align} 
    \mathbf{T}_1&=-\frac{\pi (X)^{-\tilde{p}_1}}{2\nu {\rm sinh}(2\pi \nu)}\frac{\Gamma(n{+}\tilde{p}_1)}{(n{+}p_2{+}\text{i}\nu)n!} \qquad\qquad\qquad \qquad\qquad\qquad\qquad\qquad\qquad\qquad\qquad\qquad \qquad\qquad\qquad\qquad\qquad\qquad   \nonumber\\
&\ _2F_1\left({-}n,\frac12{+}\text{i}\nu;1{+}2\text{i}\nu,{-}\frac1Y\right)\ _2F_1\left(\tilde{p}_1{+}n,\frac12{-}\text{i}\nu;1{-}2\text{i}\nu,\frac1X\right)\left({-}\frac{Y}{X}\right)^{n}, 
\end{align}
\begin{align}
\mathbf{T}_2&={-}\frac{(X)^{-\tilde{p}_1}\Gamma(\text{i}\nu)(2X)^{2\text{i}\nu}(-1)^{-2\text{i}\nu}}{2\sqrt{\pi}}\frac{\Gamma(\frac12{+}n{-}\text{i}\nu)\Gamma(n{+}\tilde{p}_1{-}2\text{i}\nu)\Gamma({-}n{+}2\text{i}\nu)}{(n{+}p_2{-}\text{i}\nu)n!} \qquad\qquad\qquad \qquad\qquad\qquad\qquad\qquad\qquad \nonumber\\
&\ _2F_1\left({-}n,{-}n{+}2\text{i}\nu,\frac12{-}n{+}\text{i}\nu,{-}Y\right)\ _2F_1\left(\tilde{p}_1{+}n{-}2\text{i}\nu,\frac12{-}\text{i}\nu;1{-}2\text{i}\nu,\frac1X\right)\left(\frac{1}{X}\right)^{n}, 
\end{align}
\begin{align}
\mathbf{T}_3&=-\frac{(X)^{-\tilde{p}_1}\Gamma(\frac12{+}\text{i}\nu)\Gamma({-}\text{i}\nu)\Gamma(-2\text{i}\nu)}{2\sqrt{\pi}(2X)^{2\text{i}\nu}}\frac{\Gamma(n{+}\tilde{p}_1{+}2\text{i}\nu)}{(n{+}p_2{+}\text{i}\nu)n!} \nonumber\\& _2F_1\left({-}n,\frac12{+}\text{i}\nu;1{+}2\text{i}\nu,{-}\frac1Y\right) \ _2F_1\left(\tilde{p}_1{+}n{+}2\text{i}\nu,\frac12{+}\text{i}\nu;1{+}2\text{i}\nu,\frac1X\right)\left({-}\frac{Y}{X}\right)^{n},\phantom{aaaaaaaaaaa} 
\end{align}
\begin{align}
    \mathbf{T}_4&=-\frac{(X)^{-\tilde{p}_1}\Gamma(-\text{i}\nu)(-1)^{-2\text{i}\nu}}{2^{2\text{i}\nu+1}\sqrt{\pi}}\frac{\Gamma(\frac12{+}n{-}\text{i}\nu)\Gamma(n{+}\tilde{p}_1)\Gamma({-}n{+}2\text{i}\nu)}{(n{+}p_2{-}\text{i}\nu)n!}\qquad\qquad\qquad \qquad\qquad\qquad\qquad\qquad\qquad \nonumber\\
    &\ _2F_1\left({-}n,{-}n{+}2\text{i}\nu,\frac12{-}n{+}\text{i}\nu,{-}Y\right)\ _2F_1\left(\tilde{p}_1{+}n,\frac12{+}\text{i}\nu;1{+}2\text{i}\nu,\frac1X\right)\left(\frac{1}{X}\right)^{n}.
\end{align}
These four series are just the four series $\{\mathbf{T}_{0,0},\cdots,\mathbf{T}_{1,1}\}$ in \eqref{eq:twositefirst}, after we sum over two of the indices $\{m_2,n_{1,2}^{(1)}\}$.  As a series representation of $\mathcal{I}_2$, we present numerical results from these four series in $R_1$ region as diagram Fig.\ref{fig:R2region}. Note that when drawing the diagrams, we omit the overall factor $(2\text{i}k)^{-\tilde{p}_1}$ for both Euler-Mellin integral and our solutions. Besides the convergent requirement that $X>Y>0$, we have constraint $|\Re(-\text{i}\nu)|<\frac12$ from the Euler-Mellin representation \eqref{eq:1}. Moreover, we choose $q_i=p_i-\frac12$ to be small real numbers, which corresponds to an inflationary background.

\paragraph{$R_2$-region} 
We can also work out result for region $R_2$ which reads
\begin{equation}\label{eq:R2region}
\mathcal{I}_2=(2\text{i}k)^{-\tilde{p}_1}C\times\left(B\ \sum_{i=1}^4 \mathbf{S}_i+\sum_{i=5}^8 \sum_{m=0}^{\infty}{\mathbf{S}_i} \right),
\end{equation} 
with 
\begin{align}
\mathbf{S}_1{=}& (-1)^{p_2+ \text{i}\nu}  \Gamma \left(\frac{1}{2}{-}\text{i}\nu\right)^2 \Gamma (2 \text{i}\nu)^2 \Gamma (p_1{-}\text{i}\nu) \Gamma (p_2{-}\text{i}\nu)
   X^{-p_1+\text{i}\nu}  (-Y)^{-p_2+\text{i}\nu} \qquad\qquad\qquad\qquad\nonumber\\& \
   _2F_1\left(\frac{1}{2}{-}\text{i}\nu,p_1{-}\text{i}\nu;1{-}2 \text{i}\nu;\frac{1}{X}\right) \,
   _2F_1\left(\frac{1}{2}{-}\text{i}\nu,p_2{-}\text{i}\nu;1{-}2 \text{i}\nu;-\frac{1}{Y}\right),
\end{align}

\begin{align}
   \mathbf{S}_2{=}& (-1)^{p_2{+}\text{i}\nu}  \Gamma \left(\frac{1}{2}{-}\text{i}\nu\right) \Gamma \left(i
   v{+}\frac{1}{2}\right) \Gamma (-2 \text{i}\nu) \Gamma (2 \text{i}\nu)\Gamma
   (p_1{+}\text{i}\nu)\Gamma (p_2{-}\text{i}\nu) X^{-p_1-\text{i}\nu}  (-Y)^{-p_2+\text{i}\nu}  \qquad\nonumber\\& \ _2F_1\left(i
   \nu{+}\frac{1}{2},p_1{+}\text{i}\nu;2 \text{i}\nu{+}1;\frac{1}{X}\right) \, _2F_1\left(\frac{1}{2}{-}i
   \nu,p_2{-}\text{i}\nu;1{-}2 \text{i}\nu;-\frac{1}{Y}\right),
\end{align}

\begin{align}
    \mathbf{S}_3{=}& (-1)^{p_2{+}\text{i}\nu} \Gamma \left(\frac{1}{2}{-}\text{i}\nu\right) \Gamma \left(\text{i}\nu{+}\frac{1}{2}\right) \Gamma (-2 \text{i}\nu) \Gamma (2 \text{i}\nu) \Gamma
   (p_1{-}\text{i}\nu) \Gamma (p_2{+}\text{i}\nu) X^{-p_1+\text{i}\nu}  ({-}Y)^{-p_2-\text{i}\nu} \qquad \qquad \qquad \qquad\nonumber\\& \ _2F_1\left(\frac{1}{2}{-}i
   \nu,p_1{-}\text{i}\nu;1{-}2 \text{i}\nu;\frac{1}{X}\right) \, _2F_1\left(\text{i}\nu{+}\frac{1}{2},p_2{+}i
   v;2 \text{i}\nu{+}1;-\frac{1}{Y}\right),
\end{align}

\begin{align}
    \mathbf{S}_4{=}& (-1)^{p_2{+}\text{i}\nu} \Gamma \left(\text{i}\nu{+}\frac{1}{2}\right)^2 \Gamma (-2 \text{i}\nu)^2 \Gamma (p_1{+}\text{i}\nu) \Gamma (p_2{+}\text{i}\nu) 
   X^{-p_1-\text{i}\nu}  (-Y)^{-p_2-\text{i}\nu} \qquad\qquad\qquad \qquad \qquad \qquad\nonumber\\&
   _2F_1\left(\text{i}\nu{+}\frac{1}{2},p_1{+}\text{i}\nu;2 \text{i}\nu{+}1;\frac{1}{X}\right)  \ _2F_1\left(i
   \nu{+}\frac{1}{2},p_2{+}\text{i}\nu;2 \text{i}\nu{+}1;-\frac{1}{Y}\right),
\end{align}
\begin{align}
    \mathbf{S}_5{=}& -\frac{(-1)^m 2^{-1+2 \text{i}\nu} (-1)^{-2\text{i}\nu} \Gamma (\text{i}\nu) \Gamma \left(m{+}i
   \nu{+}\frac{1}{2}\right) \Gamma (-m{-}2 \text{i}\nu)\Gamma (m{+}\tilde{p}_1)}{\sqrt{\pi } m! (m{+}p_1{+}\text{i}\nu)} Y^{-m-\tilde{p}_1}  \qquad \qquad \qquad\nonumber\\& \, _2F_1\left({-}m,{-}m{-}2 \text{i}\nu;{-}m{-}i
   \nu{+}\frac{1}{2};X\right)\  
   _2F_1\left(m{+}\tilde{p}_1,\frac{1}{2}{-}\text{i}\nu;1{-}2 i
   v;-\frac{1}{Y}\right),
\end{align}
\begin{align}
    \mathbf{S}_6{=}& -\frac{(-1)^m 2^{-1-2 \text{i}\nu} \Gamma (-\text{i}\nu) \Gamma \left(m{-}\text{i}\nu{+}\frac{1}{2}\right) \Gamma (2
   \text{i}\nu{-}m) \Gamma (m{+}\tilde{p}_1)}{\sqrt{\pi } m! (m{+}p_1{-}\text{i}\nu)}Y^{{-}m{-}\tilde{p}_1}
    \qquad \qquad \qquad \qquad \qquad \nonumber\\&  _2F_1\left({-}m,2 \text{i}\nu{-}m;{-}m{+}\text{i}\nu{+}\frac{1}{2};X\right)\  _2F_1\left(m{+}\tilde{p}_1,\text{i}\nu{+}\frac{1}{2};2 \text{i}\nu{+}1;-\frac{1}{Y}\right),
\end{align}
\begin{align}
    \mathbf{S}_7{=}& -\frac{(-1)^m 2^{-1+2 \text{i}\nu} (-1)^{-2\text{i}\nu} \Gamma (\text{i}\nu) \Gamma \left(m{-}i
   \nu{+}\frac{1}{2}\right) \Gamma (2 \text{i}\nu{-}m) \Gamma
   (m{+}\tilde{p}_1{-}2 \text{i}\nu) }{\sqrt{\pi }m! (m{+}p_1{-}\text{i}\nu)}\,Y^{-m-\tilde{p}_1+2 \text{i}\nu} \qquad\nonumber\\& _2F_1\left(-m,2 \text{i}\nu{-}m;{-}m{+}i
   \nu{+}\frac{1}{2};X\right)  \ _2F_1\left(\frac{1}{2}{-}\text{i}\nu,m{+}\tilde{p}_1{-}2 \text{i}\nu;1{-}2 \text{i}\nu;-\frac{1}{Y}\right), 
\end{align}
\begin{align}
    \mathbf{S}_8{=}& -\frac{(-1)^m 2^{-1-2 \text{i}\nu} \Gamma (-\text{i}\nu) \Gamma \left(m{+}\text{i}\nu{+}\frac{1}{2}\right) \Gamma
   ({-}m{-}2 \text{i}\nu)\Gamma (m{+}\tilde{p}_1{+}2 \text{i}\nu)}{\sqrt{\pi }m! (m{+}p_1{-}\text{i}\nu)} Y^{{-}m{-}\tilde{p}_1{-}2 \text{i}\nu}\qquad\qquad
   \,\nonumber\\& _2F_1\left({-}m,{-}m{-}2 \text{i}\nu;{-}m{-}\text{i}\nu+\frac{1}{2};X\right)
     \ _2F_1\left(i
   \nu{+}\frac{1}{2},m{+}\tilde{p}_1{+}2 \text{i}\nu;2 \text{i}\nu{+}1;-\frac{1}{Y}\right).
\end{align}
Similar to the $R_1$-region result, its correctness can be checked by a numerical evaluation as well (Fig.\ref{fig:R2region}). Series $\mathbf{S}_5,\cdots,\mathbf{S}_8$ converge rapidly in $R_2$-region, making it very efficient for numerical evaluation.

A point we want to emphasize here is that, remarkably, as in the massless model, MoB series solutions for massive family trees also encode the nested-time structures. One can check that the sum of first four building blocks $\mathbf{S}_1,\cdots,\mathbf{S}_4$ are actually products of two massive contact functions i.e.,
\begin{equation}
    (2\text{i}k)^{-\tilde{p}_1}(BC)\times\sum_{i=1}^{4}\mathbf{S}_i=\mathcal{F}^{(2)}(p_1,w_1,k)\times\mathcal{F}^{(1)}(p_2,w_2,k),
\end{equation}
according to the previous calculation \eqref{eq:massivecontact2}. Following the definition \eqref{eq:I11}, these four blocks in fact offer the MoB solution for the factorized massive family tree as
\begin{equation}
    \raisebox{-1em}{\begin{tikzpicture}[scale=0.75]
		\coordinate (X1) at (0,0);
		\coordinate (X2) at (1.5,0);
		\node[below] at (X1) {\small{$1$}};
		\node[below] at (X2) {\small{$2$}};
		\draw[line width=1.5pt,dashed,->] (X1)--(0.9,0);
        \draw[line width=1.5pt,dashed] (0.75,0)--(X2);
		\path[fill=black] (X1) circle[radius=0.11];
		\path[fill=black] (X2) circle[radius=0.11];
\end{tikzpicture}}=(2\text{i}k)^{-\tilde{p}_1}(BC)\times\sum_{i=1}^{4}\mathbf{S}_i.
\end{equation}
The sum of series $\mathbf{S}_5,\cdots,\mathbf{S}_8$, on the other hand, corresponds to a massive family tree with gray edge and the arrow reversed
\begin{equation}
    \raisebox{-1em}{\begin{tikzpicture}[scale=0.75]
		\coordinate (X1) at (0,0);
		\coordinate (X2) at (1.5,0);
		\node[below] at (X1) {\small{$2$}};
		\node[below] at (X2) {\small{$1$}};
		\draw[line width=1.5pt,->,gray] (X1)--(0.9,0);
        \draw[line width=1.5pt,gray] (0.75,0)--(X2);
		\path[fill=black] (X1) circle[radius=0.11];
		\path[fill=black] (X2) circle[radius=0.11];
\end{tikzpicture}}=-(2\text{i}k)^{-\tilde{p}_1}C\times\sum_{i=5}^8 \sum_{m=0}^{\infty}{\mathbf{S}_i},
\end{equation}
which can be shown by a similar MoB calculation from its Euler-Mellin integral representation in $\{Y>X>1\}$ region as
\begin{align}\label{eq:twositeMellin2}
    \raisebox{-1em}{\begin{tikzpicture}[scale=0.75]
		\coordinate (X1) at (0,0);
		\coordinate (X2) at (1.5,0);
		\node[below] at (X1) {\small{$2$}};
		\node[below] at (X2) {\small{$1$}};
		\draw[line width=1.5pt,->,gray] (X1)--(0.9,0);
        \draw[line width=1.5pt,gray] (0.75,0)--(X2);
		\path[fill=black] (X1) circle[radius=0.11];
		\path[fill=black] (X2) circle[radius=0.11];
\end{tikzpicture}}{=}-\frac{C\ \Gamma{(\tilde{p}_1)}}{(2\text{i}k)^{\tilde{p}_1}}\int_0^\infty ds_1&ds_2ds_3 s_1^{\text{i}\nu{-}1/2}s_2^{{-}\text{i}\nu{-}1/2}(1{+}s_1)^{\text{i}\nu{-}1/2}(1{+}s_2)^{{-}\text{i}\nu{-}1/2}\\
    &s_3^{p_2 - i \nu-1}(1{+}s_3)^{p_1{+}\text{i}\nu{-}1}
    {\left((1{+}s_3)(Y{-}s_1){+}s_3(X{+}s_2)\right)^{-\tilde{p}_1}}.\nonumber
\end{align}
 Correspondingly, our result in $R_2$-region \eqref{eq:R2region} in fact encodes the identity as
\begin{equation}
    \raisebox{-1em}{\begin{tikzpicture}[scale=0.75]
		\coordinate (X1) at (0,0);
		\coordinate (X2) at (1.5,0);
		\node[below] at (X1) {\small{$1$}};
		\node[below] at (X2) {\small{$2$}};
		\draw[line width=1.5pt,->] (X1)--(0.9,0);
        \draw[line width=1.5pt] (0.75,0)--(X2);
		\path[fill=black] (X1) circle[radius=0.11];
		\path[fill=black] (X2) circle[radius=0.11];
\end{tikzpicture}}=\raisebox{-1em}{\begin{tikzpicture}[scale=0.75]
		\coordinate (X1) at (0,0);
		\coordinate (X2) at (1.5,0);
		\node[below] at (X1) {\small{$1$}};
		\node[below] at (X2) {\small{$2$}};
		\draw[line width=1.5pt,dashed,->] (X1)--(0.9,0);
        \draw[line width=1.5pt,dashed] (0.75,0)--(X2);
		\path[fill=black] (X1) circle[radius=0.11];
		\path[fill=black] (X2) circle[radius=0.11];
\end{tikzpicture}}-\raisebox{-1em}{\begin{tikzpicture}[scale=0.75]
		\coordinate (X1) at (0,0);
		\coordinate (X2) at (1.5,0);
		\node[below] at (X1) {\small{$2$}};
		\node[below] at (X2) {\small{$1$}};
		\draw[line width=1.5pt,->,gray] (X1)--(0.9,0);
        \draw[line width=1.5pt,gray] (0.75,0)--(X2);
		\path[fill=black] (X1) circle[radius=0.11];
		\path[fill=black] (X2) circle[radius=0.11];
\end{tikzpicture}},
\end{equation}
which is an analog of the nested-time structure as \eqref{eq:masslessid}. This structure will be preserved at massless limit when we set $\nu\to\frac{\text{i}}2$, as we discussed. 

\begin{figure}[h]
\begin{subfigure}{0.5\textwidth}
\centering
\includegraphics[width=0.95\linewidth]{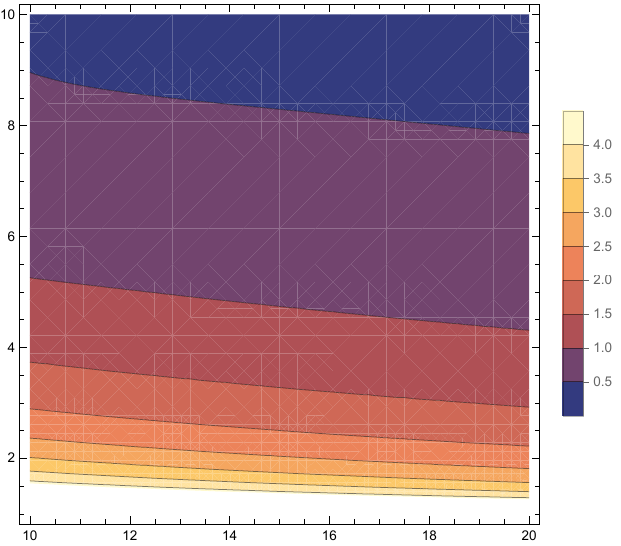}
    \caption{$\nu=\frac{\text{1}}{10}$}
     \vspace{5ex}
      \end{subfigure}
\begin{subfigure}{0.5\textwidth} 
\centering
\includegraphics[width=0.95\linewidth]{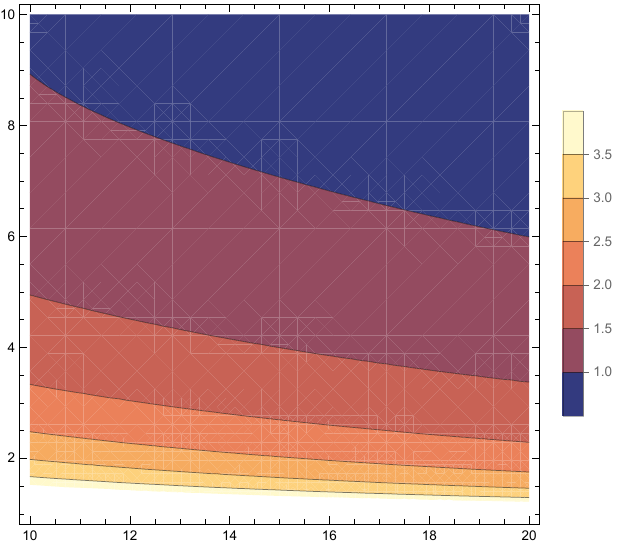}
    \caption{$\nu=\frac{\text{i}}{10}$}
     \vspace{5ex}
      \end{subfigure}
          \begin{subfigure}{0.5\textwidth}  
\centering\includegraphics[width=0.8\linewidth]{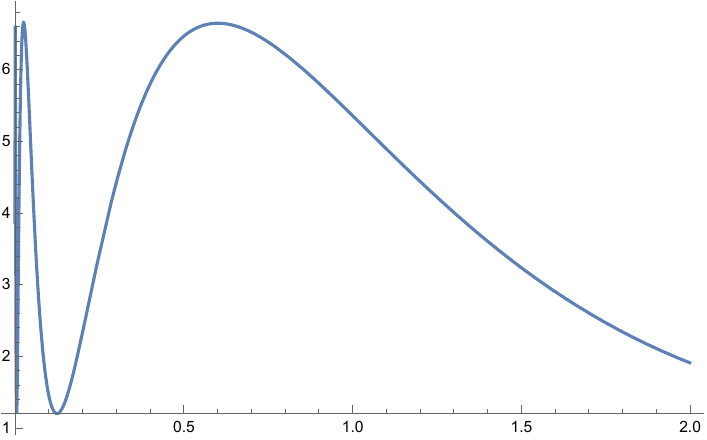}
    \caption{$\nu=1$ and $k\to 0$}
     \vspace{5ex}
      \end{subfigure}
      \begin{subfigure}{0.5\textwidth}  
\centering\includegraphics[width=0.8\linewidth]{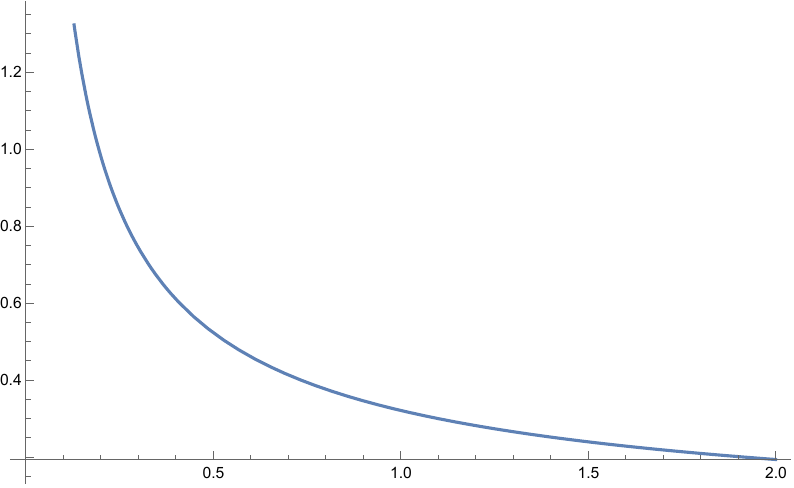}
    \caption{$\nu=\frac{\text{i}}3$ and $k\to 0$}
     \vspace{5ex}
      \end{subfigure}
    \caption{Numerical values of $\mathcal{T}_2(p_1{=}\frac34, p_2{=}\frac57,w_1,w_2,k)$ from MoB series result. In diagrams (a) and (b) we fix the mass $\nu$, $k=1$ and vary $w_1$ (x-axis) and $w_2$ (y-axis). In diagrams (c) and (d) we fix $w_1=95$, $w_2=55$, $\nu$ and vary $k$ near $k\to0$.}
    \label{fig:3}
    \vspace{3ex}
\end{figure}

\paragraph{The full results for two-site cosmological amplitudes}

Finally, we look into full results for four-point  cosmological correlator and wavefunction coefficients with one massive exchange, and give their answer by adding up our basis.

The two-site correlator we would like to consider is
\begin{equation}
\mathcal{T}_2(w_1,w_2,k)=(-\text{i})^{2}\sum_{a_1,a_2=\pm}\int_{-\infty}^0 \prod_{i=1,2}{[\rm d}\tau_i(-\tau_i)^{p_i{-}1}\text{e}^{iw_i\tau_i} a_i D_{a_i}(w_i,\tau_i)] D_{a_1,a_2}(k,\tau_1,\tau_2),
\end{equation}
which after considering the theta-identity can be recast into
\begin{equation}\label{eq:twositefull}
\begin{aligned}
    \mathcal{T}_2(w_1,w_2,k)&=\frac{\pi}4{\rm e}^{-\pi\nu}\left(\raisebox{-1em}{\begin{tikzpicture}[scale=0.75]
		\coordinate (X1) at (0,0);
		\coordinate (X2) at (1.5,0);
		\node[below] at (X1) {\small{$1$}};
		\node[below] at (X2) {\small{$2$}};
		\draw[line width=1.5pt,->] (X1)--(0.9,0);
        \draw[line width=1.5pt] (0.75,0)--(X2);
		\path[fill=black] (X1) circle[radius=0.11];
		\path[fill=black] (X2) circle[radius=0.11];
\end{tikzpicture}}{+}\raisebox{-1em}{\begin{tikzpicture}[scale=0.75]
		\coordinate (X1) at (0,0);
		\coordinate (X2) at (1.5,0);
		\node[below] at (X1) {\small{$2$}};
		\node[below] at (X2) {\small{$1$}};
		\draw[line width=1.5pt,->] (X1)--(0.9,0);
        \draw[line width=1.5pt] (0.75,0)--(X2);
		\path[fill=black] (X1) circle[radius=0.11];
		\path[fill=black] (X2) circle[radius=0.11];
\end{tikzpicture}}{-}\mathcal{F}^{(2)}(p_2,w_2,k,\nu)\times\mathcal{F}^{(1)}(p_1,{-}w_1,k,\nu)\right)+\text{c.c.}\\
&=\frac{\pi}4{\rm e}^{-\pi\nu}\left(\raisebox{-1em}{\begin{tikzpicture}[scale=0.75]
		\coordinate (X1) at (0,0);
		\coordinate (X2) at (1.5,0);
		\node[below] at (X1) {\small{$1$}};
		\node[below] at (X2) {\small{$2$}};
		\draw[line width=1.5pt,->] (X1)--(0.9,0);
        \draw[line width=1.5pt] (0.75,0)--(X2);
		\path[fill=black] (X1) circle[radius=0.11];
		\path[fill=black] (X2) circle[radius=0.11];
\end{tikzpicture}}{-}\raisebox{-1em}{\begin{tikzpicture}[scale=0.75]
		\coordinate (X1) at (0,0);
		\coordinate (X2) at (1.5,0);
		\node[below] at (X1) {\small{$1$}};
		\node[below] at (X2) {\small{$2$}};
		\draw[line width=1.5pt,->,gray] (X1)--(0.9,0);
        \draw[line width=1.5pt,gray] (0.75,0)--(X2);
		\path[fill=black] (X1) circle[radius=0.11];
		\path[fill=black] (X2) circle[radius=0.11];
\end{tikzpicture}}{+}\mathcal{F}^{(2)}(p_2,w_2,k,\nu)\right.\\
&\phantom{aaaaaaaaaaaaaaaaaaaaaaaa}\left.\times(\mathcal{F}^{(1)}(p_2,w_1,k,\nu){-}\mathcal{F}^{(1)}(p_1,{-}w_1,k,\nu))\right)+\text{c.c.},
\end{aligned}
\end{equation}
where $\text{c.c.}$ stands for the complex conjugate, and $\mathcal{F}$ are massive contact functions from \eqref{eq:massivecontact2}. For the wavefunction coefficients, a similar expansion of the basis yields the result
\begin{equation}\label{eq:twositewavefull}
\begin{aligned}
    \psi_2(w_1,w_2,k)&=\frac{\pi}{4}\text{e}^{-\pi\nu}\left(\raisebox{-1em}{\begin{tikzpicture}[scale=0.75]
		\coordinate (X1) at (0,0);
		\coordinate (X2) at (1.5,0);
		\node[below] at (X1) {\small{$1$}};
		\node[below] at (X2) {\small{$2$}};
		\draw[line width=1.5pt,->] (X1)--(0.9,0);
        \draw[line width=1.5pt] (0.75,0)--(X2);
		\path[fill=black] (X1) circle[radius=0.11];
		\path[fill=black] (X2) circle[radius=0.11];
\end{tikzpicture}}{+}\raisebox{-1em}{\begin{tikzpicture}[scale=0.75]
		\coordinate (X1) at (0,0);
		\coordinate (X2) at (1.5,0);
		\node[below] at (X1) {\small{$2$}};
		\node[below] at (X2) {\small{$1$}};
		\draw[line width=1.5pt,->] (X1)--(0.9,0);
        \draw[line width=1.5pt] (0.75,0)--(X2);
		\path[fill=black] (X1) circle[radius=0.11];
		\path[fill=black] (X2) circle[radius=0.11];
\end{tikzpicture}}{+}\text{i}\mathcal{F}^{(2)}(p_2,w_2,k,\nu)\times\mathcal{F}^{(2)}(p_1,w_1,k,\nu)\right)\\
&=\frac{\pi}4{\rm e}^{-\pi\nu}\left(\raisebox{-1em}{\begin{tikzpicture}[scale=0.75]
		\coordinate (X1) at (0,0);
		\coordinate (X2) at (1.5,0);
		\node[below] at (X1) {\small{$1$}};
		\node[below] at (X2) {\small{$2$}};
		\draw[line width=1.5pt,->] (X1)--(0.9,0);
        \draw[line width=1.5pt] (0.75,0)--(X2);
		\path[fill=black] (X1) circle[radius=0.11];
		\path[fill=black] (X2) circle[radius=0.11];
\end{tikzpicture}}{-}\raisebox{-1em}{\begin{tikzpicture}[scale=0.75]
		\coordinate (X1) at (0,0);
		\coordinate (X2) at (1.5,0);
		\node[below] at (X1) {\small{$1$}};
		\node[below] at (X2) {\small{$2$}};
		\draw[line width=1.5pt,->,gray] (X1)--(0.9,0);
        \draw[line width=1.5pt,gray] (0.75,0)--(X2);
		\path[fill=black] (X1) circle[radius=0.11];
		\path[fill=black] (X2) circle[radius=0.11];
\end{tikzpicture}}{+}\mathcal{F}^{(2)}(p_2,w_2,k,\nu)\right.\\
&\phantom{aaaaaaaaaaaaaaaaaaaaaaaa}\left.\times(\mathcal{F}^{(1)}(p_2,w_1,k,\nu){+}i\mathcal{F}^{(2)}(p_1,w_1,k,\nu))\right).
\end{aligned}
\end{equation}

Therefore, we can now express the cosmological integrals in MoB series solutions straightforwardly, and arrive at their numerical result. Let us concentrate on the region when $w_1> w_2,k$, since the non-trivial parts in \eqref{eq:twositefull} and \eqref{eq:twositewavefull} are basic massive family trees with $1$ as the 
ancestor again. Series solutions for the gray two-site family tree can be obtained from our general solution, whose explicit result can be found as \eqref{eq:graytwosite} in Appendix \ref{app:results}. In Fig.\ref{fig:3} we have plotted the above correlator fixing various parameters. In the squeezed limit, i.e., as $k\rightarrow0$, we observe the expected behavior from cosmological collider physics \cite{Arkani-Hamed:2015bza}. When the mass of the intermediate scalar is comparable to the Hubble scale (i.e., $\nu \in \text{i} ~\mathbb{R}$), we get a pure power-law decay as shown in Fig.\ref{fig:3}(d). For heavier scalars (i.e., $\nu \in \mathbb{R}$) ,  we have oscillations confined to the vicinity of $k=0$, superimposed on this power law envelope.

\section{Summary and Outlook}
\label{sec:summary}
In this work, we computed the fundamental building blocks of all tree-level cosmological correlators involving conformally-coupled and massive scalar fields with polynomial interactions in general FRW spacetimes. Our approach employed the Method of Brackets, a systematic technique for evaluating Mellin-type integrals based on a multivariable generalization of Ramanujan’s master theorem. The method is governed by a compact set of operational rules and proceeds by expanding the integrand and solving a linear system of bracket equations. This framework enabled us to obtain explicit series representations of cosmological correlators valid in various kinematic regions. Moreover, by recognizing structural patterns in the bracket equations, we were able to derive Feynman-like rules that generate the series directly from the interaction graph, offering a highly streamlined alternative to traditional integration. We then briefly showed how to obtain cosmological amplitudes from these building blocks in several examples.
These results provide both a powerful analytic tool for studying the structure of cosmological amplitudes and an efficient approach for their numerical evaluation.

We now outline several directions for future work.
\paragraph{Limits and simplifications}
Having  explicitly obtained the building blocks  for generic polynomial interactions on a generic FRW background for any tree level {\it in-in} correlator/wavefunction, a very natural question would be to investigate how our results simplify when we take various limits such as the mass to be small, objects which in certain cases have IR divergences and it would be interesting to see if we can recover some universal pieces analogous to soft-theorems in flat space. Another case of interest would be when masses differ by integers from the massless case. In these case it is known that the solutions are related by using {\it weight shifting operators} \cite{Arkani-Hamed:2018bjr} and we expect that there is drastic simplification at the level of the result.
Other interesting limits would be to special FRW background such de Sitter and flat space background where the results are expected to simplify to polylogarithms. In this work we briefly investigated our result and showed that our solutions have a smooth limit to the conformal case. We hope to report on these issues soon in a companion paper.

\paragraph{Physical interpretation of individual MoB series}

It generally true that the physical result for full {\it in-in} correlator or wavefunction coefficients are much simpler then the time integrals. We obtained series solutions for the time integrals and had to combine them to get the physical result for the for full {\it in-in} correlator or wavefunction coefficients. For instance the full result is real despite the individual building blocks not being real.  Despite having a completely generic series solution this simplification is not manifest. This is not a drawback of the method but was rather forced on us by the choice of starting point, namely the decomposition into time integrals in the first place. It is natural to wonder if there is a better starting point, such as a representation that does not involve time orderings in c.f. \cite{Werth:2024mjg}, could directly give us the simpler result rather than as a sum over large number of pieces with nested-time orderings. We hope to report on this investigation in the near future.

\paragraph{More general interactions and cosmological EFTs}
Another direction of interest would be investigating more general interactions including in particular cases with a non-trivial speed of sound \cite{Jazayeri:2022kjy,Jazayeri:2023xcj,Qin:2025xct} and derivative interactions \cite{DuasoPueyo:2023kyh}. There is no obstruction to applying MoB to these cases and it would be interesting to investigate qualitative features of the results in these cases. The nature of EFTs in curved space is also not very well understood and these investigations might shed light on certain issues \cite{Beneke:2023wmt, DuasoPueyo:2025lmq}. We leave this fascinating question to future work.

\paragraph{MoB calculation for $AdS$ correlation functions}
It would also be interesting to apply these methods to Witten diagrams for correlators in $AdS$, these are also objects that are of great interest for understanding QFT in curved spacetime and the Euclidean version has a close relation to wavefunction/correlators on the $dS$ background. We leave this investigation to future work.

\paragraph{Cosmological amplitudes at loop level}
In this work, we mainly focused on the tree-level correlators/ wavefunction coefficients. A natural direction of great interest would be to see if MoB can be applied to cosmological computations at loop level \cite{Wang:2021qez,Heckelbacher:2022hbq,Xianyu:2022jwk,Qin:2023bjk,Benincasa:2024lxe,Benincasa:2024ptf,Zhang:2025nzd,Qin:2024gtr}. Apart from being a useful computational tool to evaluate difficult integrals these computations could also shed light on various subtleties of QFT on curved spacetime. We leave this investigation to future work and expect \cite{Chowdhury:2025ohm} to help us in regard.

\section*{Acknowledgments}
It is a pleasure to thank Daniel Baumann for proposing the problem, and Iván González for help with MoB. We also thank Chandramouli Chowdhury, Harry Goodhew, Jonathan Gräfe, Song He, Jiajie Mei, Ivo Sachs, Francisco Vazão, Kamran Salehi Vaziri and Denis Werth for discussions, and especially Ivo Sachs, Francisco Vazão and Denis Werth for valuable comments on the manuscript. PR and QY are funded by the European Union (ERC, UNIVERSE PLUS, 101118787). Views and opinions expressed are however those of the authors only and do not necessarily reflect those of the European Union or the European Research Council Executive Agency. Neither the European Union nor the granting authority can be held responsible for them.

\appendix

\section{Hardy's criterion for MoB}
\label{app:Hardy}
In Sec.\ref{sec:RMT}, we mentioned that the theorem applies provided the Taylor coefficients satisfy some growth conditions. Since the theorem gives us the result by an analytic continuation of the Taylor coefficients from positive integers to complex values, one would have to worry about the uniqueness of this analytic continuation. For example, if $a(n)=n!$ for all $n=0,1,2,\dots$ then a valid analytic continuation is $a_1(s)= \Gamma[s+1]$ but another analytic continuation is given by $a_2(s)=\Gamma[s+1]+\sin(2 \pi s)$. The latter, however, is ruled out if impose growth conditions on $a(s)$, and this is precisely what Carlson's theorem states. 

The interested reader can see \cite{GH, ZB} for more details and proofs of the theorems. We will merely state them here.
\begin{tcolorbox} 
{\bf Carlson's theorem:}   Let, $g(z)$ be an analytic function on the half-plane $Re(z)\ge 0$ such that 
\[|g(z)| \le C e^{\tau |z|} \quad \forall \quad Re(z)>0\]
for some $C,\tau \in \mathbb{R}$. If there exists a constant $A<\pi $ such that 
\begin{equation} \label{HC}
 |g(i y)| \le C e^{A |y|} \quad \forall y \in \mathbb{R}
\end{equation} 
and if $g(z)$ vanishes on the non-negative integers, then $g(z)$ is identically zero.
\end{tcolorbox}

The above result ensures analytic continuation of functions from natural numbers to complex values and motivates looking at a special class of functions called the \emph{Hardy type}.\\

\noindent {\bf Definition:} An analytic function $f:\mathbb{C}\rightarrow \mathbb{C}$ is of Hardy type if $\exists ~C,P,A \in \mathbb{R}$ with $A<\pi$ such that 
\[|f(z)|\le C e^{P x +A |y|}  \quad \forall  z=x +i y \in  \mathbb{C}\quad {\rm with} \quad x >0\]

We can now state the theorem Hardy proved by imposing  growth conditions on Ramanujan's master theorem \cite{GH}
\begin{tcolorbox} 
{\bf Hardy-Ramanujan Master theorem:} Let, $a(z)$ be analytic in the upper half-plane \[H(\delta)=\{z\in \mathbb{C}| Re(z) \ge -\delta \}\quad {\rm for}~{\rm some}~0<\delta <1\] that belongs to the Hardy type then,
\begin{equation}\label{HRT}
    \int_{0}^{\infty}dx~ x^{s-1}~\sum\limits_{n=0}^\infty \frac{\left(-x\right)^n}{n!} a(n) = a(-s) \Gamma[s] 
\end{equation}
holds for all $0<Re(s)<\delta$.
\end{tcolorbox}
The above theorem can be proved by Mell{\it in-in}version theorem see \cite{GH,ZB}. We can also state the multi-dimensional version which can also be proved multi-variate Mellin inversion theorem \cite{ZB}.
\begin{tcolorbox}
{\bf Generalized Ramanujan Master theorem:} Let, $a(z_1,\cdots,z_n)$ be holomorphic on 
\begin{align*}
H^n(\delta_1,\cdots,\delta_n)=\{z=(z_1,\cdots,z_n)\in \mathbb{C}^n| Re(z_i) \ge -\delta_i ~\forall~ i=1,\dots,n \}\quad \\
{\rm for}~{\rm some}~0<\delta_i <1
\end{align*} 
that belongs to the Hardy type i.e., there exists constants $C,P_1,\dots,P_n$ and $A_1,\dots,A_n<\pi$ such that
\[|f(v_1+i w_1,\dots,v_n+i w_n) \le C \exp{\left(\sum_j (P_j v_j +A_j |w_j|)\right)}\]
for all $(z_1,\dots,z_n)=(v_1+i w_1,\dots,v_n+i w_n) \in H^{n}(\delta_1,\dots,\delta_n)$. Then,
\begin{align*}
\int_{0}^{\infty}\dots \int_{0}^{\infty}dx~ x_1^{s_1-1} \dots x_n^{s_n-1}~&\sum\limits_{k_1,\dots,k_n}^\infty \frac{\left(-x\right)^{k_1+\dots+k_n}}{k_1!\dots k_n!} a(k_1,\dots,k_n) x_1^{k_1}\dots x_n^{k_n} \\
&= a(-s_1,\dotsm-s_n) \prod\limits_{j=1}^n\Gamma[s_j] 
\end{align*}
holds for all $0<Re(s_i)<\delta_i$.
\end{tcolorbox}

\section{Deriving solutions for massive family trees via MoB}
\label{sec:basic massive solution derivation}
In this Appendix, we record details of deriving result \eqref{eq:resultmass}, \eqref{eq:T000} and \eqref{eq:T111}. The procedure is a tedious but straightforward application of MoB. For simplicity of writing the indices, we firstly show an explicit computation for integral
\begin{align}\label{eq:A1}
    \mathcal{I}[(12\cdots N)]=&\raisebox{-1em}{\begin{tikzpicture}[scale=0.75]
		\coordinate (X1) at (0,0);
		\coordinate (X2) at (1.5,0);
        \coordinate (X3) at (3,0);
            \coordinate (X4) at (4.5,0);
		\node[below] at (X1) {\small{$1$}};
		\node[below] at (X2) {\small{$2$}};
        \node[below] at (X4) {\small{$N$}};
        \node[below] at (X3) {\small{$N{-}1$}};
		\draw[line width=1.5pt,->] (X1)--(0.9,0);
        \draw[line width=1.5pt] (0.75,0)--(X2);
        \draw[line width=1.5pt,dashed] (X2)--(X3);
        \draw[line width=1.5pt,->] (X3)--(3.9,0);
        \draw[line width=1.5pt] (3.75,0)--(X4);
		\path[fill=black] (X1) circle[radius=0.11];
		\path[fill=black] (X2) circle[radius=0.11];
        \path[fill=black] (X3) circle[radius=0.11];
        \path[fill=black] (X4) circle[radius=0.11];
\end{tikzpicture}}\nonumber\\
    =&\int_{-\infty}^0\prod_{i=1}^{N}[{\rm d}\tau_i(-\tau_i)^{p_i{-}1}{\rm e}^{\text{i}w_i\tau_i}]\prod_{i=1}^{N{-}1}[\text{H}_{-\text{i}\nu}^{(2)}({-}k_{i,i{+}1}\tau_i)\text{H}_{\text{i}\nu}^{(1)}({-}k_{i,i{+}1}\tau_{i{+}1})\theta(\tau_{i{+}1}{-}\tau_i)]
\end{align}
and generalization for the procedure to any trees will be then commented, like what we have done for massless cases. Furthermore, we will mainly focus on the following physical region
\begin{equation}\label{eq:A11}
    0<\left\{\frac{k_{i,j}}{w_1{+}k_{1,2}}, \frac{w_i{-}k_{i{-}1,i}{+}k_{i,i{+}1}}{w_1{+}k_{1,2}}\right\}_{i=2,\cdots, n}<1,\ \ k_{n,n{+}1}:=0 
\end{equation}
which is an analog of $R_1$-region we discussed in two-site chain example. As we have emphasized, series solutions for other physical regions also arise from MoB, and encode the nested-time structure simply.

As we have mentioned in the beginning part of section 4, identity \eqref{eq:1} helps us to rewrite an $N$-site massive family tree as $2(N{-}1)$-fold integration over a massless one, with the energies and twists deformed. Following this logic we introduce one fold integration $s_{i,i{+}1}^{(t)}$ for each $\text{H}_{\pm \text{i}\nu}^{(t)}(-k_{i,i{+}1}\tau)$, and finally have
\begin{equation}\label{eq:deformed}
    \mathcal{I}[(1\cdots N)]{=}C^{N{-}1}\int_0^\infty\prod_{i=1}^{N{-}1}{\rm d}s_{i,i{+}1}^{(1)}{\rm d}s_{i,i{+}1}^{(2)}(s_{i,i{+}1}^{(1)}(s_{i,i{+}1}^{(1)}{+}1))^{\text{i}\nu{-}\frac12}(s_{i,i{+}1}^{(2)}(s_{i,i{+}1}^{(2)}{+}1))^{{-}\text{i}\nu{-}\frac12}\hat{P}[(1\cdots N)]
\end{equation}
where $C=\frac{4{\rm e}^{2\pi \nu}{\rm cosh}(\pi \nu)}{\pi^2}$, $\hat{P}[(1\cdots N)]$ stands for a deformed massless chain \eqref{eq:masleschain} with energies
\begin{align}\label{eq:defenergy}
    &\hat\omega_1=w_1{+}k_{1,2}(1{+}2s_{1,2}^{(2)}),\ \  \hat{\omega}_N=w_N{-}k_{N{-}1,N}(1{+}2s_{N{-}1,N}^{(1)})\\
    &\hat{\omega}_i=w_{i}{-}k_{i{-}1,i}(1{+}2s_{i{-}1,i}^{(1)}){+}k_{i,i{+}1}(1{+}2s_{i,i{+}1}^{(2)}), \ \ \ i=2,\cdots, N{-}1
\end{align}
and twists
\begin{equation}\label{eq:deftwist}
    \hat{p}_1=p_1{-}\text{i}\nu,\  \hat{p}_N=p_N{+}\text{i}\nu,\ \hat{p}_i=p_i\ \ \ i=2,\cdots, N{-}1
\end{equation}
Similar to the two-site chain calculation, we adopt Euler-Mellin representation for the deformed massless chain as
\begin{equation}
   \hat{P}[(1\cdots N)]=(-\text{i})^{N+\tilde{p}_{1}}\Gamma(\tilde{p}_{1})\int_0^\infty \prod_{i=2}^N\left({\rm d}r_i r_i^{\tilde{p}_i{+}\text{i}\nu{-}1}(1{+}r_i)^{\tilde{p}_{1}{-}\tilde{p}_i{-}\text{i}\nu{-}1}\right)\ \left(\sum_{i=1}^N\hat{\omega}_i\prod_{j=2}^i r_j\prod_{j=i{+}1}^N(1{+}r_j)\right)^{-\tilde{p}_{1}} 
\end{equation}
Here  $\tilde{p_i}=\sum_{j=i}^{N}p_i$ is a sum of twists accordingly. Therefore, $\mathcal{I}[(1\cdots N)]$ is written purely in Euler-Mellin integral with $3(N{-}1)$-folds integration, and this is the beginning point for us to apply MoB.

\paragraph{Deriving bracket equations and summand}
We firstly introduce indices and bracket equations from each polynomial factors in the integrand following \eqref{eq:bracketrule2}, which will finally result in summand. Note that we omit all factor $\sum_{n}\phi_{n}$ for all expressions in the following for simplicity.

Firstly, we introduce $2(N{-}1){+}N=3N{-}2$ indices 
\begin{equation}
    \{n_{i,i{+}1}^{(1)},n_{i,i{+}1}^{(2)}\}_{i=1,\cdots, N{-}1}\cup\{m_i\}_{i=1,\cdots,N}
\end{equation}
and one bracket equation as
\begin{align}\label{eq:massbra0}
    &\left(\sum_{i=1}^N\hat{\omega}_i\prod_{j=2}^i r_j\prod_{j=i{+}1}^N(1{+}r_j)\right)^{-\tilde{p}_{1}}=\frac{\langle\tilde{p}_1{+}\tilde{m}_1{+}\sum_{i=1}^{N{-}1}\sum_{s=1,2}n_{i,i{+}1}^{(s)}\rangle}{\Gamma(\tilde{p}_1)} \prod_{i=1}^N(w_i{-}k_{i{-}1,i}{+}k_{i,i{+}1})^{m_i}\nonumber\\    &\phantom{aaaaaaaaaaaaa}\times\prod_{i=1}^{N{-}1}\left((-1)^{n_{i,i{+}1}^{(1)}}\prod_{s=1,2}(2k_{i,i{+}1}s_{i,i{+}1}^{(s)})^{n_{i,i{+}1}^{(s)}}\right)\prod_{i=2}^N\left(r_i^{d_{i,1}}(1{+}r_i)^{d_{i,2}}\right)
\end{align}
with $k_{0,1}=k_{N,N{+}1}=0$, and the shorthand notation
\begin{align}
    &d_{i,1}:=n_{i{-}1,i}^{(1)}{+}\sum_{j=i}^{N{-}1}\sum_{s=1,2}n_{j,j{+}1}^{(s)}+\tilde{m}_i\\    &d_{i,2}:=\sum_{j=1}^{N{-}1}\sum_{s=1,2}n_{j,j{+}1}^{(s)}+\tilde{m}_1-d_{i,1}
\end{align}
Secondly, for $(1{+}s_{i,i{+}1}^{(s)})$, we introduce $4(N{-}1)$ indices
\begin{equation}
    \{l_{i,i{+}1}^{(s)},m_{i,i{+}1}^{(s)}\}_{i=1,\cdots,N{-}1\ ;\ s=1,2}
\end{equation}
and $2(N{-}1)$ bracket equations as
\begin{equation}\label{eq:massbra1}
    (s_{i,i{+}1}^{(1)}{+}1)^{\text{i}\nu{-}\frac12}\sim\frac{1}{\Gamma({-}\text{i}\nu{+}\frac12)}(s_{i,i+1}^{(1)})^{m_{i,i{+}1}^{(1)}}\langle l_{i,i{+}1}^{(1)}{+}m_{i,i{+}1}^{(1)} {-}\text{i}\nu{+}\frac12\rangle
\end{equation}
\begin{equation}\label{eq:massbra2}
    (s_{i,i{+}1}^{(2)}{+}1)^{-\text{i}\nu{-}\frac12}=\frac{1}{\Gamma(\text{i}\nu{+}\frac12)}(s_{i,i+1}^{(2)})^{m_{i,i{+}1}^{(2)}}\langle l_{i,i{+}1}^{(2)}{+}m_{i,i{+}1}^{(2)} {+}\text{i}\nu{+}\frac12\rangle
\end{equation}
Furthermore, for $(1{+}r_i)$, we have $2(N{-}1)$ more indices
\begin{equation}
    \{u_i,v_i\}_{i=2,\cdots,N}
\end{equation}
and $N{-}1$ more bracket equations as
\begin{equation}\label{eq:massbra4}
    (1+r_i)^{d_{i,2}{+}\tilde{p}_1{-}\tilde{p}_i{-}\text{i}\nu{-}1}=\frac{r_i^{u_i}}{\Gamma(1{+}\text{i}\nu{-}d_{i,2}{-}\tilde{p}_1{+}\tilde{p}_i)}\langle 1{+}\text{i}\nu{-}d_{i,2}{-}\tilde{p}_1{+}\tilde{p}_i{+}u_i{+}v_i\rangle
\end{equation}
Finally, gathering powers of all variables $s_{i,i{+}1}^{(s)}$ and $r_i$, we have extra $3(N{-}1)$ bracket equations
\begin{equation}\label{eq:massbra3}
\begin{aligned}
    &s_{i,i{+}1}^{(1)}:\ \langle \text{i}\nu{+}\frac12{+}n_{i,i{+}1}^{(1)}{+m_{i,i{+}1}^{(1)}}\rangle,\\ \ &s_{i,i{+}1}^{(2)}:\ \langle {-}\text{i}\nu{+}\frac12{+}n_{i,i{+}1}^{(2)}{+m_{i,i{+}1}^{(2)}}\rangle,\\
    &r_i:\phantom{aaa} \langle d_{i,1}{+}u_i{+}\tilde{p}_i{+}\text{i}\nu\rangle
\end{aligned}
\end{equation}
Based on these ingredients, the integrand \eqref{rmt3} except the bracket equations finally reads
\begin{equation}\label{eq:generalS}
    (-\text{i})^{N{+}\tilde{p}_1}(BC)^{N{-}1}\frac{(-1)^{\sum_{i=1}^{N{-}1}n_{i,i{+}1}^{(1)}}\prod_{i=2}^{N}(2k_{i{-}1,i})^{n_{i-1,i}^{(1)}{+}n_{i,i{+}1}^{(2)}}\prod_{i=1}^{N}(w_i{-}k_{i-1,i}{+}k_{i,i{+}1})^{m_i}}{\prod_{i=2}^{N}\Gamma(1{+}\text{i}\nu{-}d_{i,2}{-}\tilde{p}_1{+}\tilde{p}_i)}
\end{equation}
with $B$ and $C$ following \eqref{eq:BC}

\paragraph{Solving bracket equations}
During the procedure, we introduce $9N{-}8$ indices and $6N{-}5$ brackets equations, so therefore generally a $N$-site graph is of rank $3(N{-}1)$. We will mainly focus on the region with $w_1{+}k_{1,2}>w_i,k_{i,j}$ for other energies in the chain, which is also an analog of \eqref{eq:chainres}. Therefore, we would mainly focus on the bracket solutions, such that independent arguments  are
\begin{equation}
    0<\left\{\frac{k_{i,j}}{w_1{+}k_{1,2}}, \frac{w_i{-}k_{i{-}1,i}{+}k_{i,i{+}1}}{w_1{+}k_{1,2}}\right\}<1
\end{equation}
in the power series\footnote{Here is a subtlety that number of the variables we list in \eqref{eq:A11}, or scales of this massive correlators, is only $2N{-}1$, which is smaller than the number of independent indices from the bracket equations. As a result, it is not straightforward to disentangle valid series solutions from the spurious ones for a physical region, since there may be some coincident degeneration of the arguments and series. To solve this problem, one can firstly deform $(1{+}2s_{i,i{+}1}^{(t)})\to (1{+}2 A_{i,i{+}1}^{(t)}s_{i,i{+}1}^{(t)})$ in the computation, treat $A_{i,i{+}1}^{(t)}$ as extra $(N{-}1)$ independent variables, and then do the bracket computation. Correspondingly, all arguments in each MoB series will always be ratio of kinematical variables. Finally we set $A_{i,i{+}1}^{(t)}\to1$, and discard unphysical ones. This is also the reason why only 12 out of 40 series are physical in two-site chain discussion (Sec.\ref{sec:two site massive}). We thank Iván González for pointing out this to us.}. For this special physical region, simple linear algebraic computation shows that the independent $3(N{-}1)$ indices $\sigma$ should be chosen from
\[\{m_i\}_{i=2,\cdots, N}\ \cup \{\text{one of}\ n_{i,i{+}1}^{(s)} \text{or}\ l_{i,i{+}1}^{(s)}\}_{i=2,\cdots N,\ s=1,2}\]
and since we can choose $n_{i,i{+}1}^{(s)}$ or $l_{i,i{+}1}^{(s)}$ for each $s=1,2$ and all $i$, there are $2^{2(N{-}1)}$ choices for this specific region. From bracket equations  \eqref{eq:massbra1}, \eqref{eq:massbra2}, and \eqref{eq:massbra3}, it can be easily seen that each pair of these two indices enjoys simple relation
\begin{equation}\label{eq:shift}   l_{i,i{+}1}^{(1)}=n_{i,i{+}1}^{(1)}+2\text{i}\nu,\ l_{i,i{+}1}^{(2)}=n_{i,i{+}1}^{(2)}-2\text{i}\nu
\end{equation}
And finally, other indices can be expressed by linear combinations of these indices as
\begin{align}
    &m_1^*=-\sum_{i=1}^{N{-}1}(n_{i,i{+}1}^{(1)}{+}n_{i,i{+}1}^{(2)})-\tilde{m}_2-\tilde{p}_1\nonumber\\
    &(m_{i,i{+}1}^{(1)})^*={-}n_{i,i{+}1}^{(1)}{-}\text{i}\nu{-}\frac12,\ \ (m_{i,i{+}1}^{(2)})^*={-}n_{i,i{+}1}^{(2)}{+}\text{i}\nu{-}\frac12\\
    &u_i^*=-{d}_{i,1}{-}\tilde{p}_i{-}\text{i}\nu,\ \ v_i^*=-1\nonumber
\end{align}
For the solutions with $\sigma=\{m_i\}_{i=2,\cdots,N}\cup\{n_{i,i{+}1}^{(s)}\}_{s=1,2,\ i=2,\cdots,N}$ , the summand finally reads (we retain the $\phi_n$ factors for independent indices now)
\begin{align}\label{eq:T0001}  
(-\text{i})^{N{+}\tilde{p}_1}&\phi_{m_i,n_{i,i{+}1}^{(s)}}(BC)^{N{-}1}\frac{(-1)^{\sum_in_{i,i{+}1}^{(1)}}}{(w_1{+}k_{1,2})^{\tilde{p}_1}}\Gamma(\sum_{i=1}^{N{-}1}n_{i,i{+}1}^{(1)}{+}n_{i,i{+}1}^{(2)}{+}\tilde{m}_2{+}\tilde{p}_1)\\   &\prod_{i=1}^{N{-}1}f(n_{i,i{+}1}^{(1)})g(n_{i,i{+}1}^{(2)})\prod_{i=2}^{N}\left(\frac{w_i{-}k_{i{-}1,i}{+}k_{i,i{+}1}}{w_1{+}k_{1,2}}\right)^{m_i}\left(\frac1{\hat{n}_i{+}\tilde{p}_i{+}\text{i}\nu}\right)\left(\frac{2k_{i{-}1,i}}{w_1{+}k_{1,2}}\right)^{n_{i{-}1,i}^{(1)}{+}n_{i{-}1,i}^{(2)}}\nonumber
\end{align}
where we denote $k_{N,N{+}1}:=0$ again,
and
\begin{equation}
    \hat{n}_k=\sum_{i=k}^Nm_i{+}\sum_{i=k}^{N{-}1}(n_{i,i{+}1}^{(1)}{+}n_{i,i{+}1}^{(2)}){+}n_{k{-}1,k}^{(1)},
\end{equation}
following \eqref{eq:deformn}. Each of the indices is summed from 0 to $\infty$, as definition. This is actually the general solution \eqref{eq:T000} when applied to massive family chain integrals with only black thick edges.

Especially, indices $m_i$ in the expression can be generally summed over, by first performing change of variables $n_{i{-}1,i}^{(1)}\to n_{i{-}1,i}^{(1)}{-}m_i$, and then sum over $m_i$ from $0$ to $n_{i{-}1,i}^{(1)}$ following the equation
\begin{align}    &\sum_{m_i=0}^{n_{i{-}1,i}^{(1)}}\left(-\frac{w_i{-}k_{i{-}1,i}{+}k_{i,i{+1}}}{2k_{i{-}1,i}}\right)^{m_i}\frac{\Gamma(n_{i{-}1,i}^{(1)}{-}m_i{+}\text{i}\nu{+}\frac12)\Gamma(m_i{-}n_{i{-}1,i}{-}2\text{i}\nu)}{\Gamma(m_i)\Gamma(n_{i{-}1,i}^{(1)}{-}m_i)}\nonumber\\
&=\frac{f(n_{i{-}1,i}^{(1)})}{\Gamma(n_{i{-}1,i}^{(1)})}\ _2F_1\left({-}n_{i{-}1,i}^{(1)}, {-}n_{i{-}1,i}^{(1)}{-}2\text{i}\nu,\ \frac12{-}n_{i{-}1,i}^{(1)}{+}\text{i}\nu,{-}\frac{w_i{-}k_{i{-}1,i}{+}k_{i,i{+1}}}{2k_{i{-}1,i}}\right)
\end{align}
we can also arrive at the series solution over only $2(N{-}1)$ indices $n_{i,i{+}1}^{(s)}$
\begin{align} \label{gnsr}
    &\mathbf{T}_{0,\cdots,0}{=}(-\text{i})^{N{+}\tilde{p}_1}\left(BC\right)^{N{-}1}\sum_{n_{i,i{+}1}^{(s)}}\phi_{n_{i,i{+}1}^{(s)}}\frac{(-1)^{\sum_i n_{i,i{+}1}^{(1)}}}{(w_1{+}k_{1,2})^{\tilde{p}_1}}f(n_{i,i{+}1}^{(1)})g(n_{i,i{+}1}^{(2)})\Gamma(\sum_{i=1}^{N{-}1}n_{i,i{+}1}^{(1)}{+}n_{i,i{+}1}^{(2)}{+}\tilde{p}_1)\nonumber\\    &\prod_{i=2}^N\left(\frac1{\hat{n}_i^\prime{+}\tilde{p}_i{+}\text{i}\nu}\right)\left(\frac{2k_{i{-}1,i}}{w_1{+}k_{1,2}}\right)^{n_{i{-}1,i}^{(1)}{+}n_{i{-}1,i}^{(2)}}\ _2F_1\left({-}n_{i{-}1,i}^{(1)}, {-}n_{i{-}1,i}^{(1)}{-}2\text{i}\nu,\ \frac12{-}n_{i{-}1,i}^{(1)}{+}\text{i}\nu,{-}\frac{w_i{-}k_{i{-}1,i}{+}k_{i,i{+1}}}{2k_{i{-}1,i}}\right)
\end{align}
where $\hat{n}_i^\prime=\hat{n}_i-\tilde{m}_i$.

As we have mentioned, besides the primary series, we still have $2^{2(N{-}1)}-1$ other series for this physical region, whose result can be obtained by solving $n_{i,i{+}1}^{(s)}$ by $l_{i,i{+}1}^{(s)}$ instead following relation \eqref{eq:shift}, {\it i.e.}, we need to do the replacement
\[\phi_{n_{i,i{+}1}^{(s)}}\Gamma(-l_{i,i{+}1}^{(s)})\to \phi_{l_{i,i{+}1}^{(s)}}\Gamma(-n_{i,i{+}1}^{(s)})\]
in the summand, and then substitute indices by their solutions from bracket equation. Practically, it can be achieved by shifting $n_{i,i{+}1}^{(t)}\to n_{i,i{+}1}^{(t)}{-}(-1)^t 2\text{i}\nu$, and then time a factor
\[\frac{(-1)^{(-)^{t{-}1}2\text{i}\nu}\Gamma(-n_{i,j}^{(t)}{-}(-1)^t2\text{i}\nu)\Gamma(n_{i,j}^{(t)}{+}(-1)^t2\text{i}\nu{+}1)}{\Gamma(n_{i,j}^{(t)}{+}1)\Gamma(-n_{i,j}^{(t)})}\]
in front of the series. These actions are exactly \eqref{eq:T111} in the main text . By acting this operation $2^{2(N{-}1)}{-}1$ times for all indices $n_{i,i{+}1}^{(s)}$, we finally arrive at $2^{2(N{-}1)}$ series, whose summation yields the finally result for \eqref{eq:A1}.

\paragraph{More details of \eqref{eq:R1region} and \eqref{eq:R2region}}
We present some of the details for deriving \eqref{eq:R1region} and \eqref{eq:R2region} from MoB.

Similar to general solutions for arbitrary basic massive family tree in physical region \eqref{eq:A11}, we refer to the procedure from eq.\eqref{eq:A1} to eq.\eqref{eq:generalS} for derivation of all bracket equations and the summand. According to the general procedure, for the special case of two-site chain graph, we should introduce $10$ indices as
\begin{equation}
    \{n_{1,2}^{(1)},n_{1,2}^{(2)},m_1,m_2,l_{1,2}^{(1)},l_{1,2}^{(2)},m_{1,2}^{(1)},m_{1,2}^{(2)},u_2,v_2\}
\end{equation}
satisfying $7$ bracket equations
\begin{equation}
\left\{\begin{aligned}
    &m_1{+}m_2{+}n_{1,2}^{(1)}{+}n_{1,2}^{(2)}{+}\tilde{p}_1=0,\ \ m_{1,2}^{(1)}{+}l_{1,2}^{(1)}{+}\text{i}\nu{+}\frac12=0,\ \ m_{1,2}^{(2)}{+}l_{1,2}^{(2)}{-}\text{i}\nu{-}\frac12=0\\    &\phantom{aaaaaaaa}1{-}m_1{-}n_{1,2}^{(2)}{-}p_1{+}u_2{+}v_2+\text{i}\nu=0,\ \ m_{1,2}^{(1)}{+}n_{1,2}^{(2)}{+}\frac12{-}\text{i}\nu=0,\\
    &\phantom{aaaaaaaaaaa}m_{1,2}^{(2)}{+}n_{1,2}^{(1)}{+}\frac12{+}\text{i}\nu=0,\ \ m_2{+}n_{1,2}^{(1)}{+}u_2{+}p_2{+}\text{i}\nu=0
\end{aligned}\right\}
\end{equation}
and the summand is
\begin{equation}
    \phi_{n,m,l,u,v}(BC)\frac{(-1)^{n_{1,2}^{(1)}}(2k)^{n_{1,2}^{(1)}{+}n_{1,2}^{(2)}}(w_1{+}k)^{m_1}(w_2{-}k)^{m_2}}{\Gamma(1{-}m_1{-}n_{1,2}^{(2)}{-}p_1{+}\text{i}\nu)}
\end{equation}
with $B:=\frac{\cosh{\pi \nu}}{\pi}$. Since the basis series has rank three, we should choose three indices as the independent ones and solve the others from bracket equations. Naively there are $\left(\substack{10\\3}\right)=120$ choices, but only $40$ of these choices lead to solutions of bracket equations above, as can be checked. Moreover, not all series correspond to physical regions. For instance, if consider  $\sigma=\{u_2,m_{1,2}^{(1)},m_{1,2}^{(2)}\}$ to be free indices, arguments in the power series will be $X^{m_{1,2}^{(1)}}Y^{m_{1,2}^{(2)}}(X/Y)^{u_2}$. Then when we sum over these three indices, physical condition $X>1$ leads to divergent result, so this series should be discarded in any physical region. Only $12$ combinations among all $40$ triples yield convergent series in region $R_1$ or $R_2$. For the region $R_1=\{0<Y<X\}$, we have four choices for the independent indices, which are
\begin{equation}\label{eq:R1indices}
    \{m_2,n_{1,2}^{(1)},n_{1,2}^{(2)}\}\ \ ,\ \ \{m_2,n_{1,2}^{(1)},l_{1,2}^{(2)}\}\ \ ,\ \ \{m_2,l_{1,2}^{(1)},n_{1,2}^{(2)}\}\ \ ,\ \ \{m_2,l_{1,2}^{(1)},l_{1,2}^{(2)}\}
\end{equation}
In each of the cases, as shown in \eqref{eq:twositefirst}, the arguments of the power series are always 
\begin{equation}
    \left(\frac{Y}{X}\right)^{e_1}\left(\frac{1}{X}\right)^{e_2{+}e_3}
\end{equation}
satisfying the convergent condition in $R_1$, with $\{e_1,e_2,e_3\}$ being any of the triple in \eqref{eq:R1indices}. So we take the sum of these four series as result for $R_1$-region, which is just the sum of four $\mathbf{T}_i$ in \eqref{eq:R1region} after two of the indices have been summed over.

On the other hand, in region $R_2=\{1<X<Y\}$, we have instead eight choices which can be divided into two groups as
\begin{equation}\label{eq:R2indices1}
    \{u_2,n_{1,2}^{(1)},n_{1,2}^{(2)}\}\ \ ,\ \ \{u_2,n_{1,2}^{(1)},l_{1,2}^{(2)}\}\ \ ,\ \ \{u_2,l_{1,2}^{(1)},n_{1,2}^{(2)}\}\ \ ,\ \ \{u_2,l_{1,2}^{(1)},l_{1,2}^{(2)}\}
\end{equation}
and
\begin{equation}\label{eq:R2indices2}
    \{m_1,n_{1,2}^{(1)},n_{1,2}^{(2)}\}\ \ ,\ \ \{m_1,n_{1,2}^{(1)},l_{1,2}^{(2)}\}\ \ ,\ \ \{m_1,l_{1,2}^{(1)},n_{1,2}^{(2)}\}\ \ ,\ \ \{m_1,l_{1,2}^{(1)},l_{1,2}^{(2)}\}
\end{equation}
For the first four choices, arguments of the series always read
\begin{equation}
    \left(\frac{X}{Y}\right)^{e_1}\left(\frac{1}{Y}\right)^{e_2}\left(\frac{1}{X}\right)^{e_3}
\end{equation}
For instance, for the first choice, the summand finally reads
\begin{align} 
(-1)^{n_{1,2}^{(1)}}\phi_{n_{1,2}^{(1)},n_{1,2}^{(2)},u_2} &(BC)\left(\frac{1}{X}\right)^{n_{1,2}^{(2)}-p_1{+}\text{i}\nu}\left(\frac{1}{Y}\right)^{n_{1,2}^{(1)}-p_2{-}\text{i}\nu}\left(\frac{X}{Y}\right)^{u_2}f(n_{1,2}^{(1)})g(n_{1,2}^{(2)})\nonumber\\
&\times(2\text{i}k)^{-\tilde{p}_1}\Gamma(n_{1,2}^{(1)}{+}p_2{+}u_2{+}\text{i}\nu)\Gamma(n_{1,2}^{(2)}{+}p_1{+}u_2{-}\text{i}\nu)\frac{\sin(\pi u_2)}{\pi u_2}
\end{align}
with shorthand notations \eqref{eq:fg}. We see that due to the factor $\frac{\sin (u_2\pi)}{u_2\pi}$ the summand localizes to $u_2=0$, this in turn leads the series to factorize, and the rest two indices can be summed over separately, leading to the building block $\mathbf{S}_1$. Similar things also happen for other three triples in \eqref{eq:R2indices1}, and summation of $u_2$ is always trivial in each case, yielding $\mathbf{S}_2,\mathbf{S}_3,\mathbf{S}_4$. The latter four choices, on the other hand, always result in arguments
\begin{equation}
    \left(\frac{X}{Y}\right)^{e_1}\left(\frac{1}{Y}\right)^{e_2{+}e_3}
\end{equation}
in the series, and give us four series $\mathbf{S}_5,\cdots, \mathbf{S}_8$. Finally, result of the integral in $R_2$ turns out to be a sum of eight series.

\paragraph{Generalization to a general tree topology with both black and gray edges}
Now we discuss the derivation for more general cases. 

Firstly, for a chain with both black and gray edges, the only difference in the calculation is in the deformation relation \eqref{eq:deformed}. To be more precise, we still introduce $\{s_{i,i{+}1}^{(s)}\}$ for each edge in the chain from $\text{H}_{\pm \text{i}\nu}^{(s)}$. A change of black edge to gray edge only leads to a change of the deformation in energies \eqref{eq:defenergy} and twist \eqref{eq:deftwist} as
\begin{equation}
    \hat{\omega}_i=w_i\left\{
    \begin{aligned}&{-}k_{i{-}1,i}(1+2s_{i{-}1,i}^{(1)}){+}k_{i,i{+}1}(1+2s_{i,i{+}1}^{(2)})\ \ (i{-}1,i)\text{ is black}\ \&\ (i,i{+}1)\text{ is black}\\
    &{+}k_{i{-}1,i}(1+2s_{i{-}1,i}^{(2)}){+}k_{i,i{+}1}(1+2s_{i,i{+}1}^{(2)})\ \ (i{-}1,i)\text{ is gray}\ \&\ (i,i{+}1)\text{ is black}\\
    &{-}k_{i{-}1,i}(1+2s_{i{-}1,i}^{(1)}){-}k_{i,i{+}1}(1+2s_{i,i{+}1}^{(1)})\ \ (i{-}1,i)\text{ is black}\ \&\ (i,i{+}1)\text{ is gray}\\
    &{+}k_{i{-}1,i}(1+2s_{i{-}1,i}^{(2)}){-}k_{i,i{+}1}(1+2s_{i,i{+}1}^{(1)})\ \ (i{-}1,i)\text{ is gray}\ \&\ (i,i{+}1)\text{ is gray}\end{aligned}
    \right . 
\end{equation}
and
\begin{equation}
    \hat{q}_i=p_i\left\{
    \begin{aligned}&{+}\text{i}\nu{-}\text{i}\nu\ \ \ \ \ (i{-}1,i)\text{ is black}\ \&\ (i,i{+}1)\text{ is black}\\
    &{-}\text{i}\nu{-}\text{i}\nu\ \ \ \ \ (i{-}1,i)\text{ is gray}\ \&\ (i,i{+}1)\text{ is black}\\
    &{+}\text{i}\nu{+}\text{i}\nu\ \ \ \ \ (i{-}1,i)\text{ is black}\ \&\ (i,i{+}1)\text{ is gray}\\
    &{-}\text{i}\nu{+}\text{i}\nu\ \ \ \ \ (i{-}1,i)\text{ is gray}\ \&\ (i,i{+}1)\text{ is gray}\end{aligned}
    \right . 
\end{equation}
All the procedures after \eqref{eq:deformed} and relation \eqref{eq:defenergy} and \eqref{eq:deftwist} will be exactly the same.

Secondly, for general tree topologies $\mathcal{E}$, the generalization is also straightforward. We will then introduce two folds of integrations $s_{i,j}^{(1)}$ and $s_{i,j}^{(2)}$ for each of the edge $(i,j)$ and the Hankel functions, which may be either black or gray. A massive family tree integral will then be able to expressed by integration over massless tree as well like \eqref{eq:deformed}, and deformation of the energies and twist follows similar rule we talk about for chain. In another word, a Hankel function $\text{H}^{(1)}_{\text{i}\nu}(-k\tau_i)$ attached at node $i$ and carrying energy $k$ will deform the energy $\omega_i$ by ${-}k(1+2s_{e}^{(1)})$ and the twist by ${+}i{\nu}$, while function $\text{H}^{(2)}_{-\text{i}\nu}(-k\tau_i)$ deform the energy $\omega_i$ by ${+}k(1+2s_{e}^{(2)})$ and the twist by ${-}i{\nu}$. Finally, every summation $\sum_{i=k}^{N}$ in the calculation of chain integral will be naturally generalized to a sum of all descendants for node $k$, like what we have seen in massless family trees. As a result, we summarize the most general series solutions for basic massive family tree integral in section \ref{sec:massive solution}.

\section{Necessary building blocks for  massive cosmological amplitudes}
In this Appendix, we present basic building blocks that are necessary for massive cosmological correlators with three nodes. Besides, we will give two illustrating examples for basic massive family trees with four nodes. The full results for any four-site cosmological amplitudes become too lengthy to be presented here, but it is of no essential difficulties to calculate their full results from MoB and massive family tree basis. For all examples in this section, we only give their primary solutions, since the descendants can be obtained by shifting indices following \eqref{eq:T111}.
\label{app:results}

The full result of three-site chain correlator can be presented as
\begin{equation}  
\begin{aligned}
&\mathcal{T}_3(w_1,w_2,w_3,k_{12},k_{23}){=}\frac{\pi^2}{16}{\rm e}^{-2\pi\nu}\times\\
&\left(\raisebox{-1em}
{\begin{tikzpicture}
				\coordinate (X1) at (0,0);
				\coordinate (X2) at (1,0);
				\coordinate (X3) at (2,0);;
				\node[below] at (X1) {\small{$1$}};
				\node[below] at (X2) {\small{$2$}};
				\node[below] at (X3) {\small{$3$}};
				\draw[line width=1.5pt,->] (X1)--(0.6,0);
				\draw[line width=1.5pt] (0.5,0)--(X2);
				\draw[line width=1.5pt,->] (X2)--(1.6,0);
				\draw[line width=1.5pt] (1.5,0)--(X3);
				\path[fill=black] (X1) circle[radius=0.1];
				\path[fill=black] (X2) circle[radius=0.1];
				\path[fill=black] (X3) circle[radius=0.1];
		\end{tikzpicture}}+\raisebox{-1em}{\begin{tikzpicture}
		\coordinate (X1) at (0,0);
		\coordinate (X2) at (1,0);
		\coordinate (X3) at (2,0);;
		\node[below] at (X1) {\small{$1$}};
		\node[below] at (X2) {\small{$2$}};
		\node[below] at (X3) {\small{$3$}};
		\draw[line width=1.5pt,->] (X1)--(0.6,0);
		\draw[line width=1.5pt] (0.5,0)--(X2);
		\draw[line width=1.5pt,->] (X3)--(1.4,0);
		\draw[line width=1.5pt] (1.5,0)--(X2);
		\path[fill=black] (X1) circle[radius=0.1];
		\path[fill=black] (X2) circle[radius=0.1];
		\path[fill=black] (X3) circle[radius=0.1];
	\end{tikzpicture}}+\raisebox{-1em}{\begin{tikzpicture}
		\coordinate (X1) at (0,0);
		\coordinate (X2) at (1,0);
		\coordinate (X3) at (2,0);;
		\node[below] at (X1) {\small{$1$}};
		\node[below] at (X2) {\small{$2$}};
		\node[below] at (X3) {\small{$3$}};
\draw[line width=1.5pt,->] (X2)--(0.4,0);
\draw[line width=1.5pt] (0.5,0)--(X1);
		\draw[line width=1.5pt,->] (X2)--(1.6,0);
		\draw[line width=1.5pt] (1.5,0)--(X3);
		\path[fill=black] (X1) circle[radius=0.1];
		\path[fill=black] (X2) circle[radius=0.1];
		\path[fill=black] (X3) circle[radius=0.1];
\end{tikzpicture}} +
\raisebox{-1em}{\begin{tikzpicture}
		\coordinate (X1) at (0,0);
		\coordinate (X2) at (1,0);
		\coordinate (X3) at (2,0);;
		\node[below] at (X1) {\small{$1$}};
		\node[below] at (X2) {\small{$2$}};
		\node[below] at (X3) {\small{$3$}};
\draw[line width=1.5pt,->] (X2)--(0.4,0);
\draw[line width=1.5pt] (0.5,0)--(X1);
		\draw[line width=1.5pt,->] (X3)--(1.4,0);
		\draw[line width=1.5pt] (1.5,0)--(X2);
		\path[fill=black] (X1) circle[radius=0.1];
		\path[fill=black] (X2) circle[radius=0.1];
		\path[fill=black] (X3) circle[radius=0.1];
\end{tikzpicture}}\right.\\
&\left.{+}\raisebox{-1em}{\begin{tikzpicture}[scale=0.75]
		\coordinate (X1) at (0,0);
		\coordinate (X2) at (1.5,0);
		\node[below] at (X1) {\small{$1$}};
		\node[below] at (X2) {\small{$2_{2,k_{23}}$}};
		\draw[line width=1.5pt,->] (X1)--(0.9,0);
        \draw[line width=1.5pt] (0.75,0)--(X2);
		\path[fill=black] (X1) circle[radius=0.11];
		\path[fill=black] (X2) circle[radius=0.11];
\end{tikzpicture}}\times \mathcal{F}^{(1)}(p_3,{-}w_3,k_{23},\nu){+}\raisebox{-1em}{\begin{tikzpicture}[scale=0.75]
		\coordinate (X1) at (0,0);
		\coordinate (X2) at (1.5,0);
		\node[below] at (X1) {\small{$2_{2,k_{23}}$}};
		\node[below] at (X2) {\small{$1$}};
		\draw[line width=1.5pt,->] (X1)--(0.9,0);
        \draw[line width=1.5pt] (0.75,0)--(X2);
		\path[fill=black] (X1) circle[radius=0.11];
		\path[fill=black] (X2) circle[radius=0.11];
\end{tikzpicture}}\times \mathcal{F}^{(1)}(p_3,{-}w_3,k_{23},\nu)\right.\\
&\left.{-}\raisebox{-1em}{\begin{tikzpicture}[scale=0.75]
		\coordinate (X1) at (0,0);
		\coordinate (X2) at (1.5,0);
		\node[below] at (X1) {\small{$3$}};
		\node[below] at (X2) {\small{$2_{2,k_{12}}$}};
		\draw[line width=1.5pt,->] (X1)--(0.9,0);
        \draw[line width=1.5pt] (0.75,0)--(X2);
		\path[fill=black] (X1) circle[radius=0.11];
		\path[fill=black] (X2) circle[radius=0.11];
\end{tikzpicture}}\times \mathcal{F}^{(1)}(p_1,{-}w_1,k_{12},\nu){-}\raisebox{-1em}{\begin{tikzpicture}[scale=0.75]
		\coordinate (X1) at (0,0);
		\coordinate (X2) at (1.5,0);
		\node[below] at (X1) {\small{$2_{2,k_{12}}$}};
		\node[below] at (X2) {\small{$3$}};
		\draw[line width=1.5pt,->] (X1)--(0.9,0);
        \draw[line width=1.5pt] (0.75,0)--(X2);
		\path[fill=black] (X1) circle[radius=0.11];
		\path[fill=black] (X2) circle[radius=0.11];
\end{tikzpicture}}\times \mathcal{F}^{(1)}(p_1,{-}w_1,k_{12},\nu)\right.\\
&\phantom{aaaaaaaaaaa}\left.{-}\mathcal{F}^{(1)}(p_1,-w_1,k_{12},\nu)\mathcal{F}^{(1)}(p_3,-w_3,k_{23},\nu)\mathcal{F}^{(2,2)}(p_2,w_2,k_{12},k_{23},\nu)\right){+} \text{c.c.}
\end{aligned}
\end{equation}
If we focus on the physical region we discussed for general $n$-site massive tree, i.e. $w_1>>w_2,w_3,k_{1,2},k_{2,3}$, then its result can be  rewritten as
\begin{equation}\label{eq:fullthreesite}  
\begin{aligned}
&\mathcal{T}_3(w_1,w_2,w_3,k_{12},k_{23}){=}\frac{\pi^2}{16}{\rm e}^{-2\pi\nu}\times\\
&\left(\raisebox{-1em}
{\begin{tikzpicture}
				\coordinate (X1) at (0,0);
				\coordinate (X2) at (1,0);
				\coordinate (X3) at (2,0);;
				\node[below] at (X1) {\small{$1$}};
				\node[below] at (X2) {\small{$2$}};
				\node[below] at (X3) {\small{$3$}};
				\draw[line width=1.5pt,->] (X1)--(0.6,0);
				\draw[line width=1.5pt] (0.5,0)--(X2);
				\draw[line width=1.5pt,->] (X2)--(1.6,0);
				\draw[line width=1.5pt] (1.5,0)--(X3);
				\path[fill=black] (X1) circle[radius=0.1];
				\path[fill=black] (X2) circle[radius=0.1];
				\path[fill=black] (X3) circle[radius=0.1];
		\end{tikzpicture}}-\raisebox{-1em}{\begin{tikzpicture}
				\coordinate (X1) at (0,0);
				\coordinate (X2) at (1,0);
				\coordinate (X3) at (2,0);;
				\node[below] at (X1) {\small{$1$}};
				\node[below] at (X2) {\small{$2$}};
				\node[below] at (X3) {\small{$3$}};
				\draw[line width=1.5pt,->] (X1)--(0.6,0);
				\draw[line width=1.5pt] (0.5,0)--(X2);
				\draw[line width=1.5pt,->,gray] (X2)--(1.6,0);
				\draw[line width=1.5pt,gray] (1.5,0)--(X3);
				\path[fill=black] (X1) circle[radius=0.1];
				\path[fill=black] (X2) circle[radius=0.1];
				\path[fill=black] (X3) circle[radius=0.1];
		\end{tikzpicture}}-\raisebox{-1em}{\begin{tikzpicture}
				\coordinate (X1) at (0,0);
				\coordinate (X2) at (1,0);
				\coordinate (X3) at (2,0);;
				\node[below] at (X1) {\small{$1$}};
				\node[below] at (X2) {\small{$2$}};
				\node[below] at (X3) {\small{$3$}};
				\draw[line width=1.5pt,->,gray] (X1)--(0.6,0);
				\draw[line width=1.5pt,gray] (0.5,0)--(X2);
				\draw[line width=1.5pt,->] (X2)--(1.6,0);
				\draw[line width=1.5pt] (1.5,0)--(X3);
				\path[fill=black] (X1) circle[radius=0.1];
				\path[fill=black] (X2) circle[radius=0.1];
				\path[fill=black] (X3) circle[radius=0.1];
		\end{tikzpicture}} +
\raisebox{-1em}{\begin{tikzpicture}
				\coordinate (X1) at (0,0);
				\coordinate (X2) at (1,0);
				\coordinate (X3) at (2,0);;
				\node[below] at (X1) {\small{$1$}};
				\node[below] at (X2) {\small{$2$}};
				\node[below] at (X3) {\small{$3$}};
				\draw[line width=1.5pt,->,gray] (X1)--(0.6,0);
				\draw[line width=1.5pt,gray] (0.5,0)--(X2);
				\draw[line width=1.5pt,->,gray] (X2)--(1.6,0);
				\draw[line width=1.5pt,gray] (1.5,0)--(X3);
				\path[fill=black] (X1) circle[radius=0.1];
				\path[fill=black] (X2) circle[radius=0.1];
				\path[fill=black] (X3) circle[radius=0.1];
		\end{tikzpicture}}\right.\\
&\left.{-}\raisebox{-1em}{\begin{tikzpicture}[scale=0.75]
		\coordinate (X1) at (0,0);
		\coordinate (X2) at (1.5,0);
		\node[below] at (X1) {\small{$1$}};
		\node[below] at (X2) {\small{$2_{1,k_{23}}$}};
		\draw[line width=1.5pt,->,gray] (X1)--(0.9,0);
        \draw[line width=1.5pt,gray] (0.75,0)--(X2);
		\path[fill=black] (X1) circle[radius=0.11];
		\path[fill=black] (X2) circle[radius=0.11];
\end{tikzpicture}}\times \mathcal{F}^{(2)}(p_3,w_3,k_{23},\nu){-}\raisebox{-1em}{\begin{tikzpicture}[scale=0.75]
		\coordinate (X1) at (0,0);
		\coordinate (X2) at (1.5,0);
		\node[below] at (X1) {\small{$2_{2,k_{12}}$}};
		\node[below] at (X2) {\small{$3$}};
		\draw[line width=1.5pt,->,gray] (X1)--(0.9,0);
        \draw[line width=1.5pt,gray] (0.75,0)--(X2);
		\path[fill=black] (X1) circle[radius=0.11];
		\path[fill=black] (X2) circle[radius=0.11];
\end{tikzpicture}}\times \mathcal{F}^{(1)}(p_1,w_1,k_{12},\nu)\right.\\  
&\left.{+}\raisebox{-1em}{\begin{tikzpicture}[scale=0.75]
		\coordinate (X1) at (0,0);
		\coordinate (X2) at (1.5,0);
		\node[below] at (X1) {\small{$1$}};
		\node[below] at (X2) {\small{$2_{1,k_{23}}$}};
		\draw[line width=1.5pt,->] (X1)--(0.9,0);
        \draw[line width=1.5pt] (0.75,0)--(X2);
		\path[fill=black] (X1) circle[radius=0.11];
		\path[fill=black] (X2) circle[radius=0.11];
\end{tikzpicture}}\times \mathcal{F}^{(2)}(p_3,w_3,k_{23},\nu){+}\raisebox{-1em}{\begin{tikzpicture}[scale=0.75]
		\coordinate (X1) at (0,0);
		\coordinate (X2) at (1.5,0);
		\node[below] at (X1) {\small{$2_{2,k_{12}}$}};
		\node[below] at (X2) {\small{$3$}};
		\draw[line width=1.5pt,->] (X1)--(0.9,0);
        \draw[line width=1.5pt] (0.75,0)--(X2);
		\path[fill=black] (X1) circle[radius=0.11];
		\path[fill=black] (X2) circle[radius=0.11];
\end{tikzpicture}}\times \mathcal{F}^{(1)}(p_1,w_1,k_{12},\nu)\right.\\  
&\left.{+}\raisebox{-1em}{\begin{tikzpicture}[scale=0.75]
		\coordinate (X1) at (0,0);
		\coordinate (X2) at (1.5,0);
		\node[below] at (X1) {\small{$1$}};
		\node[below] at (X2) {\small{$2_{2,k_{23}}$}};
		\draw[line width=1.5pt,->] (X1)--(0.9,0);
        \draw[line width=1.5pt] (0.75,0)--(X2);
		\path[fill=black] (X1) circle[radius=0.11];
		\path[fill=black] (X2) circle[radius=0.11];
\end{tikzpicture}}\times \mathcal{F}^{(1)}(p_3,{-}w_3,k_{23},\nu){-}\raisebox{-1em}{\begin{tikzpicture}[scale=0.75]
		\coordinate (X1) at (0,0);
		\coordinate (X2) at (1.5,0);
		\node[below] at (X1) {\small{$1$}};
		\node[below] at (X2) {\small{$2_{2,k_{23}}$}};
		\draw[line width=1.5pt,->,gray] (X1)--(0.9,0);
        \draw[line width=1.5pt,gray] (0.75,0)--(X2);
		\path[fill=black] (X1) circle[radius=0.11];
		\path[fill=black] (X2) circle[radius=0.11];
\end{tikzpicture}}\times \mathcal{F}^{(1)}(p_3,{-}w_3,k_{23},\nu)\right.\\
&\left.{+}\raisebox{-1em}{\begin{tikzpicture}[scale=0.75]
		\coordinate (X1) at (0,0);
		\coordinate (X2) at (1.5,0);
		\node[below] at (X1) {\small{$2_{2,k_{12}}$}};
		\node[below] at (X2) {\small{$3$}};
		\draw[line width=1.5pt,->,gray] (X1)--(0.9,0);
        \draw[line width=1.5pt,gray] (0.75,0)--(X2);
		\path[fill=black] (X1) circle[radius=0.11];
		\path[fill=black] (X2) circle[radius=0.11];
\end{tikzpicture}}\times \mathcal{F}^{(1)}(p_1,{-}w_1,k_{12},\nu){-}\raisebox{-1em}{\begin{tikzpicture}[scale=0.75]
		\coordinate (X1) at (0,0);
		\coordinate (X2) at (1.5,0);
		\node[below] at (X1) {\small{$2_{2,k_{12}}$}};
		\node[below] at (X2) {\small{$3$}};
		\draw[line width=1.5pt,->] (X1)--(0.9,0);
        \draw[line width=1.5pt] (0.75,0)--(X2);
		\path[fill=black] (X1) circle[radius=0.11];
		\path[fill=black] (X2) circle[radius=0.11];
\end{tikzpicture}}\times \mathcal{F}^{(1)}(p_1,{-}w_1,k_{12},\nu)\right.\\
&\left. {+}\left(\mathcal{F}^{(1)}(p_1,w_1,k_{12},\nu){-}\mathcal{F}^{(1)}(p_1,-w_1,k_{12},\nu)\right)\times\right.\\
&\left.\left(\mathcal{F}^{(2,2)}(p_2,w_2,k_{12},k_{23},\nu)\mathcal{F}^{(1)}(p_3,-w_3,k_{2,3},\nu){-}\mathcal{F}^{(2,1)}(p_2,w_2,k_{12},k_{23},\nu)\mathcal{F}^{(2)}(p_3,w_3,k_{2,3},\nu)\right)\right)\\
&\phantom{aaaaaaaaaaaaaaaaaaaaa}+\text{c.c.}
\end{aligned}
\end{equation}
Solutions for these basis, either in MoB series or explicit provided by hypergeometric functions, can be found throughout the main text and as the following

\paragraph{Contact function}
In \eqref{eq:fullthreesite}, there is one more type of contact function needed, whose definition is
\begin{equation}
    \mathcal{F}^{(2,1)}(p,w,k_1,k_2,\nu):=(-\text{i})\int_{-\infty}^0{\rm d}\tau\ (-\tau)^{p{-}1}\text{e}^{\text{i}w\tau}\ \text{H}_{-\text{i}\nu}^{(2)}(-k_1\tau)\text{H}_{\text{i}\nu}^{(1)}(-k_2\tau)
\end{equation}
Based on the rule discussed in \ref{sec:generl massive trees}, we can present its result depending on the simplest contact function \eqref{eq:massivecontact3}. It reads a sum over four series,
\begin{equation}
    \mathcal{F}^{(2,1)}(p,w,k_1,k_2,\nu)=(-\text{i})^{1{+}p} C\left(\frac{k_2}{k_1}\right)^{\text{i}\nu}\sum_{e_{1}^{(2)},e_{2}^{(1)}\in\{0,1\}}\sum_{n_{1}^{(2)}n_{2}^{(1)}}^{\infty}\mathbf{T}_{e_{1}^{(2)},e_{2}^{(1)}}
\end{equation}
and the primary series should be
\begin{equation}
    \mathbf{T}_{0,0}=\frac{\phi_{n_{1}^{(2)},n_{2}^{(1)}}\Gamma(n_{1}^{(2)}{+}n_{2}^{(1)}{+}p)}{\Gamma(\text{i}\nu{+}\frac12)\Gamma({-}\text{i}\nu{+}\frac12)} g(n_1^{(2)})f(n_2^{(1)})(2k_1)^{n_1^{(2)}}(2k_2)^{n_2^{(1)}}(w{+}k_1{-}k_2)^{{-}n_{1}^{(2)}{-}n_{2}^{(1)}{-}p}
\end{equation}

\paragraph{Two-site trees} In \eqref{eq:twositefull} and \eqref{eq:fullthreesite}, there are several basic and general massive family tree basis are involved, besides the $\mathcal{I}_2$ we discuss in the section \ref{sec:two site massive}. Here we present explicit results for their primary solutions. 

Firstly, we have one more piece of two-site basic massive family tree, which can be expressed by a sum over four series, and its primary series reads
\begin{equation}\label{eq:graytwosite}
    \raisebox{-1em}{\begin{tikzpicture}[scale=0.75]
		\coordinate (X1) at (0,0);
		\coordinate (X2) at (1.5,0);
		\node[below] at (X1) {\small{$1$}};
		\node[below] at (X2) {\small{$2$}};
		\draw[line width=1.5pt,->,gray] (X1)--(0.9,0);
        \draw[line width=1.5pt,gray] (0.75,0)--(X2);
		\path[fill=black] (X1) circle[radius=0.11];
		\path[fill=black] (X2) circle[radius=0.11];
\end{tikzpicture}}: \mathbf{T}_{0,0}=\frac{\phi_{m_2,n_{1,2}^{(1)},n_{1,2}^{(2)}}\Gamma(m_2{+}n_{1,2}^{(1)}{+}n_{1,2}^{(2)}+\tilde{p}_1)}{(m_2{+}n_{1,2}^{(2)}{+}p_2{-}\text{i}\nu)}\mathcal{E}_{1,2}\mathcal{V}_1\mathcal{V}_2
\end{equation}
with the factors $\mathcal{E}_{i,j}$ from \eqref{eq:E} and $\mathcal{V}_j$ as
\begin{equation}
    \mathcal{V}_1{=}(w_1{-}k_{1,2})^{-m_2{-}n_{1,2}^{(1)}{-}n_{1,2}^{(2)}-\tilde{p}_1}, \ \mathcal{V}_2{=}(w_2{+}k_{1,2})^{m_2}
\end{equation}
Besides, in \eqref{eq:fullthreesite} there are $5$ more pieces of general massive family trees. Each of them is a sum over $8$ series, and their primary series are recorded here as
\begin{equation}
    \raisebox{-1em}{\begin{tikzpicture}[scale=0.75]
		\coordinate (X1) at (0,0);
		\coordinate (X2) at (1.5,0);
		\node[below] at (X1) {\small{$1$}};
		\node[below] at (X2) {\small{$2_{2, k_{2,3}}$}};
		\draw[line width=1.5pt,->,gray] (X1)--(0.9,0);
        \draw[line width=1.5pt,gray] (0.75,0)--(X2);
		\path[fill=black] (X1) circle[radius=0.11];
		\path[fill=black] (X2) circle[radius=0.11];
\end{tikzpicture}}:\ \mathbf{T}_{0,0,0}{=}\frac{\phi_{n,m}(2k_{2,3})^{n_{2,3}^{(2)}}\Gamma(m_2{+}n_{1,2}^{(1)}{+}n_{1,2}^{(2)}{+}n_{2,3}^{(2)}{+}\tilde{p}_1{-}\text{i}\nu) }{(m_2{+}n_{1,2}^{(2)}{+}n_{2,3}^{(2)}{+}p_2{-}2\text{i}\nu)\Gamma(\text{i}\nu{+}\frac12)}g(n_{2,3}^{(2)}) \mathcal{E}_{1,2}\mathcal{V}_1{\mathcal{V}}_2
\end{equation}
with the factors $\mathcal{E}_{i,j}$ from \eqref{eq:E} and $\mathcal{V}_j$ as
\begin{equation}
    \mathcal{V}_1=(w_1{-}k_{1,2})^{-m_2-n_{1,2}^{(1)}{-}n_{1,2}^{(2)}{-}n_{2,3}^{(2)}{-}\tilde{p}_1{+}\text{i}\nu},\ \mathcal{V}_2=(w_2{+}k_{1,2}{+}k_{2,3})^{m_2}
\end{equation}
Similarly, for other four extra general two-site massive trees, we have the expressions
\begin{equation}
    \raisebox{-1em}{\begin{tikzpicture}[scale=0.75]
		\coordinate (X1) at (0,0);
		\coordinate (X2) at (1.5,0);
		\node[below] at (X1) {\small{$1$}};
		\node[below] at (X2) {\small{$2_{1,k_{2,3}}$}};
		\draw[line width=1.5pt,->] (X1)--(0.9,0);
        \draw[line width=1.5pt] (0.75,0)--(X2);
		\path[fill=black] (X1) circle[radius=0.11];
		\path[fill=black] (X2) circle[radius=0.11];
\end{tikzpicture}}:\ \mathbf{T}_{0,0,0}{=}\frac{\phi_{n,m}(2k_{2,3})^{n_{2,3}^{(1)}}\Gamma(m_2{+}n_{1,2}^{(1)}{+}n_{1,2}^{(2)}{+}n_{2,3}^{(1)}{+}\tilde{p}_1{+}\text{i}\nu) }{(m_2{+}n_{1,2}^{(1)}{+}n_{2,3}^{(1)}{+}p_2{+}2\text{i}\nu)\Gamma({-}\text{i}\nu{+}\frac12)}f(n_{2,3}^{(1)}) \mathcal{E}_{1,2}\mathcal{V}_1{\mathcal{V}}_2
\end{equation}
with the factors $\mathcal{E}_{i,j}$ from \eqref{eq:E} and $\mathcal{V}_j$ as
\begin{equation}
    \mathcal{V}_1=(w_1{+}k_{1,2})^{-m_2-n_{1,2}^{(1)}{-}n_{1,2}^{(2)}{-}n_{2,3}^{(1)}{-}\tilde{p}_1{-}\text{i}\nu},\ \mathcal{V}_2=(w_2{-}k_{1,2}{+}k_{2,3})^{m_2}
\end{equation}
\begin{equation}
    \raisebox{-1em}{\begin{tikzpicture}[scale=0.75]
		\coordinate (X1) at (0,0);
		\coordinate (X2) at (1.5,0);
		\node[below] at (X1) {\small{$1$}};
		\node[below] at (X2) {\small{$2_{1,k_{2,3}}$}};
		\draw[line width=1.5pt,->,gray] (X1)--(0.9,0);
        \draw[line width=1.5pt,gray] (0.75,0)--(X2);
		\path[fill=black] (X1) circle[radius=0.11];
		\path[fill=black] (X2) circle[radius=0.11];
\end{tikzpicture}}:\ \mathbf{T}_{0,0,0}{=}\frac{\phi_{n,m}(2k_{2,3})^{n_{2,3}^{(1)}}\Gamma(m_2{+}n_{1,2}^{(1)}{+}n_{1,2}^{(2)}{+}n_{2,3}^{(1)}{+}\tilde{p}_1{+}\text{i}\nu) }{(m_2{+}n_{1,2}^{(2)}{+}n_{2,3}^{(1)}{+}p_2)\Gamma({-}\text{i}\nu{+}\frac12)}f(n_{2,3}^{(1)}) \mathcal{E}_{1,2}\mathcal{V}_1{\mathcal{V}}_2
\end{equation}
with the factors $\mathcal{E}_{i,j}$ from \eqref{eq:E} and $\mathcal{V}_j$ as
\begin{equation}
    \mathcal{V}_1=(w_1{-}k_{1,2})^{-m_2-n_{1,2}^{(1)}{-}n_{1,2}^{(2)}{-}n_{2,3}^{(1)}{-}\tilde{p}_1{-}\text{i}\nu},\ \mathcal{V}_2=(w_2{+}k_{1,2}{+}k_{2,3})^{m_2}
\end{equation}
\begin{equation}
    \raisebox{-1em}{\begin{tikzpicture}[scale=0.75]
		\coordinate (X1) at (0,0);
		\coordinate (X2) at (1.5,0);
		\node[below] at (X1) {\small{$2_{2,k_{1,2}}$}};
		\node[below] at (X2) {\small{$3$}};
		\draw[line width=1.5pt,->] (X1)--(0.9,0);
        \draw[line width=1.5pt] (0.75,0)--(X2);
		\path[fill=black] (X1) circle[radius=0.11];
		\path[fill=black] (X2) circle[radius=0.11];
\end{tikzpicture}}:\ \mathbf{T}_{0,0,0}{=}\frac{\phi_{n,m}(2k_{1,2})^{n_{1,2}^{(2)}}\Gamma(m_2{+}n_{1,2}^{(2)}{+}n_{2,3}^{(1)}{+}n_{2,3}^{(2)}{+}\tilde{p}_2{-}\text{i}\nu)}{(m_2{+}n_{2,3}^{(1)}{+}p_3{+}\text{i}\nu)\Gamma(\text{i}\nu{+}\frac12) }g(n_{1,2}^{(2)}) \mathcal{E}_{2,3}\mathcal{V}_2{\mathcal{V}}_3
\end{equation}
with the factors $\mathcal{V}_j$ as
\begin{equation}
    \mathcal{V}_2=(w_2{+}k_{1,2}{+}k_{2,3})^{-m_3-n_{1,2}^{(2)}{-}n_{2,3}^{(1)}{-}n_{2,3}^{(2)}{-}\tilde{p}_2{+}\text{i}\nu},\ \mathcal{V}_3=(w_3{-}k_{2,3})^{m_3}
\end{equation}
and
\begin{equation}
    \raisebox{-1em}{\begin{tikzpicture}[scale=0.75]
		\coordinate (X1) at (0,0);
		\coordinate (X2) at (1.5,0);
		\node[below] at (X1) {\small{$2_{2,k_{1,2}}$}};
		\node[below] at (X2) {\small{$3$}};
		\draw[line width=1.5pt,->,gray] (X1)--(0.9,0);
        \draw[line width=1.5pt,gray] (0.75,0)--(X2);
		\path[fill=black] (X1) circle[radius=0.11];
		\path[fill=black] (X2) circle[radius=0.11];
\end{tikzpicture}}:\ \mathbf{T}_{0,0,0}{=}\frac{\phi_{n,m}(2k_{1,2})^{n_{1,2}^{(2)}}\Gamma(m_2{+}n_{1,2}^{(2)}{+}n_{2,3}^{(1)}{+}n_{2,3}^{(2)}{+}\tilde{p}_2{-}\text{i}\nu) }{(m_2{+}n_{2,3}^{(2)}{+}p_3{-}\text{i}\nu)\Gamma(\text{i}\nu{+}\frac12)}g(n_{1,2}^{(2)}) \mathcal{E}_{2,3}\mathcal{V}_2{\mathcal{V}}_3
\end{equation}
with the factors $\mathcal{V}_j$ as
\begin{equation}
    \mathcal{V}_2=(w_2{+}k_{1,2}{-}k_{2,3})^{-m_3-n_{1,2}^{(2)}{-}n_{2,3}^{(1)}{-}n_{2,3}^{(2)}{-}\tilde{p}_2{+}\text{i}\nu},\ \mathcal{V}_3=(w_3{+}k_{2,3})^{m_3}
\end{equation}
Note that for each of these five case, extra prefactors $Ak^{-\text{i}\nu}$ or $\bar{A}(-k)^{\text{i}\nu}$ are still needed when summing over the eight series.

\paragraph{Three-site trees}
Besides \eqref{eq:3sitemassivechain1}, in \eqref{eq:fullthreesite} there are three extra basic massive family trees at three sites are needed. Each of them can be expressed by a sum of $16$ series, and the primary solutions are recorded here as
\begin{align}
    \raisebox{-1em}{\begin{tikzpicture}
				\coordinate (X1) at (0,0);
				\coordinate (X2) at (1,0);
				\coordinate (X3) at (2,0);;
				\node[below] at (X1) {\small{$1$}};
				\node[below] at (X2) {\small{$2$}};
				\node[below] at (X3) {\small{$3$}};
				\draw[line width=1.5pt,->] (X1)--(0.6,0);
				\draw[line width=1.5pt] (0.5,0)--(X2);
				\draw[line width=1.5pt,->,gray] (X2)--(1.6,0);
				\draw[line width=1.5pt,gray] (1.5,0)--(X3);
				\path[fill=black] (X1) circle[radius=0.1];
				\path[fill=black] (X2) circle[radius=0.1];
				\path[fill=black] (X3) circle[radius=0.1];
		\end{tikzpicture}}:\  \mathbf{T}_{0,0,0,0}=\frac{\phi_{n_{i,i{+}1}^{(s)},m_i}\Gamma(\tilde{m}_2{+}\sum_{i=1}^2(n_{i,i{+}1}^{(1)}{+}n_{i,i{+}1}^{(2)}){+}\tilde{p}_1)}{(\tilde{m}_2{+}n_{1,2}^{(1)}{+}n_{2,3}^{(1)}{+}n_{2,3}^{(2)}{+}p_2{+}p_3{+}\text{i}\nu)(m_3{+}n_{2,3}^{(2)}{+}p_3{-}\text{i}\nu)}\mathcal{E}_{1,2}\mathcal{E}_{2,3}\mathcal{V}_1\mathcal{V}_2\mathcal{V}_3
\end{align}
with the factors $\mathcal{E}_{i,j}$ from \eqref{eq:E} and $\mathcal{V}_j$ as
\begin{equation}
    \mathcal{V}_1{=}(w_1{+}k_{1,2})^{-m_2{-}m_3{-}n_{1,2}^{(1)}{-}n_{1,2}^{(2)}{-}n_{2,3}^{(1)}{-}n_{2,3}^{(2)}-\tilde{p}_1}, \ \mathcal{V}_2{=}(w_2{-}k_{1,2}{-}k_{2,3})^{m_2},\ \mathcal{V}_3{=}(w_3{+}k_{2,3})^{m_3}
\end{equation}
\begin{align}
    \raisebox{-1em}{\begin{tikzpicture}
				\coordinate (X1) at (0,0);
				\coordinate (X2) at (1,0);
				\coordinate (X3) at (2,0);;
				\node[below] at (X1) {\small{$1$}};
				\node[below] at (X2) {\small{$2$}};
				\node[below] at (X3) {\small{$3$}};
				\draw[line width=1.5pt,->,gray] (X1)--(0.6,0);
				\draw[line width=1.5pt,gray] (0.5,0)--(X2);
				\draw[line width=1.5pt,->] (X2)--(1.6,0);
				\draw[line width=1.5pt] (1.5,0)--(X3);
				\path[fill=black] (X1) circle[radius=0.1];
				\path[fill=black] (X2) circle[radius=0.1];
				\path[fill=black] (X3) circle[radius=0.1];
		\end{tikzpicture}}:\ \mathbf{T}_{0,0,0,0}=\frac{\phi_{n_{i,i{+}1}^{(s)},m_i}\Gamma(\tilde{m}_2{+}\sum_{i=1}^2(n_{i,i{+}1}^{(1)}{+}n_{i,i{+}1}^{(2)}){+}\tilde{p}_1)}{(\tilde{m}_2{+}n_{1,2}^{(2)}{+}n_{2,3}^{(1)}{+}n_{2,3}^{(2)}{+}p_2{+}p_3{-}\text{i}\nu)(m_3{+}n_{2,3}^{(1)}{+}p_3{+}\text{i}\nu)}\mathcal{E}_{1,2}\mathcal{E}_{2,3}\mathcal{V}_1\mathcal{V}_2\mathcal{V}_3 
\end{align}
with the factors $\mathcal{V}_j$ as
\begin{equation}
    \mathcal{V}_1{=}(w_1{-}k_{1,2})^{-m_2{-}m_3{-}n_{1,2}^{(1)}{-}n_{1,2}^{(2)}{-}n_{2,3}^{(1)}{-}n_{2,3}^{(2)}-\tilde{p}_1}, \ \mathcal{V}_2{=}(w_2{+}k_{1,2}{+}k_{2,3})^{m_2},\ \mathcal{V}_3{=}(w_3{-}k_{2,3})^{m_3}
\end{equation}
and
\begin{align}
\raisebox{-1em}{\begin{tikzpicture}
				\coordinate (X1) at (0,0);
				\coordinate (X2) at (1,0);
				\coordinate (X3) at (2,0);;
				\node[below] at (X1) {\small{$1$}};
				\node[below] at (X2) {\small{$2$}};
				\node[below] at (X3) {\small{$3$}};
				\draw[line width=1.5pt,->,gray] (X1)--(0.6,0);
				\draw[line width=1.5pt,gray] (0.5,0)--(X2);
				\draw[line width=1.5pt,->,gray] (X2)--(1.6,0);
				\draw[line width=1.5pt,gray] (1.5,0)--(X3);
				\path[fill=black] (X1) circle[radius=0.1];
				\path[fill=black] (X2) circle[radius=0.1];
				\path[fill=black] (X3) circle[radius=0.1];
		\end{tikzpicture}}:\ \mathbf{T}_{0,0,0,0}=\frac{\phi_{n_{i,i{+}1}^{(s)},m_i}\Gamma(\tilde{m}_2{+}\sum_{i=1}^2(n_{i,i{+}1}^{(1)}{+}n_{i,i{+}1}^{(2)}){+}\tilde{p}_1)}{(\tilde{m}_2{+}n_{1,2}^{(2)}{+}n_{2,3}^{(1)}{+}n_{2,3}^{(2)}{+}p_2{+}p_3{-}\text{i}\nu)(m_3{+}n_{2,3}^{(2)}{+}p_3{-}\text{i}\nu)}\mathcal{E}_{1,2}\mathcal{E}_{2,3}\mathcal{V}_1\mathcal{V}_2\mathcal{V}_3 
\end{align}
with the factors $\mathcal{V}_j$ as
\begin{equation}
    \mathcal{V}_1{=}(w_1{-}k_{1,2})^{-m_2{-}m_3{-}n_{1,2}^{(1)}{-}n_{1,2}^{(2)}{-}n_{2,3}^{(1)}{-}n_{2,3}^{(2)}-\tilde{p}_1}, \ \mathcal{V}_2{=}(w_2{+}k_{1,2}{-}k_{2,3})^{m_2},\ \mathcal{V}_3{=}(w_3{+}k_{2,3})^{m_3}
\end{equation}

\paragraph{Four-site trees}
Finally, we present two more examples at four sites to illustrate our solution furthermore, which are
\begin{align}
    \mathcal{I}_4^{(a)}&=\raisebox{-1em}{\begin{tikzpicture}[scale=0.75]
		\coordinate (X1) at (0,0);
		\coordinate (X2) at (1.5,0);
        \coordinate (X3) at (3,0);
            \coordinate (X4) at (4.5,0);
		\node[below] at (X1) {\small{$1$}};
		\node[below] at (X2) {\small{$2$}};
        \node[below] at (X4) {\small{$4$}};
        \node[below] at (X3) {\small{$3$}};
		\draw[line width=1.5pt,->] (X1)--(0.9,0);
        \draw[line width=1.5pt] (0.75,0)--(X2);
        \draw[line width=1.5pt,gray,->] (X2)--(2.4,0);
        \draw[line width=1.5pt,gray] (2.25,0)--(X3);
        \draw[line width=1.5pt,->] (X3)--(3.9,0);
        \draw[line width=1.5pt] (3.75,0)--(X4);
		\path[fill=black] (X1) circle[radius=0.11];
		\path[fill=black] (X2) circle[radius=0.11];
        \path[fill=black] (X3) circle[radius=0.11];
        \path[fill=black] (X4) circle[radius=0.11];
\end{tikzpicture}}\nonumber\\
&=(-\text{i})^4\int_{-\infty}^0\prod_{i=1}^4\left({\rm d}\tau_i(-\tau_i)^{p_i{-}1}{\rm e}^{\text{i}w_i\tau_i}\right){\text H}_{-\text{i}\nu}^{(2)}(-k_{1,2}\tau_1){\text H}_{\text{i}\nu}^{(1)}(-k_{1,2}\tau_2)\theta_{2,1}\\
&\phantom{aaaaaaaaa}\times{\text H}_{\text{i}\nu}^{(1)}(-k_{2,3}\tau_2){\text H}_{-\text{i}\nu}^{(2)}(-k_{2,3}\tau_3)\theta_{3,2}\times{\text H}_{-\text{i}\nu}^{(2)}(-k_{3,4}\tau_3){\text H}_{\text{i}\nu}^{(1)}(-k_{3,4}\tau_4)\theta_{4,3}\nonumber
\end{align}
and
\begin{align}
\mathcal{I}_4^{(b)}&=\raisebox{-25pt}{\begin{tikzpicture}[scale=0.6]
		\coordinate (X4) at (0,0);
		\coordinate (X1) at (0,1.5);
        \coordinate (X2) at (1.3,-0.75);
        \coordinate (X3) at (-1.3,-0.75);
		\node[above] at (X1) {\small{$1$}};
		\node[below] at (X2) {\small{$3$}};
        \node[below] at (X3) {\small{$4$}};
        \node[below] at (X4) {\small{$2$}};
		\draw[line width=1.5pt,gray,->] (X1)--(0,0.5);
        \draw[line width=1.5pt,gray] (0,1)--(X4);
        \draw[line width=1.5pt,->] (X4)--(0.87,-0.5);
        \draw[line width=1.5pt] (0.65,-0.375)--(X2);
        \draw[line width=1.5pt,->] (X4)--(-0.87,-0.5);
        \draw[line width=1.5pt] (-0.65,-0.375)--(X3);
		\path[fill=black] (X1) circle[radius=0.11];
		\path[fill=black] (X2) circle[radius=0.11];
        \path[fill=black] (X3) circle[radius=0.11];
        \path[fill=black] (X4) circle[radius=0.11];
\end{tikzpicture}}\nonumber\\
&=(-\text{i})^4\int_{-\infty}^0\prod_{i=1}^4\left({\rm d}\tau_i(-\tau_i)^{p_i{-}1}{\rm e}^{\text{i}w_i\tau_i}\right){\text H}_{\text{i}\nu}^{(1)}(-k_{1,2}\tau_1){\text H}_{-\text{i}\nu}^{(2)}(-k_{1,2}\tau_2)\theta_{2,1}\\
&\phantom{aaaaaaaaa}\times{\text H}_{-\text{i}\nu}^{(2)}(-k_{2,3}\tau_2){\text H}_{\text{i}\nu}^{(1)}(-k_{2,3}\tau_3)\theta_{3,2}\times{\text H}_{-\text{i}\nu}^{(2)}(-k_{2,4}\tau_2){\text H}_{\text{i}\nu}^{(1)}(-k_{2,4}\tau_4)\theta_{4,2}\nonumber
\end{align}
Each of them reads a sum over $64$ series with $9$ independent indices. For the four-site chain graph with both black and gray edges, its primary solution $\mathbf{T}_{0,0,0,0,0,0}^{(a)}$ reads 
\begin{equation}
\begin{aligned}    \frac{\phi_{n_{i,i{+}1}^{(s)},m_i}\Gamma(\tilde{m}_2{+}\sum_{i=1}^3(n_{i,i{+}1}^{(1)}{+}n_{i,i{+}1}^{(1)}){+}\tilde{p}_1)\mathcal{E}_{1,2}\mathcal{E}_{2,3}\mathcal{E}_{3,4}\mathcal{V}_1 \mathcal{V}_2 \mathcal{V}_3 \mathcal{V}_4}{(\tilde{m}_2{+}n_{1,2}^{(1)}{+}\sum_{i=2}^3(n_{i,i{+}1}^{(1)}{+}n_{i,i{+}1}^{(2)}){+}\tilde{p}_2{+}\text{i}\nu)(\tilde{m}_3{+}n_{2,3}^{(2)}{+}n_{3,4}^{(1)}{+}n_{3,4}^{(2)}{+}\tilde{p}_3{-}\text{i}\nu)(m_4{+}n_{3,4}^{(1)}{+}p_4{+}\text{i}\nu)}
\end{aligned}
\end{equation}
with basic building blocks from nodes as
\begin{equation}
\begin{aligned}
   &\mathcal{V}_{1}=(w_1{+}k_{1,2})^{-\tilde{m}_i{-}\sum_i(n_{i,i{+}1}^{(1)}+n_{i,i{+}1}^{(2)})-\tilde{p}_1}, \mathcal{V}_2=(w_2{-}k_{1,2}{-}k_{2,3})^{m_2}\\
   &\mathcal{V}_3=(w_3{+}k_{2,3}{+}k_{3,4})^{m_3},\ \ \mathcal{V}_4=(w_4{-}k_{3,4})^{m_4}
   \end{aligned}
\end{equation}
While for the four-site star graph, its primary solution $\mathbf{T}_{0,0,0,0,0,0}^{(b)}$ reads
\begin{equation}
    \frac{\phi_{n_{i,j}^{(s)},m_i}\Gamma(\tilde{m}_2{+}\sum_{i,j}(n_{i,j}^{(1)}{+}n_{i,j}^{(1)}){+}\tilde{p}_1)\mathcal{E}_{1,2}\mathcal{E}_{2,3}\mathcal{E}_{2,4}\mathcal{V}_1 \mathcal{V}_2 \mathcal{V}_3 \mathcal{V}_4}{(\tilde{m}_2{+}n_{1,2}^{(2)}{+}\sum_{i=3}^4(n_{2,i}^{(1)}{+}n_{2,i}^{(2)}){+}\tilde{p}_2{-}\text{i}\nu)(m_3{+}n_{2,3}^{(1)}{+}p_3{+}\text{i}\nu)(m_4{+}n_{2,4}^{(1)}{+}p_4{+}\text{i}\nu)}
\end{equation}
with all nodes factors
\begin{equation}
\begin{aligned}
   &\mathcal{V}_{1}=(w_1{-}k_{1,2})^{-\tilde{m}_i{-}\sum_i(n_{i,i{+}1}^{(1)}+n_{i,i{+}1}^{(2)})-\tilde{p}_1}, 
   \ \mathcal{V}_3=(w_3{-}k_{2,3})^{m_3}\\
   &\mathcal{V}_2=(w_2{+}k_{1,2}{+}k_{2,3}{+}k_{2,3})^{m_2},\ \ \mathcal{V}_4=(w_4{-}k_{2,4})^{m_4}
   \end{aligned}
\end{equation}

\bibliographystyle{utphys} 
\bibliography{ref.bib}

\end{document}